\documentclass[twocolumn, prb, showpacsm, floatfix]{revtex4-1}

\usepackage{graphicx}
\usepackage{dcolumn}
\usepackage{bm}
\usepackage{physics}
\usepackage{verbatim}
\usepackage{amsmath}
\usepackage{mathtools}
\usepackage{placeins}
\usepackage[usenames,dvipsnames]{color}

\begin{document}

\title{Dynamical configuration interaction: Quantum embedding that combines wave functions and Green's functions}

\author{Marc Dvorak}
\email{marc.dvorak@aalto.fi}
\affiliation{Department of Applied Physics, Aalto University School of Science, 00076-Aalto, Finland}
\author{Patrick Rinke}
\affiliation{Department of Applied Physics, Aalto University School of Science, 00076-Aalto, Finland}

\date{\today}

\newcommand{\sd}{$\mathcal{D}\;$}
\newcommand{\sr}{$\mathcal{R}\;$}
\newcommand{\md}{\mathcal{D}}
\newcommand{\mr}{\mathcal{R}}

\let\oldhat\hat
\renewcommand{\hat}[1]{\mathbf{\oldhat{\text{$#1$}}}}
\let\oldtilde\tilde
\renewcommand{\tilde}[1]{\mathbf{\oldtilde{\text{$#1$}}}}

\newcommand{\id}{\mathbf{I}}

\definecolor{aaltoOrange}{RGB}{255,121,0}%

\newcommand{\PRc}[1]{\textcolor{aaltoOrange}{{\bf PR: #1 }}}

\newcommand{\PR}[1]{\textcolor{blue}{{\bf PR: #1 }}}

\begin{abstract}
We present the concept, derivation, and implementation of dynamical configuration interaction, a quantum embedding theory that combines Green's function methodology with the many-body wave function. In a strongly-correlated active space, we use full configuration interaction (CI) to describe static correlation exactly. We add energy dependent corrections to the CI Hamiltonian which, in principle, include all remaining correlation derived from the bath space surrounding the active space. Next, we replace the exact Hamiltonian in the bath with one of excitations defined over a correlated ground state. This transformation is naturally suited to the methodology of many-body Green's functions. In this space, we use a modified $GW$/Bethe-Salpeter equation procedure to calculate excitation energies. Combined with an estimate of the ground state energy in the bath, we can efficiently compute the energy dependent corrections, which correlate the full set of orbitals, for very low computational cost. We present dimer dissociation curves for H$_2$ and N$_2$ in good agreement with exact results. Additionally, excited states of N$_2$ and C$_2$ are in excellent agreement with benchmark theory and experiment. By combining the strengths of two disciplines, we achieve a balanced description of static and dynamic correlation in a fully \textit{ab-initio}, systematically improvable framework.
\end{abstract}

\maketitle



Accurate theoretical predictions for systems with strongly-correlated electrons remains one of the major challenges in condensed matter physics and quantum chemistry. Even though this topic has been intensely studied for decades, research continues because of its great importance to both fundamental physics and technological applications. In principle, quantum many-body theory\cite{fetter_quantum,altland_field_theory}  allows one to study electronic properties at the most fundamental level. Of course, the exact theory is hopelessly complex for realistic systems. Developing less expensive, approximate methods which predict spectra of correlated systems is therefore essential for the theoretical discovery of new phenomena or new materials. 

Of particular interest in condensed matter physics are systems which couple a small number of strongly-interacting electrons to a much larger bath of weakly-interacting electrons. $d$- or $f$-electron atoms on surfaces  or point defects in solids \cite{bockstedte_npjqm_3} are possible examples of such impurities. In methods designed for impurity problems, the important impurity states are described with a theory that is much more accurate and computationally expensive than the theory treating the bath. Strongly-interacting electrons on the impurity are referred to as statically correlated, which means they are energetically near each other, while the continuum of states in the bath is dynamically correlated. The multi-reference character of open-shell molecules in quantum chemistry also requires a balanced treatment of static and dynamic correlation. Developing a theory which can treat both regimes of correlation on equal footing within one unified framework is difficult.

Such impurity problems are so extensively studied and present such a major reduction in computational cost that it can be advantageous to construct an artificial impurity from an otherwise homogeneous system. This is the central idea behind quantum embedding or active space (AS) theories $-$ an effective impurity is selected from the complete system and treated with high accuracy. If the relevant physics are determined primarily by the impurity states, then such a partitioning gives a good overall description of the system. $d$-electron levels in solids or a chemical active space in an otherwise uniform molecule can be considered an effective impurity cut out from a homogeneous system. With a sensible and physically motivated choice of active space, the impurity concept can be applied to any system, not only those with an obvious physical defect. Quantum embedding theories\cite{chan_acr_49} are therefore powerful methods for strongly-correlated electrons.

By construction, electrons in the impurity and bath exist in different regimes of correlation: static and dynamic. Static correlation on the impurity is best described by brute force exact diagonalization of the many-body Hamiltonian. While the method describes all levels of static correlation, it is far too expensive for large systems. Correlation among high energy degrees of freedom in the bath is described efficiently by the many-body Green's function (GF). When the long-lived quasiparticle concept applies, approximate Green's function methods can be applied to much larger systems than wave function methods and give excellent results for comparatively low cost. 

In this work, we derive and test a new quantum embedding theory that treats static correlation with the electronic wave function (WF) and dynamic correlation with the many-body Green's function. Unlike previous embedding theories which treat a single central object (either the WF or GF) described at different levels of theory in each space, we consider different quantities in either space. The conceptual key to our approach is to selectively rewrite the many-body Hamiltonian, depending on the chosen perspective of the physics. The direct construction of the Hamiltonian in the WF picture is based on bare electrons in the vacuum. An equivalent representation is to describe excitations above a correlated ground state. By choosing different representations of the Hamiltonian in either space, we can choose to work with either the WF or GF.

\section{Introduction}
\subsection{Brief survey of current methods}
We very briefly introduce many-body WF and GF methods that are relevant to the current work. Many modern approaches to strong correlation in physics study the many-body Green's function.\cite{nolting_quantum,fetter_quantum} We refer to such applications of quantum field theory in condensed matter as many-body perturbation theory (MBPT), which is distinct from the chemists' MBPT based on perturbation in the residual potential  (as in M{\o}ller-Plesset perturbation theory or related methods). The GF is a direct link to spectral information about the system. Particle addition/removal spectra, as well as the neutral excitation spectrum, are accessible by solving for the single-particle $G$ or two-particle $L$. In contrast, quantum chemistry focuses on efficient approximations to the many-body wave function. All of the same spectral information as with GFs is accessible if one knows the many-body wave functions for all eigenstates of the system.

The $GW$ approximation \cite{hedin_pr_139,aryasetiawan_rpp_61,louie_prb_34} and its extension to optical properties, the Bethe-Salpeter equation \cite{salpeter_pr_84,rohlfing_prb_62,blase_csr_47,albrecht_prl_80} (BSE), are very successful GF theories for predicting spectra of weakly- to moderately-correlated systems. $GW$ and the common implementation of the BSE have their limitations, however. BSE with a static kernel cannot describe multiple excitations.\cite{marini_jcp_134} BSE also has a tendency to underestimate triplet excitation energies and suffers from self-screening error.\cite{bruneval_jcp_142} There is not yet a widely adopted route to improve these shortcomings of the BSE, though developing extensions to both $GW$ and BSE is a very active area of research.\cite{romaniello_jcp_130,marini_jcp_134,attaccalite_prb_84,zhang_jcp_139,lischner_prl_109,kresse_jctc_13,kresse_prb_85,louie_prb_94,pavlyukh_prl_117,isobe_pra_97,kwahara_prb_94,gruneis_prl_112,romaniello_jcp_131,peng_jctc_14,berkelbach_jctc_14}

Dynamical mean-field theory (DMFT) is a quantum embedding theory based on the electronic GF. In combination with the local density approximation (LDA), LDA+DMFT \cite{georges_prb_45,held_jpcm_20,kotliar_rmp_78} has been very successful in predicting spectral properties of strongly-correlated solids. By including cluster extensions to DMFT,\cite{kotliar_prl_87,maier_rmp_77} one can treat even non-local correlation with high accuracy. However, LDA+DMFT is based on a model Hamiltonian, rather than a true \textit{ab-initio} Hamiltonian, and does not recover the exact Hamiltonian as the embedded region increases in size. The $GW$+DMFT method\cite{biermann_prl_90,tomczak_epl_100,biermann_jpcm_26,boehnke_prb_94} solves many of these issues. Other Green's function embedding methods,\cite{rinke_prb_93,aryasetiawan_prl_102} including self-energy embedding theory \cite{zgid_njp_19,zgid_jcp_143,zgid_prb_91,lan_jctc_12} (SEET), also improve upon the shortcomings of LDA+DMFT. While they are a great success overall, all GF embedding theories have a complicated frequency structure. They require the memory consuming storage of frequency dependent quantities (the GF and self-energy) and the evaluation of difficult frequency integrals.

In contrast with such GF techniques, quantum chemistry methods based on calculating the electronic wave function are free of frequency dependence but suffer from a combinatorial explosion in the basis. \cite{szabo_qchem,helgaker_molecular} High level coupled cluster \cite{loos_jctc_14} or configuration interaction (CI), including their multi-reference variants,\cite{evangelista_jcp_134,szalay_cr_112} are very accurate but cannot be applied to systems larger than a few electrons. A family of multi-configuration self-consistent-field methods (MC-SCF), including the complete active space self-consistent-field method (CASSCF),\cite{olsen_ijqc_111,alavi_jctc_12} simultaneously optimize single-particle orbitals to a multi-configurational wave function at the same time as diagonalizing the many-body Hamiltonian. CASSCF optimizations are state dependent and include dynamic correlation beyond the orbital active space, which itself is limited to $\sim \hspace{-0.1cm} 20$ orbitals.

Perturbation theory applied to either a single- or multi-reference wave function can also give excellent results for strongly-correlated molecules. Methods based on the Rayleigh-Schr\"odinger (RS) variant of perturbation theory avoid the explicit energy dependence of the perturbation expansion, a feature common to GF methods or Brillouin-Wigner techniques, but suffer from their own issues like intruder states.\cite{choe_jcp_114} The complete active space perturbation theory to second order (CASPT2) method (and similar methods based on restricted active spaces\cite{ma_jctc_12,gagliardi_jcp_134,gagliardi_jctc_9}) has enjoyed widespread success as a reasonable compromise between computational cost and accuracy. CASPT2 begins with orbital optimization simultaneous with full configuration interaction in an AS, followed by perturbative corrections to second order.  Confusingly, these methods are commonly referred to as MBPT in the quantum chemistry community but are different formalisms than quantum field theory, which is commonly called MBPT in condensed matter physics.

\subsection{Motivation}
Our goal is to construct a quantum embedding theory which combines the best features of quantum chemistry, GF embedding, and $GW$/BSE theory. To recover the exact Hamiltonian as the embedded region increases in size, we use exact diagonalization (ED) or CI for the electronic wave function in the embedded region, which we denote $\md$. Both ED and CI are based on the same, exact Hamiltonian, with their only difference being the choice of basis. With this choice, the accuracy of the theory can be systematically improved by increasing the size of $\md$. In practice, a large embedded region is numerically intractable, but this limit is a useful theoretical consideration to test the theory. We must also choose a theory in which to embed the wave function calculation. Motivated by the success of GF theories for weakly-correlated systems, we describe surrounding degrees of freedom with the many-body GF. Because the WF and GF are fundamentally different quantities, the embedding framework must serve as a bridge between two disciplines that are essentially disconnected.



\begin{figure}
\begin{centering}
\includegraphics[width=0.8\columnwidth]{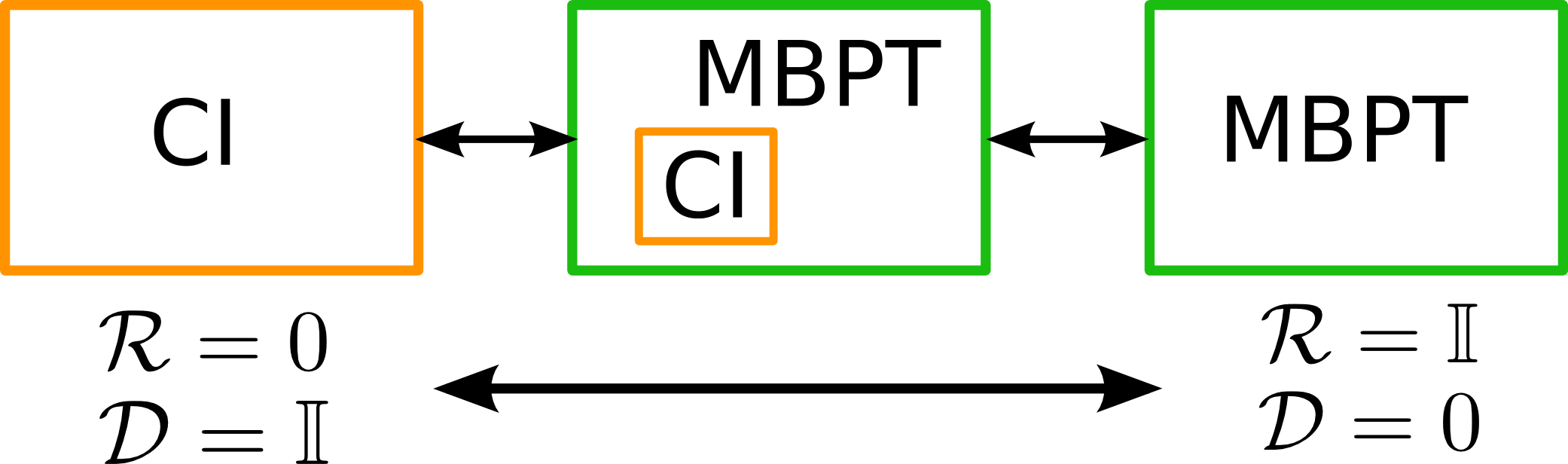}
\caption{A schematic showing the proposed embedding theory as an interpolation between two exact theories, CI and MBPT. At intermediate sizes of $\md$, a configuration interaction problem is embedded inside of MBPT.\label{embedding}}
\end{centering}
\end{figure}

At a high, conceptual level, such an embedding scheme is shown in Fig. \ref{embedding}. By changing the size of the embedded region, one can interpolate between a normal CI calculation ($\md = \id$) or a standard calculation with MBPT ($\md = 0$). In these two limits, we enforce that the embedding must match these two theories. This requirement is the major guiding principle behind our approach. At intermediate sizes of $\md$, a CI problem is embedded inside of MBPT. Ideally, the final result should be invariant to the choice of embedded region. This is extremely difficult in practice but is another useful consideration for constructing the theory.

The choice of embedding CI inside of MBPT is motivated by their different strengths and weaknesses. High energy degrees of freedom in $\mr$, the complementary space to $\md$, are dominated by dynamic correlation which is included even with simple approximations in MBPT. Such algorithms, including those based on the $GW$ approximation, include some dynamic correlation with polynomial scaling and are size consistent. These methods include correlation among the full set of orbitals through sums over intermediate states in the diagrammatic expansion. CI, however, very efficiently captures static correlation among degenerate configurations or low energy degrees of freedom with a single matrix diagonalization. The CI Hamiltonian is also frequency independent, making it conceptually and computationally simpler than frequency dependent quantities in MBPT. Such correlation is also accessible with Green's function methods, in principle, but requires a challenging self-consistent solution with difficult frequency integrals. A sensible partitioning of the Hamiltonian with an embedding theory could capture most static correlation in a system with a modest sized CI calculation and avoid the difficulties of describing static correlation with the GF. Motivated by these considerations, our goal is to treat static correlation with CI and dynamic correlation with MBPT.

To clearly distinguish our theory from previous work, we point out that we do not use CI as a high accuracy calculation of the self-energy or vertex in $\md$.\cite{zgid_prb_86,pavlyukh_prb_75} Our goal is to keep the electronic WF in \sd and never compute a vertex in the strongly-correlated subspace. Treating an embedded vertex in \sd would introduce the issues we are trying to avoid $-$ storage of the GF and vertex, frequency dependent impurity solvers, etc. Our theory is also not an application of perturbation theory based on the residual potential $U=v-v^{\mathrm{MF}}$ for mean-field potential $v^{\mathrm{MF}}$.

\section{Theory}
\subsection{Subspace partitioning}
In this work, the relevant Hamiltonian is the non-relativistic, Born-Oppenheimer electronic Hamiltonian,
\begin{equation}
H = \sum_{ij}^N t_{ij} a_{i}^{\dagger} a_j + \frac{1}{2} \sum_{ijkl}^N v_{ijkl} a_{i}^{\dagger} a_{j}^{\dagger} a_l a_k. \label{secondquant}
\end{equation}
for one-body matrix elements $t_{ij}=\bra{i}\frac{-\nabla^2}{2}+U_{\mathrm{ext}}(\mathbf{r})\ket{j}$ and Coulomb matrix elements $v_{ijkl}=\bra{ij}v(\mathbf{r},\mathbf{r'})\ket{kl}$ for two-body interaction $v(\mathbf{r},\mathbf{r}')$. The Hilbert space for this Hamiltonian is all Slater determinants generated from orbitals of the non-interacting Hamiltonian.

Partitioning the Hilbert space is intrinsic to any embedding theory. Because Eq. \ref{secondquant} is defined in a many-body Hilbert space, we first partition the many-body space and then make a connection to the single-particle picture. We define projection operators \sd and \sr that project onto the strongly- and weakly-correlated portions of the many-body Hilbert space.
\begin{equation}
\md = \sum_{I} \ket{I} \bra{I}  \;\; ; \;\; \mr = \sum_{J} \ket{J} \bra{J} \;\; ; \;\;   \id = \md + \mr.
\end{equation}

We consider only configurations $\ket{I}$ and $\ket{J}$ with fixed particle number $N$. We define configurations in \sd as the low-energy excitations of the system. These are most easily defined in the single-particle picture with an orbital active space. The AS can be constructed with an energy cutoff around the Fermi energy (above and below) or based on chemical intuition of the problem. An excitation that promotes any number of AS occupied orbitals to AS virtual orbitals is considered low-energy, and its corresponding $N$-particle configuration is in $\md$. Note that this definition includes all excitation levels, $\ket{S_{\md}},\ket{D_{\md}}, \ket{T_{\md}},...$, for configurations with single, double, triple, and all higher excitations in $\md$. In quantum chemistry terms, \sd is generated as in complete active space (CAS) theories. \sd and the orbital AS are related but not identical: \sd is a space of many-body configurations, while the AS is a collection of single-particle states.

All other configurations are placed in $\mr$. Our separation implies that transitions which mix AS orbitals with orbitals outside the AS are placed in $\mr$. Different projection techniques to define the AS and \sd are a topic for further study. Defining \sd based solely on low-energy transitions is not mandatory. Here, we focus on the formalism instead of details of the projection method.
\begin{figure}
\begin{centering}
\includegraphics[width=0.9\columnwidth]{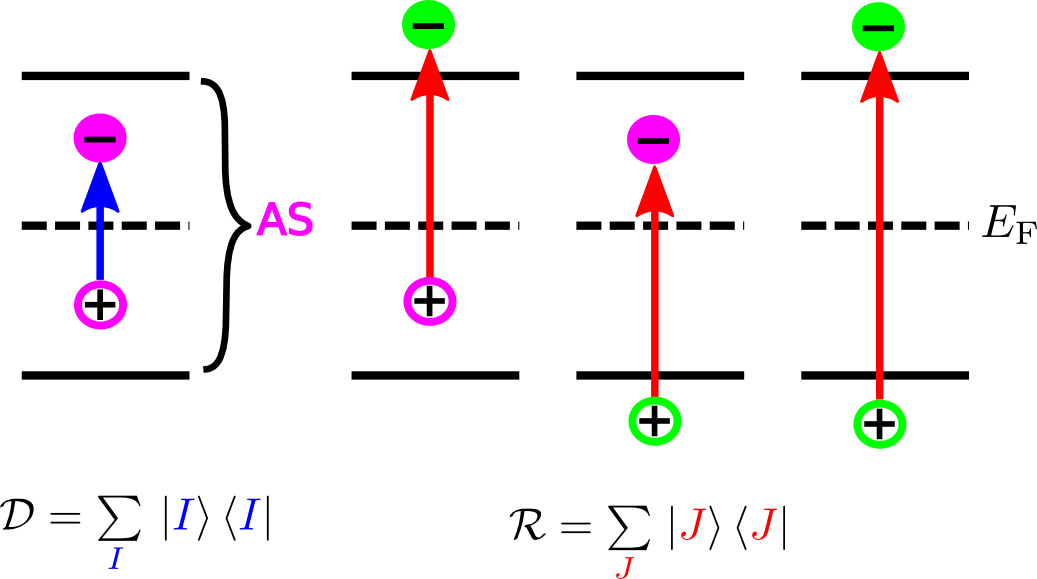}
\caption{Orbitals that fall within an energy cutoff around $E_{\mathrm{F}}$ are placed in an orbital AS, shown here in magenta. Configurations with AS excitations, shown in blue, belong to the many-body subspace $\md$. Configurations that contain any other transition belong to $\mr$, shown in red. We place configurations of the mixed transition type from AS to outside AS in $\mr$. Excitations include all excitation levels as in complete active space (CAS) theories of quantum chemistry.\label{configurations}}
\end{centering}
\end{figure}

There are two different classifications shown in Fig. \ref{configurations}: one defining the single-particle AS and one based on configurations. In this partitioning, one cannot say if any given \textit{orbital} belongs to \sd or $\mr$ as one might do when embedding the single-particle GF. In this projection scheme, \sr can be considered a high energy bath, though the bath is a set of many-particle configurations or transitions, not individual orbitals.

\begin{figure}
\begin{centering}
\vspace{0.5cm}
\includegraphics[width=1.0\columnwidth]{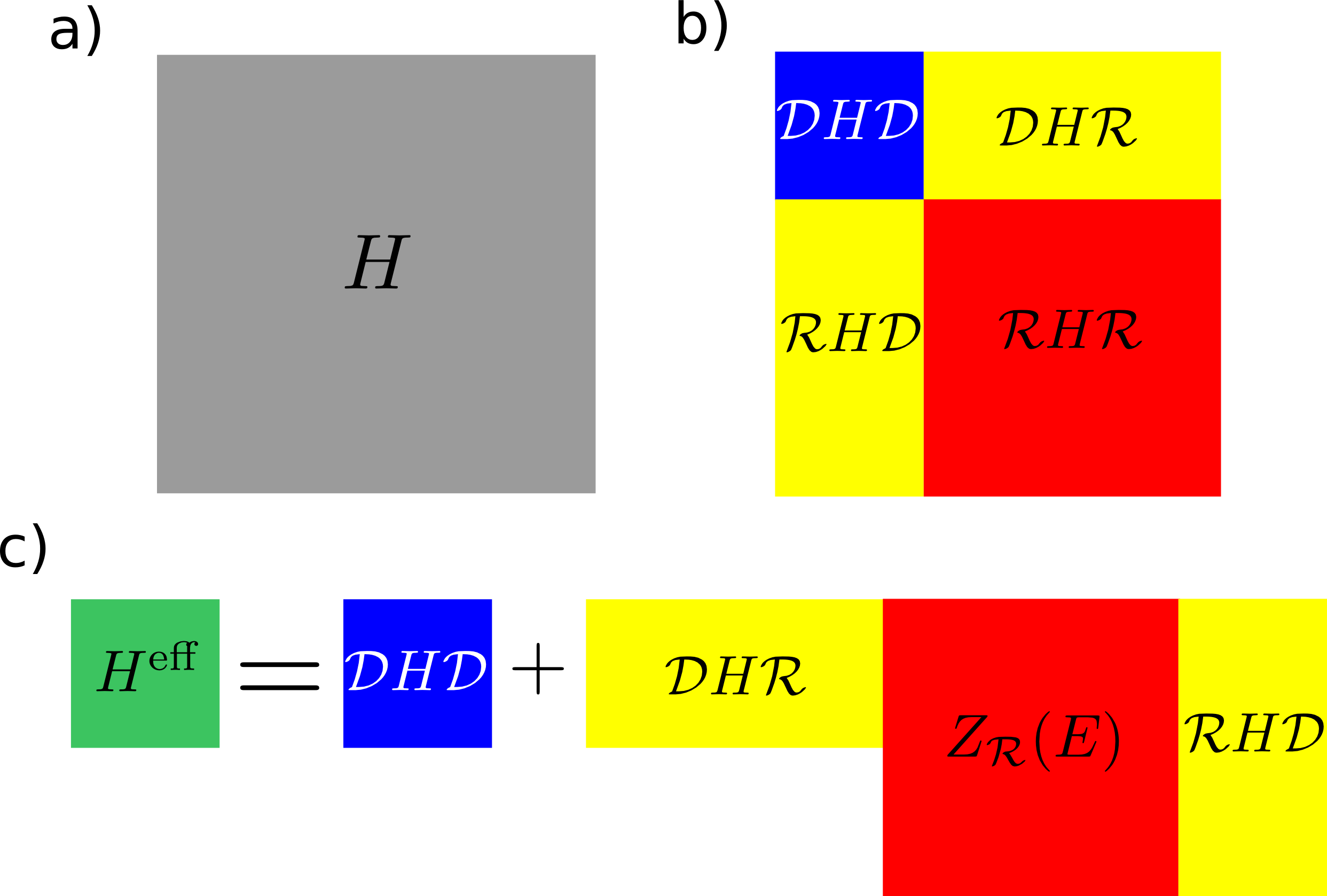}
\caption{Schematic showing the block structure of the Hamiltonian and the matrix multiplication involved in the downfolding. The full Hamiltonian (a) is partitioned into spaces \sd and \sr (b). The effect of the L\"owdin downfolding is to add an energy dependent correction to $\md H \md$ (c). The resulting effective Hamiltonian is of size $\dim \md$. In practice, \sr can be several orders of magnitude larger than $\md$.\label{downfolding}}
\end{centering}
\end{figure}

We use the Feshbach-Schur decomposition,\cite{dzuba_pra_54,dzuba_pra_95,pavlyukh_prb_91} shown in Fig.~\ref{downfolding}, to transform the exact Hamiltonian to an effective, downfolded Hamiltonian in the \sd space. This decomposition is also known as the L\"owdin projection.\cite{gagliardi_jcp_134,gagliardi_jctc_9,lowdin_jmp_3,lowdin_ijqc_2} The Schr\"odinger equation can be written in block form based on the two projectors
\begin{equation}
\begin{bmatrix}
\md H \md & \md H \mr  \\
\mr H \md & \mr H \mr 
\end{bmatrix}
\begin{bmatrix}
\phi \\
\chi
\end{bmatrix}
=E
\begin{bmatrix}
\phi \\
\chi
\end{bmatrix} \, .
\end{equation}
Here, $\md H \md$ is the Hamiltonian projected only into the \sd space. This block of the Hamiltonian corresponds to the frozen core approximation in configuration interaction, where \sr excitations are completely ignored. The component of any given eigenstate of the full $H$ in the \sd space, $\phi$, depends on the solution in \sr through the offdiagonal blocks of the Hamiltonian. The energy $E$ is the total electronic energy of the system. The second line of the equation reads
\begin{equation}
\left[ \mr H \md \right] \phi + \left[ \mr H \mr \right] \chi = E \chi.
\end{equation}
We can solve this equation for the \sr solution, $\chi$, and insert it into the first line of the Schr\"odinger equation.
\begin{eqnarray}
Z^{\mr}(E) &=& \frac{1}{E - \mr H \mr}   \nonumber  \\
\chi &=& Z^{\mr}(E) \left[ \mr H \md \right] \phi    \label{resolvent}
\end{eqnarray}
We have introduced the resolvent defined by Eq.~\ref{resolvent}, $Z^{\mr}(E)$. We insert Eq.~\ref{resolvent} for $\chi$ into the first line of the Hamiltonian,
\begin{eqnarray}
M(E) &=& \left[ \md H \mr \right] Z^{\mr}(E) \left[ \mr H \md \right]   \nonumber  \\
H^{\mathrm{eff}}(E) \phi &=& \left[ \md H \md + M(E) \right] \phi = E \phi.     \label{effective}
\end{eqnarray}
The downfolding procedure therefore gives an effective Hamiltonian $H^{\mathrm{eff}}$ of the size $\dim\md$. Eq. \ref{effective} is the effective Schr\"odinger equation in the \sd space and is an exact rewriting of the eigenvalue problem. While the size of the matrix to diagonalize is much smaller than the original $H$, $H^{\mathrm{eff}}$ is now energy dependent, and the energy eigenvalue $E$ must be found self-consistently. The exact solution (equivalent to full CI or ED) requires inverting the $\mr H \mr$ block of the Hamiltonian to compute $Z^{\mr}(E)$. This is an extremely large space for any realistic problem and the inversion is numerically intractable. The remainder of our theory is to find a suitable approximation for $\mr H \mr$ to simplify the inversion.

In matrix notation, $H^{\mathrm{eff}}$ is constructed with the matrix elements
\begin{eqnarray}
Z_{\mr,JJ'} (E) &=& \left[ E \; \delta_{JJ'} - \bra{J} H \ket{J'} \right]^{-1}  \nonumber  \\
M_{II'} (E) &=& \sum_{J,J' \in \mr} \bra{I} H \ket{J} Z_{\mr,JJ'} (E) \bra{J'} H \ket{I'}  \nonumber  \\
H^{\mathrm{eff}}_{II'} (E) &=& \bra{I} H \ket{I'} + M_{II'}(E).
\end{eqnarray}
The indices $I$ and $I'$ refer to configurations in $\md$, while $J$ and $J'$ denote \sr configurations. Matrix elements of $H$ are evaluated with the Slater-Condon rules, \cite{slater_pr_34,condon_pr_36,szabo_qchem} which are briefly presented in Appendix \ref{slater_condon}. Based on the Slater-Condon rules, there are certain selection rules for the matrix elements $H_{II'}$ and $M_{II'}$ which depend on the differences in occupation numbers between states $I$ and $I'$. The matrix elements $\bra{I} H \ket{I'}$ are the CI matrix in the frozen core approximation. The elements $M_{II'}$, therefore, contain all the correlation beyond the frozen core approximation and can be considered a dynamical core correction or a configuration self-energy.

\subsection{Combining Green's function- and wave function-based theories}
The most important element of our theory is the combination of different methods based on wave functions and Green's functions. To demonstrate how we connect these two theories, we discuss the separation of any total energy into a ground state and excitation energy. While this is trivial for eigenstates of the full Hamiltonian, we discuss it in detail since subtle aspects of this separation for a subspace of the full Hamiltonian are important for the embedding procedure.

First, consider exactly diagonalizing the full Hamiltonian in Eq.~\ref{secondquant}. For total energy eigenvalues $E_i$ and excitation energies $\Omega_i = E_i - E_0$, the Hamiltonian can be written in the eigenbasis as
\begin{equation}
H = 
\begin{bmatrix}
E_0 & 0 & 0 & 0 \\
0 & E_1 & 0 & 0 \\
0 & 0 & E_i & 0 \\
0 & 0 & 0 & ...
\end{bmatrix}
=
E_0 + 
\begin{bmatrix}
0 & 0 & 0 & 0 \\
0 & \Omega_1 & 0 & 0 \\
0 & 0 & \Omega_i & 0 \\
0 & 0 & 0 & ...  \label{subtract}
\end{bmatrix}.
\end{equation}
We denote the excitation matrix with eigenvalues $\Omega_i$ as simply $\Omega$. The lowest element of $\Omega$ is $0$ for two reasons: the reference energy to define the excitation energies is the correlated ground state, and the correlated ground state is the lowest eigenvalue of $H$.

All of the quantities in Eq. \ref{subtract} are accessible with either WF- or GF-based methods. In WF-based methods, the eigenvalues of the Schr\"odinger equation are total energies, and excitation energies are computed as total energy differences. In contrast with such quantum chemistry methods, MBPT computes excitation energies \textit{directly} as the response of the system to perturbation, e.g., particle addition/removal or optical excitations. In these GF theories, the degrees of freedom are quasiparticles $-$ electrons and holes $-$ propagating over the correlated ground state. The quasiparticle Hamiltonians of many-body Green's functions are effective, particle-like equations of motion with eigenvalues related to excitation energies of the system.

We use the excitation energies as the connection between these two theories. When introducing Eq.~\ref{subtract}, we assumed an excitation matrix computed as total energy differences so that the excitation matrix is $\Omega = H - E_0$. Instead, consider computing the matrix $\Omega$ with GF. If we compute the excitation matrix with GF, we effectively replace bare electronic degrees of freedom with quasiparticles. This transformation is the quasiparticle renormalization of the many-body Hamiltonian,
\begin{equation}
 H \rightarrow H^{\mathrm{QP}} = E_0 + \Omega^{\mathrm{QP}},  \label{transformation}
\end{equation}
where $\Omega^{\mathrm{QP}}$ is now a \textit{quasiparticle} (QP) excitation matrix. Note that the scalar energy on the diagonal, $E_0$, is the same in Eqs. \ref{subtract} and \ref{transformation}. By definition of the correlation functions used in MBPT, excitation energies are also defined with respect to the correlated ground state. The renormalization in Eq. \ref{transformation} is set up for the language and methodology of many-body GF.

Most importantly, the effective quasiparticle Hamiltonian of $\Omega^{\mathrm{QP}}$ is different than the ``bare'' excitation matrix defined by $\Omega = H - E_0$. Quasiparticle Hamiltonians contain sums over occupied and virtual states at intermediate times, the so-called diagrammatic expansion.  For weak coupling ($v_{ijkl}<1$), these expansions involve products of Coulomb matrix elements that effectively weaken the strength of the interaction. Certain classes of diagrams can even be summed to infinite order. The end result is that matrix elements of the quasiparticle Hamiltonian are often weaker than those based on the bare Hamiltonian. Computing $\Omega^{\mathrm{QP}}$ with a quasiparticle Hamiltonian could, therefore, converge faster with respect to the many-body basis or perform better in a diagonal approximation than the bare Hamiltonian.

MBPT does not give exactly the same eigenstates as wave function methods based on diagonalizing Eq.~\ref{secondquant}. Even exact eigenstates of MBPT have a finite lifetime derived from the decay of the bare excitation into many different states. In practice, however, quasiparticle excitations in many systems are rather long-lived. As long as excitations from MBPT have sufficiently long lifetime, it is a safe approximation to replace $\Omega = H - E_0$ with $\Omega^{\mathrm{QP}}$. This is a point we discuss in detail in later sections. Here, we point out that numerous benchmark studies comparing the two methods suggest this replacement is a reasonable approximation in weakly- to moderately-correlated systems.\cite{jacquemin_jctc_11,bruneval_jcp_142,chelikowsky_prb_73,botti_jctc_10}

Now consider diagonalizing \textit{only} the block $\mr H \mr$. This is the block of the Hamiltonian needed for the resolvent, $Z^{\mr}(E)$. The diagonalized $\mr H \mr$ is a \textit{different} matrix than the \sr block of the diagonal matrix $H$. We want to write the projected Hamiltonian $\mr H \mr$ in a way similar to Eq. \ref{subtract}. Working in the eigenbasis, we are free to subtract some scalar energy from the diagonal,
\begin{equation}
\mr H \mr
=
E_0^{\mr} + 
\begin{bmatrix}
\Omega_i^{\mr} & 0 & 0 & 0 \\
0 & \Omega_j^{\mathcal{R}} & 0 & 0 \\
0 & 0 & \Omega_k^{\mathcal{R}} & 0 \\
0 & 0 & 0 & ... \label{matrix}
\end{bmatrix},
\end{equation}
where $\Omega_J^{\mr} \equiv E_J^{\mr} - E_0^{\mr}$. Here, $E_J^{\mr}$ is the $J^{\mathrm{th}}$ eigenvalue of $\mr H \mr$ from exact diagonalization, and $E_0^{\mr}$ is an as-of-yet undefined energy.

Next, we impose physical constraints on the energy $E_0^{\mr}$ and eigenvalues $\Omega_J^{\mr}$. A fundamental requirement of the embedding is to recover a normal MBPT calculation in the limit that $\mr \rightarrow \id$. For this reason, we assign $E_0^{\mr}$ meaning as a ground state energy in \sr and relate $\Omega_J^{\mr}$ to \sr subspace excitation energies. While eigenstates of $\mr H \mr$ are not physical excitations since they exist in only a subspace of the full $H$, they can be connected to a physical excitation by enlarging \sr. This is the defining criterion for the matrix $\Omega^{\mr}$. In the limit $\mr \rightarrow \id$, $\Omega^{\mr}$ must reach the physical excitation matrix $\Omega$. Similarly, the ground state $E_0^{\mr}$ must reach the physical, correlated ground state $E_0$.

We point out that the lowest eigenvalue of $\Omega^{\mr}$, the matrix in Eq. \ref{matrix}, is not zero. Unlike in Eq. \ref{subtract}, even the lowest eigenvalue of $\mr H \mr$ is itself an excited state or, in other words, can be connected to a physical excitation. Even though no \sr eigenstate can be connected to the physical ground state, Eq.~\ref{matrix} is still \textit{exactly} true for $\Omega_J^{\mr} \equiv E_J^{\mr} - E_0^{\mr}$

In Eq.~\ref{matrix}, we now assume that the subspace excitation matrix $\Omega^{\mr}$ is computed with MBPT, $\Omega^{\mr,\mathrm{QP}}$. Assume an exact diagonalization of $\mr H \mr$ to find $E_J^{\mr}$. Then we redefine the ground state energy $E_0^{\mr}$ as the difference between the eigenvalue $E_J^{\mr}$ and the QP excitation energy $\Omega_J^{\mr,\mathrm{QP}}$.
\begin{equation}
E_0^{\mr} \equiv E_J^{\mr} - \Omega_J^{\mr,\mathrm{QP}} \label{e0rdefinition}
\end{equation}
Equation~\ref{e0rdefinition} is useful to define the problem, but computing $E_0^{\mr}$ this way requires the exact diagonalization of $\mr H \mr$, which is extremely expensive. To avoid this expense, we must develop a different strategy to compute $E_0^{\mr}$. From now on, we assume $\Omega^{\mr}$ is computed with MBPT and drop the $\mathrm{QP}$ label. If we can compute the ground state energy $E_0^{\mr}$ and excitation matrix $\Omega^{\mr}$ separately, and for less expense than exact diagonalization, we can assemble them to rewrite the \sr Hamiltonian.

The goal of our theory is to perform the same quasiparticle renormalization as in Eq. \ref{transformation} for only the weakly-correlated subspace of the full Hilbert space.
\begin{equation}
 \mr H \mr \rightarrow H^{\mr} = E_0^{\mr} + \Omega^{\mr}  \label{renormalization}
\end{equation}
The high energy space \sr is dominated by dynamic correlation, which is described very well by MBPT. In $\md$, we describe static correlation with the bare Hamiltonian; in $\mr$, we treat each configuration as a quasiparticle excitation propagating above a ground state. We construct $H^{\mr}$ so that it matches the exact $\mr H \mr$ as closely as possible. $E_0^{\mr}$ is a reference energy to define the subspace QP excitation matrix $\Omega^{\mr}$ \textit{so that both quantities connect to their physical values as $\mr \rightarrow \id$}. 

After computing the renormalized Hamiltonian $H^{\mr}$, we insert it into the denominator of the resolvent in place of $\mr H \mr$ in Eq.~\ref{effective}. We treat $H^{\mr}$ as ``exact'' so that Eq.~\ref{effective} is not solved perturbatively. We do not take the route of Brillouin-Wigner (BW) perturbation theory in the residual operator $U= v - v^{\mathrm{MF}}$ for mean-field potential $v^{\mathrm{MF}}$. Such an order-by-order construction of the resolvent is possible but introduces a difficult energy dependence at every term in the expansion. Nor do we apply the Rayleigh-Schr\"odinger (RS) variant of the perturbation expansion for $Z^{\mr}$. Our procedure to compute \sr excitation energies has its own energy dependence as in BW theory, to be discussed in the next section and in Appendix \ref{resolvents}, but it is a different energy than $E$ entering $Z^{\mr}$. We apply perturbation theory to the subspace problem of diagonalizing $\mr H \mr$, which has a separate energy dependence than $E$. If we can diagonalize $\mr H \mr$ $-$ even by MBPT $-$ the connection between \sd and \sr is automatically set up through the resolvent.

While our framework has similarities to other methods based on the L\"owdin partitioning, this is because there is just one way to exactly partition the many-body Hilbert space. Our concept and final theory are different than past work.\cite{gagliardi_jcp_134,gagliardi_jctc_9,dzuba_pra_95,dzuba_pra_54} To transform the \sr Hamiltonian, we need the excitation matrix $\Omega^{\mr}$ and ground state energy $E_0^{\mr}$.


\subsection{Excitation matrix}
To compute the excitation matrix with GF, there are three conceptual hurdles to overcome. First, we must match the basis for the excitation matrix to the CI basis. The CI Hamiltonian is naturally written up to all excitation levels, while the typical correlation function for neutral excitations is written in a basis of only single excitations. The two basis sets must match for the matrix multiplication of the resolvent $Z^{\mr}(E)$ to be meaningful.

Second, we must eliminate the frequency dependence of the GF. In general, any many-body GF and its equation of motion are frequency dependent. Using an exact GF equation of motion for $\Omega^{\mr}$ would therefore give an energy dependent matrix. However, we want to avoid complicated frequency dependencies, so we must eliminate the frequency dependence in $\Omega^{\mr}$.

Last, we must constrain the calculation of the excitation matrix to the \sr subspace. If one considers exactly diagonalizing $\mr H \mr$, it is clear that this matrix contains only intra-$\mr$ correlation. The corresponding calculation of the excitation matrix $\Omega^{\mr}$ with GF must also include only intra-$\mr$ correlation.

\subsubsection{Definition based on Green's function}
Here, for an easier discussion, we briefly abandon our subspace partitioning to discuss the formalism for the full Hamiltonian. The continuation to a subspace of the Hamiltonian will be discussed later.

We first review the standard theory in MBPT so that our method can be built as an extension to it. We will restrict ourselves to optical excitations in this section for illustrative purposes. The standard approach for computing optical excitations with Green's functions is based on the electron-hole correlation function,\cite{reining_rmp_74,fetter_quantum} $L$, defined as
\begin{eqnarray}
L(1,2 ; 1',2') &=& G(1,1')G(2,2') \label{twoparticleg} \\
&+& \bra{\Psi} T [ \psi(1) \psi(2) \psi^{\dagger}(2') \psi^{\dagger}(1') ] \ket{\Psi}. \nonumber 
\end{eqnarray}
$L$ describes the propagating portion of the two-particle GF, defined by the second line of Eq. \ref{twoparticleg}, and ignores the motion of independent pairs, given by the first line of Eq. \ref{twoparticleg}. Here, $T$ is the time ordering operator and $\ket{\Psi}$ is the $N$-particle ground state. \cite{nolting_quantum} Each number in Eq. \ref{twoparticleg} is a set of spatial, spin, and time coordinates, $1=(\mathbf{r}_1,\sigma_1,t_1)$. In order to approximate exact excitation energies from quantum chemistry, we restrict possible time orderings to simultaneous creation/annihilation of one e-h pair ($t_1=t_{1'}$) and instantaneous annihilation/creation of a second e-h pair ($t_2=t_{2'}$). $L$ can be expanded in a basis of noninteracting electron-hole pairs, also called single excitations, and its equation of motion is determined by the Bethe-Salpeter equation (BSE).\cite{salpeter_pr_84} All excitations are accessible, in principle, by solving the frequency dependent BSE.

While single excitations are clearly described by the correlation function in Eq. \ref{twoparticleg}, excitation energies of exact eigenstates which involve the creation of double or higher noninteracting excitations are hidden in the BSE. These so-called multiple excitations are contained in the vertex function, $\Gamma$, of MBPT. This is most easily demonstrated in the Lehmann representation, in which $L$ is written as
\begin{eqnarray}
L_{SS'}(\omega) &=& \sum_{N \neq 0} \frac{ \bra{\Psi} \widehat{S} \ket{N} \bra{N} \widehat{S}'^{ \dagger} \ket{\Psi} }{ \omega - (E_{N} - E_0) + i \eta} \nonumber \\
&-& \sum_{N \neq 0} \frac{ \bra{\Psi} \widehat{S}'^{ \dagger} \ket{N} \bra{N} \widehat{S} \ket{\Psi} }{ \omega + (E_{N} - E_0) - i \eta}, \label{fourierl}
\end{eqnarray}
where $\widehat{S}'^{\dagger}$ creates the single excitation $\ket{S'}$ and the sum runs over exact eigenstates $\ket{N}$ of the many-body system. The sum over $\ket{N}$ is independent of the outer indices $S,S'$ $-$ any \textit{single} matrix element of $L$, $L_{SS}$, contains poles at every, exact excitation energy of the system. In principle, one needs only the single, frequency dependent matrix element $L_{SS}$ to access \textit{all} excitations. The excitation energies can be read off as the pole positions of $L_{SS}$. In practice, however, this is extremely difficult. The amplitudes in the numerator of $L_{SS}$, which are the amplitudes for each pole, are extremely small for most eigenstates $\bra{N}$. The overlap between $\bra{N}$ and $S'^{\dagger}\ket{\Psi}$ is appreciable only for the eigenstate $\bra{N}$ which has the same principal configuration as $S'^{\dagger}\ket{\Psi}$, and perhaps a small number of additional $\bra{N}$. This makes it impossible to numerically produce most poles in $L_{SS}$ at the remaining eigenstates with weak amplitude and extract any meaning as they relate to a principal noninteracting transition.

The full e-h correlation function, $L$, has several poles with high amplitudes, with each state $S'^{\dagger} \ket{\Psi}$ having high overlap with roughly one $\bra{N}$. Compared to the single element $L_{SS}$, this makes it much easier to find excitation energies and relate poles in $L$ to a noninteracting transition. Any exact eigenstate $\bra{N}$ of strong single excitation character should have a large Lehmann amplitude for a corresponding noninteracting single transition. However, the correlation function is still in a basis of only single excitations, and excitation energies for states with strong multiple excitation character can still be difficult to compute. For example, the overlap between any $S'^{\dagger}\ket{\Psi}$ with $\bra{N}$ of strong double excitation character will be low, making the pole corresponding to $\bra{N}$ difficult to produce.


If we need excitation energies for multiple excitations, we are free to choose any correlation function which may have stronger amplitudes at the relevant poles. While it is not formally necessary to use a different correlation function to compute multiple excitations, it may make the calculation much easier. Furthermore, we are not interested in the spectrum of the correlation function, only the pole positions. Only the excitation energy is needed for the excitation matrix $\Omega^{\mr}$. This grants us freedom in choosing which correlation function to compute. Just as changing from a single matrix element $L_{SS}$ to the full $L$ makes it easier to find single excitations, we introduce a new correlation function in the $N$-particle space to make it easier to compute multiple excitations. For arbitrary excitation level $m$, we write any excitation in the Hilbert space as a string of $m$ creation and destruction operators acting on the reference configuration. We write configurations as
\begin{eqnarray}
\ket{S} &=& a^{\dagger}_{\alpha} a_{\lambda} \ket{0}    \nonumber  \\
\ket{D} &=& a^{\dagger}_{\alpha} a^{\dagger}_{\beta} a_{\mu} a_{\lambda} \ket{0}   \nonumber \\
\ket{T} &=& a^{\dagger}_{\alpha} a^{\dagger}_{\beta} a^{\dagger}_{\gamma} a_{\nu} a_{\mu} a_{\lambda} \ket{0} 
\end{eqnarray}
for reference configuration $\ket{0}$ and single-particle creation (destruction) operators $a_i^{\dagger}$ ($a_i$). Higher excitation levels follow accordingly. Define the string of excitation operators to create configuration $\ket{J}$, which can take any excitation level, as $\widehat{\Omega}_J^{\dagger}$.
\begin{eqnarray}
\widehat{\Omega}_J^{\dagger} &=& \prod_{\alpha,\beta \in J} a^{\dagger}_{\alpha} a_{\beta}   \nonumber  \\
\ket{J} &=& \widehat{\Omega}_J^{\dagger} \ket{0} \label{operatorstring}
\end{eqnarray}

The set of all $\widehat{\Omega}_J^{\dagger}$ generates the entire $N$-particle Hilbert space by acting on the reference configuration. The time dependence of $\widehat{\Omega}_J^{\dagger}$ is inherited from the time dependence of the Heisenberg operators $a_i^{\dagger}$. We take all creation and annihilation operators in the string for $\widehat{\Omega}_J^{\dagger}$ to act at the same time. In this way, $\widehat{\Omega}_J^{\dagger}$ instantaneously creates the many-body configuration $\ket{J}$, and $\widehat{\Omega}_j$ instantaneously annihilates it. We use $\Omega_J$ without the operator hat to denote the excitation energy for the creation process $\widehat{\Omega}_J^{\dagger}$.

We consider a new correlation function $\mathcal{L}$, defined in the $\{ J,J' \}$ representation and related to an $N$-particle GF,
\begin{equation}
\mathcal{L}_{JJ'}(t ;t') = \bra{\Psi} T [ \; \widehat{\Omega}_J \; \widehat{\Omega}_{J'}^{\dagger} \; ] \ket{\Psi} + \mathcal{G}_0 \label{bigl}
\end{equation}
for ground state $\ket{\Psi}$. If we expand $\mathcal{L}$ in a basis of all possible $\widehat{\Omega}_J$ and $\widehat{\Omega}_{J'}^{\dagger}$, the matrix for each time ordering covers the entire $N$-particle Hilbert space. With this definition, the outer lines of $\mathcal{L}$ can be any possible excitation, not only singles. In principle, no excitation is also an allowed state, $\widehat{\Omega}_{\mathrm{I}}^{\dagger} \ket{0} = \ket{0}$, in order to complete the Hilbert space. In Eq. \ref{bigl}, $\mathcal{G}_0$ is our generic notation for removing the non-propagating portion of the GF, $\mathcal{G}_0$, from the full $N$-particle GF, $\mathcal{G}$. This leaves only the time evolving portion or $N$-particle propagator, which we label $\mathcal{L}$. We do not make any connection between $\mathcal{L}$ and experimental spectra, but only use it for an easier calculation of multiple excitations.

Now we return to the embedding problem and consider $\mathcal{L}^{\mr}$, which tracks the propagation of only \sr excitations. Allowing the outer lines of $\mathcal{L}$ to be multiple excitations is more meaningful than just a difference of convention $-$ it allows us to directly compute multiple excitation energies with a frequency independent kernel. The connection between \sd and \sr is through matrix elements of the exact Hamiltonian, $\md H \mr$. These matrix elements are evaluated using the Slater-Condon rules (Appendix \ref{slater_condon}) for all excitation levels. In order to multiply $\md H \mr$ by the resolvent $Z^{\mr}$, they must share a common basis. If we estimate the denominator of the resolvent with MBPT, we must use MBPT to compute \textit{multiple} excitations in an efficient way. Because $\mathcal{L}^{\mr}$ is of dimension $\mathrm{dim} \mr$, every excitation is accessible with a static kernel. This way, the excitation matrix $\Omega^{\mr}$ can be calculated in a static framework, and the matrix multiplication between $\md H \mr$ and $Z^{\mr}$ is meaningful. This connection is essential for the embedding. If we base our MBPT calculation on $L^{\mr}$, which is in the basis of only single excitations, instead of $\mathcal{L}^{\mr}$, the MBPT basis does not match the CI basis. In this case, the energy dependence of $L^{\mr}$ and $H^{\mr}$ is necessary to couple to multiple excitations and greatly complicates the embedding.


Returning to the general $\mathcal{L}$, we define the excitation matrix $\Omega$ as the frequency space equation for $\mathcal{L}$ based on Eq. \ref{bigl}. $\mathcal{L}$ can be rewritten in the Lehmann representation by inserting a complete set of eigenstates and Fourier transforming as
\begin{eqnarray}
\mathcal{L}_{JJ'}(\omega) &=& \sum_{N \neq 0} \frac{ \bra{\Psi} \widehat{\Omega}_J \ket{N} \bra{N} \widehat{\Omega}_{J'}^{\dagger} \ket{\Psi} }{ \omega - (E_{N} - E_0) + i \eta} \nonumber \\
&-& \sum_{N \neq 0} \frac{ \bra{\Psi} \widehat{\Omega}_{J'}^{\dagger} \ket{N} \bra{N} \widehat{\Omega}_{J} \ket{\Psi} }{ \omega + (E_{N} - E_0) - i \eta} \label{lehmann_big_l}
\end{eqnarray}
where the sum runs over all $N$-particle eigenstates of the exact Hamiltonian. $\mathcal{L}$ has poles at the exact excitation energies of the system. This representation is not very useful, however, since one needs the exact ground state $\ket{\Psi}$ and all $N$-particle eigenstates $\ket{N}$ of the system.

In order to actually compute $\mathcal{L}$, we want to set up a perturbation expansion for $\mathcal{L}$ with MBPT. We define $\mathcal{L}_0$ in terms of independently propagating, noninteracting electrons and holes. The time dependence of the individual fermionic operators depends on only the one-body portion of $H$. The individual phases associated with each $a_i^{\dagger}$ and $a_j$ entering $\widehat{\Omega}^{\dagger}_j$ combine so that the entire excitation propagates with an energy given by
\begin{equation}
\Omega_{0,J} = \sum_{e \in \mathrm{els}}^m t_{e} - \sum_{h \in \mathrm{holes}}^m t_{h}.
\end{equation}
for excitation level $m$ and eigenvalues of the one-body Hamiltonian $t_i$. The corresponding $\omega$ representation is
\begin{equation}
\mathcal{L}_{0,JJ'}(\omega) =  \frac{ \delta_{J J'} }{ \omega - \Omega_{0,J} + i \eta} - \frac{ \delta_{J J'} }{ \omega + \Omega_{0,J} - i \eta}. \label{l0}
\end{equation}
The expansion amplitudes for $\mathcal{L}_0$ are all 1 since the non-interacting problem is diagonal in this basis. The one-body eigenvalues $t_i$ are exactly the particle addition/removal energies of the noninteracting system.

\begin{figure}
\begin{centering}
\includegraphics[width=1.0\columnwidth]{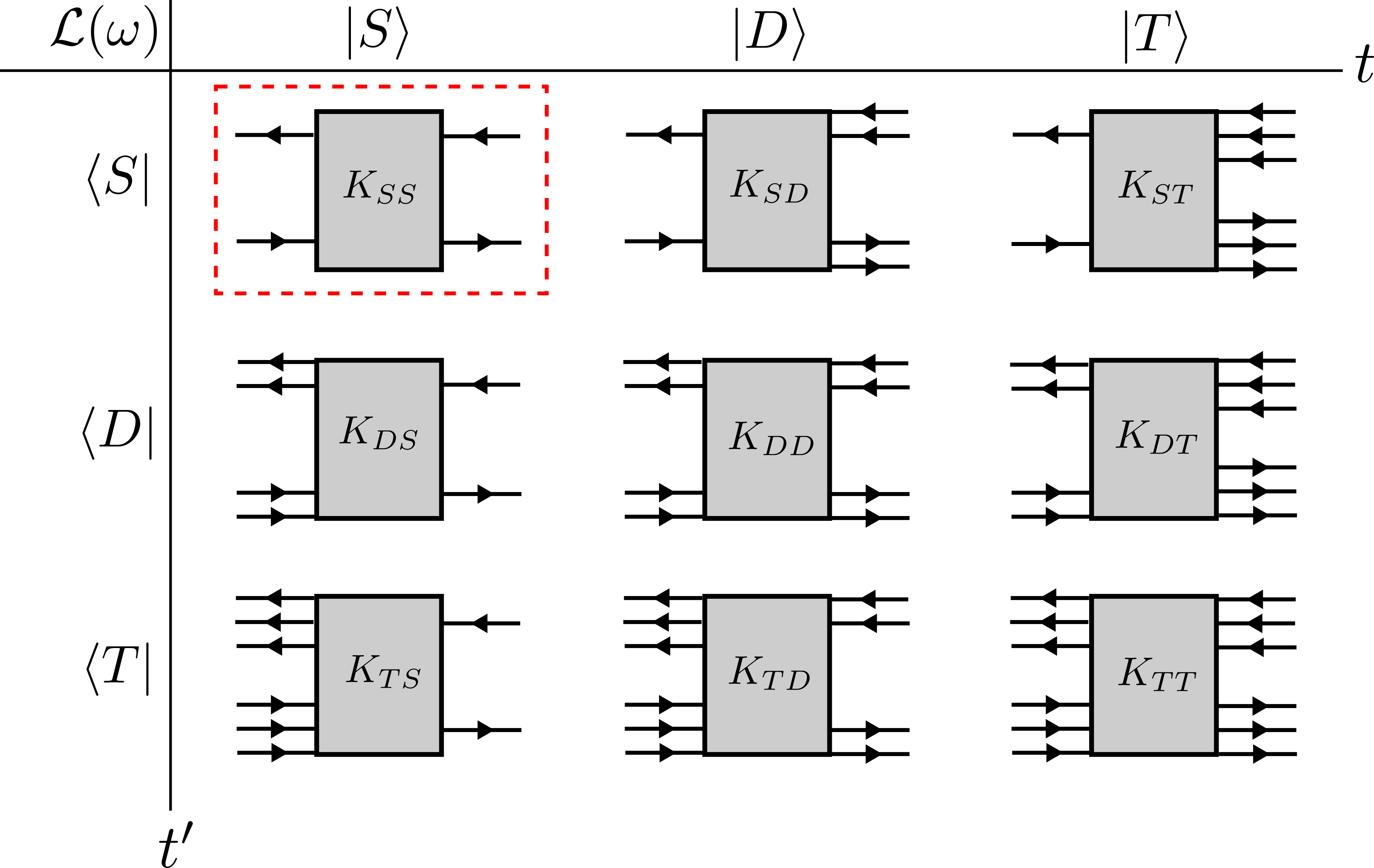}
\caption{Schematic representation of $\mathcal{L}$. In general, both $\mathcal{L}$ and the kernel $K$ are functions of frequency. The reference configuration $\ket{0}$ is also included in the basis for $\mathcal{L}$, though we omit it here for comparison to the subspace $\mathcal{L}^{\mr}$ which contains only excitations. All excitations in $\mathcal{L}$ couple to all others so that the matrix is not sparse. Each block contains both time orderings of $t,t'$. The electron-hole correlation function $L$ is the block outlined in red.\label{exc_mat}}
\end{centering}
\end{figure}

Next, we assume some frequency dependent kernel $K$ which connects the bare $\mathcal{L}_0$ to the full $\mathcal{L}$ as
\begin{eqnarray}
\mathcal{L} = \mathcal{L}_0 + \mathcal{L}_0 K \mathcal{L}_0  \nonumber  \\
\mathcal{L} = \mathcal{L}_0 + \mathcal{L}_0 K^* \mathcal{L} \label{bse}
 \label{series}
\end{eqnarray}
where the kernel $K^*$ is the irreducible version of the full kernel $K$. Diagramatically, we can represent blocks of $\mathcal{L}$ as in Fig. \ref{exc_mat}. For each block of Fig. \ref{exc_mat}, the number of external lines is determined by the excitation levels, $m$ and $m'$, of the basis states. Each block of $K$ is meant to be taken to any desired order in the time evolution operator $U(t,t')$.

$\mathcal{L}$ and $K$ are complicated objects. In this work, our interest in them is mostly utilitarian: to efficiently compute multiple excitations and match the basis for CI. Rather than exploring the exact structure of the kernel $K$ that connects initial and final states in Fig. \ref{exc_mat}, we focus on physical approximations to $K$. We explore the relationships among $\mathcal{L}$, $K$, the vertex function $\Gamma$, and the BSE in future work.

\subsubsection{MBPT for $\mathcal{L}^{\mr}$}
We can now discuss a strategy for calculating $\Omega^{\mr}$, the frequency space equation for the subspace correlation function $\mathcal{L}^{\mr}$. The subspace correlation function of interest is
\begin{equation}
\mathcal{L}_{JJ'}^{\mr}(t ;t') = \bra{\Psi^{\mr}} T [ \; \widehat{\Omega}_J^{\mr} \; \widehat{\Omega}_{J'}^{\mr \dagger} \; ] \ket{\Psi^{\mr}} + \mathcal{G}_0^{\mr} . \label{biglr}
\end{equation}
Here, $\ket{\Psi^{\mr}}$ is a fictitious $N$-particle ground state that contains only intra-$\mr$ correlation $-$ the same ground state discussed previously. The precise meaning of this state is difficult to define since we assume the reference configuration always belongs to $\md$. However, we can still construct an approximate $\mathcal{L}^{\mr}$ with physically motivated approximations and by enforcing the correct limits on $\mathcal{L}^{\mr}$ and $\ket{\Psi^{\mr}}$ as $\mathcal{R} \rightarrow \id$ or $\mr \rightarrow 0$.

The basic idea is to apply many-body perturbation theory for $\mathcal{L}^{\mr}$ in the full many-body basis from the exactly projected Hamiltonian
\begin{eqnarray}
\mr H \mr &=& \mr \Bigg( \sum_{ij}^N t_{ij} a_{i}^{\dagger} a_j \nonumber \\
&+& \frac{1}{2} \sum_{ijkl}^N v_{ijkl} a_{i}^{\dagger} a_{j}^{\dagger} a_l a_k \Bigg) \mr . \label{rhr}
\end{eqnarray}
For this subspace Hamiltonian, each interaction line carries \sr projectors that check the overall $N$-particle configuration. For each new interaction in the perturbation expansion, the configuration must belong to $\mr$, otherwise the entire diagram is killed by the projector. For an interaction at time $t$, we take the two projectors to act at times vanishingly close to $t$, $t \pm \epsilon$ for $\epsilon \rightarrow 0^{+}$. The projectors, therefore, check the configuration of the system just before and just after each interaction. Applying the projectors this way requires us to mix the two conceptual pictures: \sr acts on the $N$-particle configuration, but the diagrammatic expansion for $\mathcal{L}^{\mr}$ only shows GF lines for the excited particles. We only need to check the excitations above the reference configuration to know if a certain diagram/configuration belongs to $\mr$, so the two pictures agree.

\begin{figure}
\begin{centering}
\includegraphics[width=1.0\columnwidth]{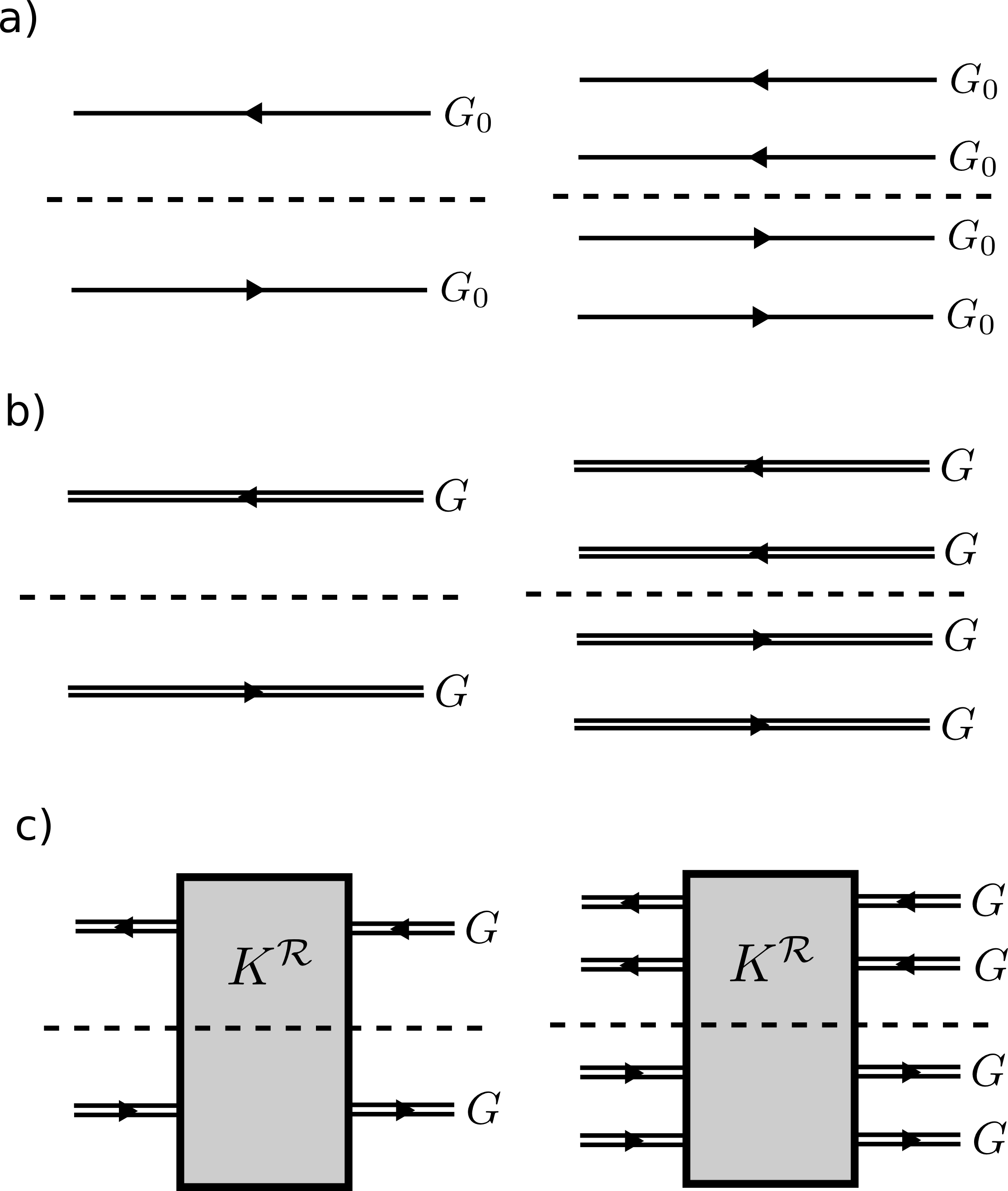}
\caption{Conceptual hierarchy of diagrams approaching the true $\mathcal{L}^{\mr}$ for single (left) and double (right) excitations. Here, outer lines can only belong to \sr excitations. Non-interacting particles (a) contributing to the excitation are given a self-energy (b), then allowed to interact with each other (c). The dashed line separate electrons from holes for the shown time ordering.\label{propagators}}
\end{centering}
\end{figure}

To $0^{\mathrm{th}}$ order in the interaction, the excitation $\widehat{\Omega}^{\mr \dagger}_j$ propagates indefinitely. This is a terrible approximation since it neglects all electron-electron interactions. Diagrammatically, each particle participating in the excitation is described by a noninteracting single-particle GF line, $G_0$. The sum of $G_0$ lines for the excitation $\widehat{\Omega}_j^{\mr \dagger}$ gives the bare $\mathcal{L}_0^{\mr}$, as shown in Fig. \ref{propagators}a. $\mathcal{L}_0^{\mr}$ is a diagonal matrix, as in Eq. \ref{l0}.

To improve this estimate, we allow each propagating particle to gain a self-energy describing its interaction with other electrons. At this level, we assume that $\mathcal{L}^{\mr}$ is still separable between different particles. Each excited particle is now described by a dressed $G$. Diagrammatically, $G$ lines do not connect to each other. Off-diagonal elements of the self-energy allow any given electron or hole to decay into any other \textit{within} a block for a fixed excitation level. However, blocks of $\Omega^{\mr}$ describing different excitation levels are still uncoupled so that the kernel is block diagonal. At this level, $\mathcal{L}^{\mr}$ is approximated by $2m$ separate dressed $G$ lines ($m$ electrons and $m$ holes) for excitation level $m$.

Finally, we allow propagating quasiparticles to interact with each other. For diagrams describing their interaction, the overall configuration must at all times belong to $\mr$. In principle, all possible time orderings and interactions among quasiparticles are allowed $-$ this generates the exact kernel $K^{\mr}$. We relate the full $\mathcal{L}^{\mr}$ to the bare $\mathcal{L}_0^{\mr}$ by some kernel $K^{\mr *}$ as
\begin{equation}
\mathcal{L}^{\mr} = \mathcal{L}_0 + \mathcal{L}_0^{\mr} K^{\mr *} \mathcal{L}^{\mr}  \label{rkernel}
\end{equation}
in analogy with Eq. \ref{series}. Here, internal frequency integrals are implied. As \sr grows to contain the entire Hilbert space, $\mr \rightarrow \id$, the subspace kernel $K^{\mr *}$ must become the full kernel $K^{*}$. Here, $K^{\mr *}$ contains both levels of improvement over $\mathcal{L}_0^{\mr}$ discussed above $-$ the dressing of bare particles by a self-energy and the interactions among all dressed $G$ lines. At this level of theory, an initial excitation of level $m$ is allowed to decay through all possible decay channels to excitation level $m'$, as in Fig. \ref{exc_mat}. The full matrix is not sparse, though elements which couple different excitation levels are expected to be small. The essential point of this subsection is that MBPT for $\mathcal{L}^{\mr}$ must begin from Eq. \ref{rhr}.

After Fourier transforming $\mathcal{L}^{\mr}$ in Eq. \ref{biglr}, it gains a frequency dependence, $\mathcal{L}^{\mr}(\omega^{\mr})$. In analogy with other subspace quantities, the frequency $\omega^{\mr}$ is not related to a physical time or energy. $\omega^{\mr}$ is the frequency variable for the Fourier transform of Eq. \ref{biglr} and defined by the auxiliary eigenvalue problem of diagonalizing $\mr H \mr$: $\omega^{\mr} \equiv E^{\mr} - E_0^{\mr}$ for the continuous variable $E^{\mr}$.

The meaning of $\omega^{\mr}$ is more obvious by considering the eigenvalues of $\mr H \mr$, as well as the meaning of the ground state defining the correlation function $\mathcal{L}^{\mr}$. In the Lehmann representation for $\mathcal{L}^{\mr}$, only energy differences of the type $\omega^{\mr} - (E_J^{\mr} - E_0^{\mr})$ enter the denominator. Therefore, the energy dependence of the perturbation expansion for $\mathcal{L}^{\mr}$ is not the physical $E$ or $\omega$ in the denominator of $Z^{\mr}$. More discussion along this line is in Appendix \ref{resolvents}. In principle, the time variables in Eq. \ref{biglr} also need to be reinterpreted. In practice, however, $\omega^{\mr}$ can be treated as a generic frequency parameter.

\subsubsection{Approximations and practical implementation for $\Omega^{\mr}$}
Now, we focus on the approximations and practical aspects of computing the excitation matrix $\Omega^{\mr}$.

We entirely skip the $\mathcal{L}_0^{\mr}$ approximation and dress each $G_0$ with a self-energy.  Here, we adopt the $GW$ approximation. In order to constrain the correlation to the \sr subspace, we restrict the polarization to the constrained random phase approximation (cRPA) instead of the full RPA. In the cRPA, only \sr transitions enter the polarization and screen the Coulomb interaction. The cRPA is already well known in strongly-correlated physics for determining effective Hubbard $U$ parameters in GF embedding theories.\cite{aryasetiawan_prb_70,sasioglu_prb_83,biermann_prb_86,biermann_prb_96,werner_prb_91} The resulting interaction $W_{\mr}$ interpolates between the fully screened $W$ and bare interaction $v$ as \sr changes size from $\mr = \id$ to $\mr = 0$. Accordingly, the self-energy $GW_{\mr}$ interpolates between the full $GW$ self-energy and bare exchange in these two limits.

Note that the internal $G$ line in $\Sigma = i GW_{\mr}$ contains all poles. Because $G$ and $W_{\mr}$ exist at the same times, it is sufficient to constrain only $W$ to satisfy the \sr projectors. Essentially, only one $G$ line in the self-energy diagram must be \textit{outside} the AS for the overall configuration to belong to $\mr$. We assign this constraint to the polarization entering the interaction. This construction also depends sensitively on starting from a true $G_0$, which has poles unambiguously assigned to single-particle states, instead of a mean-field $G^{\mathrm{MF}}$. The projectors act on non-interacting states, not mean-field particles, and cannot be applied exactly to a $G^{\mathrm{MF}}$ starting point.
\begin{figure*}
\begin{centering}
\includegraphics[width=1.0\textwidth]{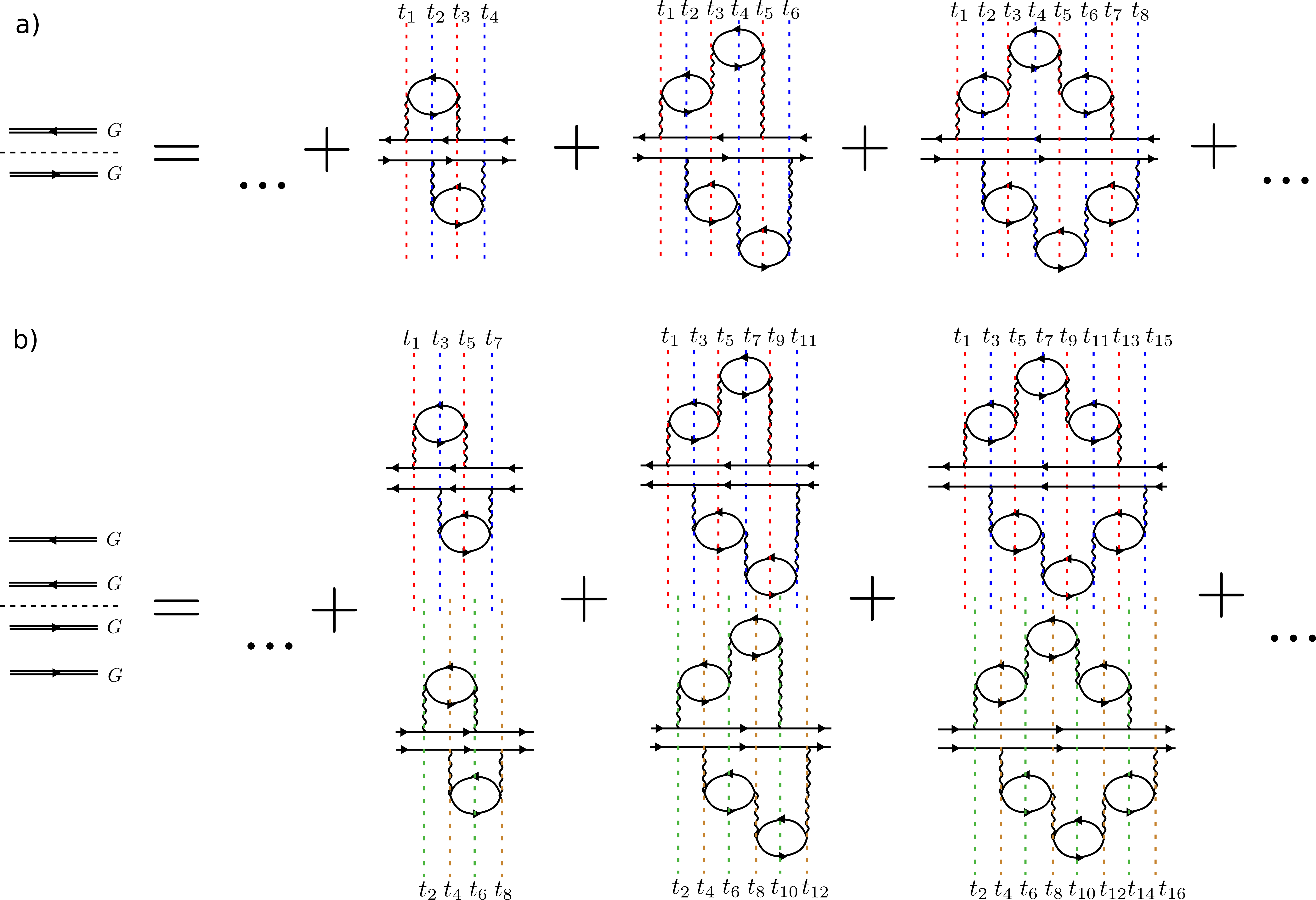}
\caption{Schematic representation of $\mathcal{L}^{\mr}_{GW}$. Each particle participating in a single (a), double (b), or higher excitation gains a self-energy based on the $GW$ series. For a single excitation, interactions at odd numbered times contribute to the hole propagator, while even numbered times contribute to the electron. The generalization to double and higher excitations follows by alternately assigning interactions at different times to different particles. The diagram demonstrates the complexity of applying the \sr projectors at every intermediate time. To avoid a complex procedure, we simply apply the cRPA at the single quasiparticle level to constrain correlation to $\mr$.\label{gw_times}}
\end{centering}
\end{figure*}

For $\mathcal{L}^{\mr}$, there are many particles propagating simultaneously. We assign a constrained self-energy to \textit{each} propagating particle, with each insertion taken at different times as in Fig. \ref{gw_times}. One can roughly argue that the self-energy insertion for any one particle should be invariant to all possible time orderings of the \textit{overall} expansion. To satisfy this invariance, it follows that the self-energy of \textit{each} particle should be constrained in the same way. This is not a rigorous proof, but a detailed exposition of all possible time orderings, excitation levels, and diagrammatic insertions for $\mathcal{L}^{\mr}$ is tedious.

Even though the application of \sr at each interaction appears to be formally well-defined, the mixing of the many-body projector \sr with a perturbation expansion based on the single-particle $G$ is extremely complicated. To circumvent the complexity of applying \sr exactly, we take advantage of a known and successful result (the cRPA) and apply it to our perturbation expansion at the \textit{single} quasiparticle level. Crucially, applying the $\Sigma = iGW_{\mr}$ approximation to each particle correctly interpolates between our two required limits. As $\mathcal{R} \rightarrow \id$, the self-energy correctly reaches the full $GW$ self-energy. As $\mathcal{R} \rightarrow 0$, the quasiparticle energies correctly approach the HF eigenvalues. The special limit of $\mr \rightarrow 0$ is discussed in detail in Appendix \ref{one_conf}.

We denote the GF described at this level of approximation by $\mathcal{L}_{GW}^{\mr}$. We assume that $G$ lines are dominated by well-defined quasiparticles with a long lifetime. By either making a diagonal approximation or working in the diagonalized-QP basis, the excitation matrix at this level of theory, $\Omega_{GW,JJ'}^{\mr}$, has only diagonal elements $\Omega_{GW,J}^{\mr}$. Neglecting the imaginary part of the self-energy, the excitation energy is the sum of these noninteracting quasiparticle energies,
\begin{equation}
\Omega_{GW,J}^{\mr} = \sum_{e \in \mathrm{els}}^m \epsilon_e^{GW_{\mr}} - \sum_{h \in \mathrm{holes}}^m \epsilon_h^{GW_{\mr}}.  \label{exc_gw}
\end{equation}
The Fourier transform of $\mathcal{L}_{GW}^{\mathcal{R}}$ is
\begin{eqnarray}
\mathcal{L}_{GW, JJ'}^{\mr}(\omega^{\mr}) &=&  \frac{ \delta_{J J'} }{ \omega^{\mr} - \Omega_{GW,J}^{\mr} + i \eta} \nonumber \\
&+&  \frac{ \delta_{J J'} }{ \omega^{\mr} - \Omega^{\mr}_{GW,J} - i \eta} \label{lgw}
\end{eqnarray}
The poles of $\mathcal{L}_{GW}^{\mr}$ are determined by the excitation energies $\Omega_{GW,J}^{\mr}$. 

Next, we must include interactions among quasiparticles. Quasiparticle interactions between the electron and hole are known to be very important, for example, in optical excitations. We also expect them to be important for the multiple excitations needed here. We assume a Dyson equation relating $\mathcal{L}^{\mr}_{GW}$ to the full $\mathcal{L}^{\mr}$ as
\begin{equation}
\mathcal{L}^{\mr} = \mathcal{L}_{GW}^{\mr} + \mathcal{L}_{GW}^{\mr} K^{\mr *}_{GW} \mathcal{L}^{\mr}. \label{dyson}
\end{equation}
Eq. \ref{dyson} can be formally solved as
\begin{equation}
\mathcal{L}^{\mr} = \frac{1}{ \omega^{\mr} - (\Omega_{GW}^{\mr} + K^{\mr *}_{GW}) }. \label{excitation_eigen}
\end{equation}
By Fourier transforming the exact $\mathcal{L}^{\mr}$ in Eq. \ref{biglr}, the poles of $\mathcal{L}^{\mr}$ are determined by the energy differences $E_J^{\mr} - E_0^{\mr}$. Equating these energy differences with the pole positions of $\mathcal{L}^{\mr}$ in Eq. \ref{excitation_eigen} sets up an effective $2m$-particle Hamiltonian \cite{martin_reining_ceperley_2016}
\begin{equation}
\left[ \Omega_{GW}^{\mr} + K^{\mr *}_{GW} \right] \ket{\psi} = \Omega_J^{\mr} \ket{\psi} \label{statich}
\end{equation}
for subspace excitation energies $\Omega_J^{\mr} = E_J^{\mr} - E_0^{\mr}$.

We can now approximate the kernel $K^{\mr *}_{GW}$. For a single excitation in $\mr$, we can use exactly the result for the electron-hole kernel based on our $\Sigma = i GW_{\mr}$ approximation. The electron-hole kernel is based on the derivative of the self-energy. In the common approximation to the BSE, the electron and hole therefore attract via $W_{\mr}$ and repel via their exchange interaction $v$ (ignoring variation of $W$ with respect to $G$). For multiple excitations, we include pairwise electron-hole interactions using the same $W_{\mr}$ interaction between all possible pairs.

The multiple excitations in \sr also include electron-electron and hole-hole interactions. For this, we extrapolate the result for electron-electron terms known for trion matrix elements.\cite{rohlfing_prl_116,deilmann_prb_96} Their interaction is similar to the electron-hole case, except that both their direct and exchange interactions are screened. We use screened direct and screened exchange interactions among all electron-electron pairs and hole-hole pairs.
\begin{figure}
\begin{centering}
\includegraphics[width=1.0\columnwidth]{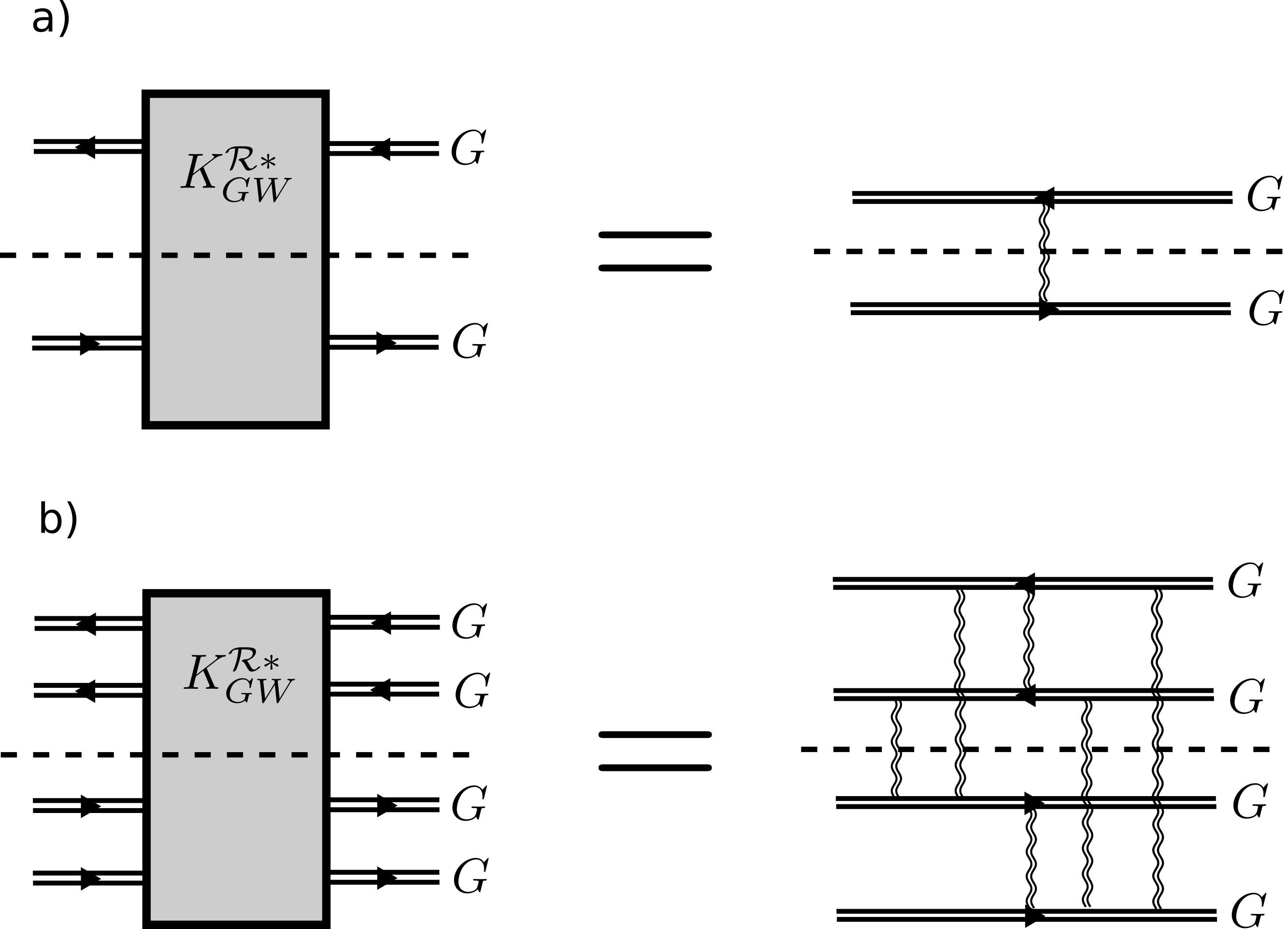}
\caption{Schematic representation of the approximate $\mathcal{L}^{\mr}$. At this level of approximation, the exact kernel $K^{\mr *}$ is approximated by separate, two-body interactions for single (a), double (b), and higher excitations in \sr. The double-wiggly line represents the partially screened Coulomb interaction, $W_{\mr}$, while corresponding exchange diagrams are implied but not shown.\label{kernel}}
\end{centering}
\end{figure}
The diagram in Fig. \ref{kernel} represents this approximation diagramatically. We approximate $K^{\mr *}_{GW}$ with separable, effective two-body interactions. We take each quasiparticle interaction to occur at different times so that $K_{GW}^{\mr *}$ can be approximated in this way.

Last, we make a global static approximation to the frequency dependent problem shown in Fig. \ref{kernel}. We evaluate $G$ lines at their corresponding quasiparticle energies, and take each quasiparticle interaction to be instantaneous. These approximations transform the frequency dependent $\mathcal{L}^{\mr}$ problem into a static renormalized Hamiltonian. The same set of approximations is very commonly applied to the BSE and quite successful for predicting optical excitation energies of weak- to moderately-correlated systems.\cite{rohlfing_prb_62,bruneval_jcp_142,bruneval_jcp_142,jacquemin_jctc_11,chelikowsky_prb_73,botti_jctc_10} By placing low energy transitions in $\md$, \sr is effectively a large band gap system, and we expect the correlation to be weak. As long as \sr has well-defined and long-lived quasiparticles, we can justify the static approximation. The full frequency dependence of $W_{\mr}$ is included in the self-energy $GW_{\mr}$ but neglected for inter-quasiparticle interactions.

\begin{figure}
\begin{centering}
\includegraphics[width=1.0\columnwidth]{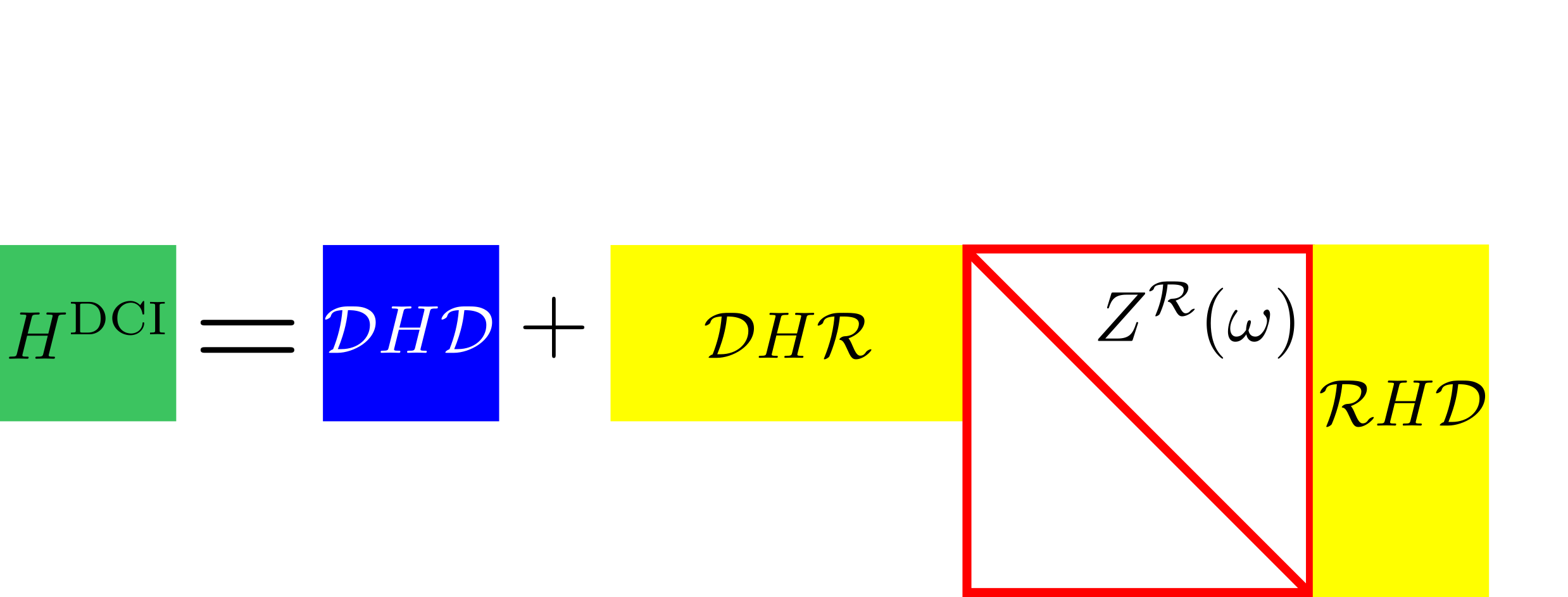}
\caption{The diagonal approximation to the transformation $\mr H \mr \rightarrow H^{\mr}$ dramatically reduces the computational cost while keeping a quasiparticle description of correlation. Coupling between the two spaces is computed in the WF picture through matrix elements of the exact $H$, $\md H \mr$ and $\mr H \md$. The WF description of hybridization avoids frequency dependent hybridization between \sd and $\mr$. The MBPT calculation, and therefore the frequency dependence of the perturbation expansion, is contained in the \sr subspace. The physical $E$ in the denominator of $Z^{\mr}$ determines the coupling strength between the two spaces, but $E$ dependence is \textit{not} part of the perturbation expansion.\label{downfolding_dci}}
\end{centering}
\end{figure}

We can finally write our approximation to the excitation matrix $\Omega^{\mr}$, which is equivalent to the Hamiltonian in Eq. \ref{statich}. We limit $\Omega^{\mr}$ to a diagonal approximation. The diagonal approximation in \sr is the major computational savings of our approach, as demonstrated in Fig. \ref{downfolding_dci}. In such an approximation, there is no coupling between forward and backward time orderings of $\mathcal{L}^{\mr}$. The Tamm-Dancoff approximation (TDA) makes connections to the resolvent of the auxiliary eigenvalue problem (Appendix \ref{resolvents}) and projected Hamiltonian $\mr H \mr$ easier, since neither quantity has any time ordering. Even in a diagonal approximation to $\mathcal{L}^{\mr}$ in the TDA, however, correlation is described at the quasiparticle level. Diagonal matrix elements of $\Omega^{\mr}$ are
\begin{eqnarray}
\Omega_J^{\mr} &=& \bra{J} \Omega^{\mr} \ket{J}  \nonumber  \\
&=& \sum_{e \in J}^m \epsilon_e^{GW_{\mr}} - \sum_{h \in J}^m \epsilon_h^{GW_{\mr}}   \nonumber  \\  
&+& \sum_{e,h \in J}^m (-W_{\mr,eheh} + \delta_{\sigma_e \sigma_h} v_{ehhe} ) \nonumber  \\
&+& \sum_{\mathclap{\substack{e \in J \\
e \neq e'}}}^m \;\; (W_{\mr,ee'ee'} - \delta_{\sigma_e \sigma_{e'}} W_{\mr, ee'e'e} ) \nonumber \\  
&+& \sum_{\mathclap{\substack{h \in J \\
h \neq h'}}}^m \;\; (W_{\mr,hh'hh'} - \delta_{\sigma_h \sigma_{h'}} W_{\mr, hh'h'h} ).   \label{excitation}
\end{eqnarray}
Here, we reintroduce the configuration notation $\ket{J}$ to make the connection to the wave function picture clear. The basis sets for the WF and GF calculations are now connected. The sums in Eq. \ref{excitation} run up to the excitation level $m$ of the configuration. $e$ and $h$ denote the excited electrons and holes of the configuration $\ket{J}$ $-$ they are exactly the creation and destruction operators in Eq. \ref{operatorstring} that define the configuration $\ket{J}$. The final result to compute excitation energies is relatively straightforward: the one-body part of the effective Hamiltonian is $GW_{\mr}$ quasiparticles, and the two-body part is their interaction via the screened interaction $W_{\mr}$. Eq. \ref{excitation} is not a diagonal approximation to the BSE, though it is BSE-like.

\subsection{Ground state energy}
To complete the transformation $\mr H \mr \rightarrow H^{\mr}$, we must also estimate the ground state energy $E_0^{\mr}$. We discuss formal limits and definitions, then our procedure for calculating $E_0^{\mr}$.

\subsubsection{Definition and embedding limits}
Consider an eigenstate of $\mr H \mr$ given by $\ket{J}$ with eigenvalue $E_J^{\mr}$. $E_J^{\mr}$ is on the order of a total energy of the system. Assume $\ket{J}$ has a corresponding long-lived excitation computed with MBPT (as outlined previously) with excitation energy $\Omega_J^{\mr}$. The ground state energy $E_0^{\mr}$ is chosen to match the total energy $E_J^{\mr}$:
\begin{equation}
E_J^{\mr} = E_0^{\mr} + \Omega_J^{\mr}.  \label{groundstate}
\end{equation}
We assume a single value $E_0^{\mr}$ exists to match Eq. \ref{groundstate} for all $J$. If we exactly diagonalize $\mr H \mr$, we can compute $E_0^{\mr}$ as the difference $E_J^{\mr} - \Omega_J^{\mr}$. Of course, this defeats the purpose of the transformation since we want to avoid the expense of diagonalizing $\mr H \mr$. We must develop a different, more efficient strategy to estimate $E_0^{\mr}$.

$E_0^{\mr}$ is difficult to define beyond Eq. \ref{groundstate}. Because \sr contains only excitations, there simply is no ground state in $\mr$ that can be related to the physical ground state. It is still possible, however, to choose a scalar value $E_0^{\mr}$ to satisfy Eq. \ref{groundstate}. While it is unintuitive that this value does not correspond to an \sr eigenstate, such a correspondence is not necessary to satisfy Eq. \ref{groundstate}. We can still place some constraints on $E_0^{\mr}$. In the limit that $\mr \rightarrow \id$, we must recover a standard MBPT calculation with the full $H$. In this case, we know that excitation energies are defined with respect to the fully correlated ground state, so $E_0^{\mr} \rightarrow E_0$.

The opposite limit $\mr \rightarrow 0$ also gives some insight. While the case $\mr = 0$ trivially recovers FCI, consider one configuration in $\mr$. This special case is considered in more detail in Appendix \ref{one_conf}. In this case, the exact diagonalization result is simply one diagonal matrix element. It can be decomposed into a clearly defined ground state, with energy equal to the energy of the reference configuration, $E^{\mathrm{ref}}$, and excitation energy. The details are presented in Appendix \ref{one_conf}, but this case provides a critical second constraint: $E_0^{\mr} \rightarrow E^{\mathrm{ref}}$ as $\mr \rightarrow 0$. At intermediate sizes of $\mr$, we enforce that $E_0^{\mr}$ must interpolate between these two limits: $E_0 < E_0^{\mr} < E^{\mathrm{ref}}$.

\subsubsection{Partitioning correlation energy}
For our estimate of $E_0^{\mr}$, we assume a nondegenerate ground state and partition the correlation energy of the true ground state into three distinct parts.
\begin{equation}
E_0 = E^{\mathrm{ref}} + C_{\md \md} + C_{\md \mr} + C_{\mr \mr} \label{correlation_energy}
\end{equation}
Here, $C_{\md \md}$ is the intra-$\md$ correlation, $C_{\md \mr}$ is the inter-space correlation, and $C_{\mr \mr}$ is the intra-$\mr$ correlation. As \sd and \sr change size, the total correlation energy ($C$) must stay constant $-$ the total energy is independent of the partitioning between \sd and $\mr$. However, the individual contributions $C_{ij}$ are allowed to change as the correlation is transferred between spaces. In the limit $\md \rightarrow \id$, $C_{\md \md} = C$; in the limit $\mr \rightarrow \id$, $C_{\mr \mr} = C$.

The change in the correlation energy $C_{\mr \mr}$ mimics the desired behavior for the energy $E_0^{\mr}$. We propose using
\begin{equation}
E_0^{\mr} = E^{\mathrm{ref}} + C_{\mr \mr} \label{e0r}
\end{equation}
to estimate the ground state energy $E_0^{\mr}$. If we can isolate the intra-$\mr$ correlation, Eq. \ref{e0r} correctly interpolates between our set limits on $E_0^{\mr}$. The limits are summarized as
\begin{eqnarray}
\mr &\rightarrow& \id \;\; \mathrm{and} \;\; C_{\mr \mr} \rightarrow C, \;\; E_0^{\mr} \rightarrow E_0  \nonumber \\
\mr &\rightarrow& 0 \;\; \mathrm{and} \;\; C_{\mr \mr} \rightarrow 0, \;\; E_0^{\mr} \rightarrow E^{\mathrm{ref}}.
\end{eqnarray}

To this end, we introduce another total energy, $\tilde{E}_0$, given by
\begin{equation}
\tilde{E}_0 = E^{\mathrm{ref}} + C_{\md \md} + C_{\md \mr}.
\end{equation}
We see that, if it is possible to compute the energy $\tilde{E}_0$, we can isolate $C_{\mr \mr}$ and calculate $E_0^{\mr}$ as
\begin{eqnarray}
C_{\mr \mr} &=& E_0 - \tilde{E}_0.  \nonumber  \\
E_0^{\mr} &=& E^{\mathrm{ref}} + C_{\mr \mr} = E^{\mathrm{ref}} + E_0 - \tilde{E}_0.  \label{rcorrelation}
\end{eqnarray}
Here, we assume that the various correlation energies $C_{ij}$ are the same for all three total energies. Generally, this is not true. It is not possible to simply ``turn off'' a given part of the correlation as it appears in the full problem. However, it is an approximation that can be aided by intuition about choosing the AS.

It will prove useful to identify where these three contributions to the correlation exist in the partitioned problem of Eq. \ref{effective}. $C_{\md \md}$ exists in the \sd projected block of $H$, the frozen core Hamiltonian. The hybridization $C_{\md \mr}$ exists in matrix elements of $\md H \mr$ (and $\mr H \md$). Finally, the intra-$\mr$ correlation is in matrix elements of $\mr H \mr$. This portion of the correlation is contained in the resolvent, which depends on the inversion of the block $\mr H \mr$.

\subsubsection{Uncorrelated resolvent}
With these considerations in mind, our strategy to estimate $C_{\mr \mr}$ depends on an estimate of the total energy $\tilde{E}_0$. To calculate the energy $\tilde{E}_0$, we self-consistently solve Eq. \ref{effective} using an \textit{uncorrelated} resolvent $\tilde{Z}^{\mr}(E)$. By using an uncorrelated Hamiltonian in place of $\mr H \mr$, we approximately remove the intra-$\mr$ correlation. It is an approximation that the correlation removed by this procedure is the same as the correlation energy of the real ground state, $C_{\mr \mr}$.

We choose the uncorrelated Hamiltonian based on particle addition or removal to the reference configuration, with no interactions among the added particles or holes. The potential must be fixed at the field created by the reference configuration. In the case of a Hartree-Fock (HF) mean-field starting point, which we use exclusively in this work, Koopman's theorem dictates that these particle addition energies are the HF eigenvalues. Matrix elements of the uncorrelated Hamiltonian $H^{\mathrm{ref}}$ (which is diagonal) are
\begin{equation}
\bra{J} H^{\mathrm{ref}} \ket{J} = E^{\mathrm{ref}} + \sum_{e \in \mathrm{els}}^m \epsilon_e^{\mathrm{HF}} - \sum_{h \in \mathrm{holes}}^m \epsilon_h^{\mathrm{HF}} .  \label{href}
\end{equation}

With the uncorrelated resolvent, matrix elements of $M(E)$ in Eq. \ref{effective} are \textit{not} zero. They still depend on the inter-space elements of $\md H \mr$ and $\mr H \md$. The purpose of this procedure is to remove the intra-$\mr$ correlation in the resolvent, so that $M(E)$ contains \textit{only} the hybridization $C_{\md \mr}$. $\tilde{E}_0$ is the self-consistent solution of Eq. \ref{effective} with the uncorrelated resolvent,
\begin{equation}
\tilde{Z}^{\mr}(E) = \frac{1}{E - H^{\mathrm{ref}}}.
\end{equation}

While this describes the procedure to estimate $E_0^{\mr}$ in this work, there are other possible routes to calculate $E_0^{\mr}$. It could be possible to estimate $C_{\mr \mr}$ with diagrammatics. For example, one could compute a correlation energy based on the cRPA. For now, we favor a more diagonalization-based approach. The RPA is known to introduce spurious bumps in dimer dissociation curves that could impact our test calculations.

\subsection{Final equations}
After calculating the excitation matrix $\Omega^{\mr}$ and energy $\tilde{E}_0$, the effective Hamiltonian becomes
\begin{eqnarray}
H^{\mr} &=& E_0^{\mr} + \Omega^{\mr}  \nonumber  \\
H^{\mr} &=& ( E^{\mathrm{ref}} + E_0 - \tilde{E}_0 ) + \Omega^{\mr}.  \label{hr}
\end{eqnarray}
With this transformed Hamiltonian, the resolvent is
\begin{eqnarray}
Z^{\mr}(E) &=& \frac{1}{E - H^{\mr}}  \nonumber  \\ 
&=& \frac{1}{E - ( E^{\mathrm{ref}} + E_0 - \tilde{E}_0 + \Omega^{\mr} )}  \nonumber \\
&=& \frac{1}{ ( E - E_0 ) - ( E^{\mathrm{ref}} - \tilde{E}_0 ) - \Omega^{\mr} }
\end{eqnarray}
Combining the different total energies, we rewrite the resolvent as
\begin{equation}
Z^{\mr}(\omega) = \frac{1}{ ( \omega - \Delta ) - \Omega^{\mr} }
\end{equation}
where $\omega \equiv E - E_0$ and $\Delta \equiv E^{\mathrm{ref}} - \tilde{E}_0$. $\omega$ now has the interpretation of an optical frequency or excitation energy. Here, $\omega$ is the physical frequency related to exciting the system. The ground state ($\omega = 0$) is effectively evaluated \textit{not} at zero frequency but at a shift $-\Delta$. Naively setting the $E$ dependence of the resolvent in Eq. \ref{effective} to $E_0$ for a ground state calculation leads to double-counting errors. $\Delta$ is a shift determined by how close the energy of the reference configuration is to the energy $\tilde{E}_0$.

$\Delta$ can also be thought of as a double-counting correction to the physical ground state. Consider an alternative definition of $E_0^{\mr}$,
\begin{equation}
E_0^{\mr} = E_0 + \mathrm{D.C.},  \label{doublecounting}
\end{equation}
where the fictitious ground state is the true ground state plus the double-counting correction $\mathrm{D.C.}$ Matching this equation with Eq. \ref{rcorrelation}, we see that the double-counting correction is
\begin{equation}
\mathrm{D.C.} = E^{\mathrm{ref}} - \tilde{E}_0 \equiv \Delta.
\end{equation}
$\Delta$ is necessarily $>0$, so that the double-counting correction in Eq. \ref{doublecounting} \textit{increases} the energy from $E_0$ to $E_0^{\mr}$. $\Delta$ corrects potential double-counting errors in the ground state, and our procedure to compute excitation energies does not double-count correlation. The overall theory is double-counting free without any adjustable parameters.

We rewrite the full set of equations for clarity:
\begin{eqnarray}
Z^{\mr}(\omega) &=& \frac{1}{ (\omega - \Delta) - \Omega^{\mr} }   \nonumber  \\
M(\omega) &=& \left[ \md H \mr \right] Z^{\mr}(\omega) \left[ \mr H \md \right]   \nonumber  \\
H^{\mathrm{DCI}}(\omega) \phi &=& \left[ \md H \md + M(\omega) \right] \phi = E \phi . \label{final}
\end{eqnarray}
The energy eigenvalue of Eq. \ref{final} is still the total electronic energy of the system. The energy dependent correction $M(\omega)$ to the frozen core Hamiltonian gives a dynamical version of configuration interaction. Because we treat the $\omega$ dependence explicitly, and to reflect the dynamical character of quasiparticles, we adopt the name dynamical configuration interaction (DCI). The final result is that an active space wave function is dynamically embedded in a bath of interacting quasiparticles.
\begin{figure}
\begin{centering}
\vspace{0.5cm}
\includegraphics[width=0.95\columnwidth]{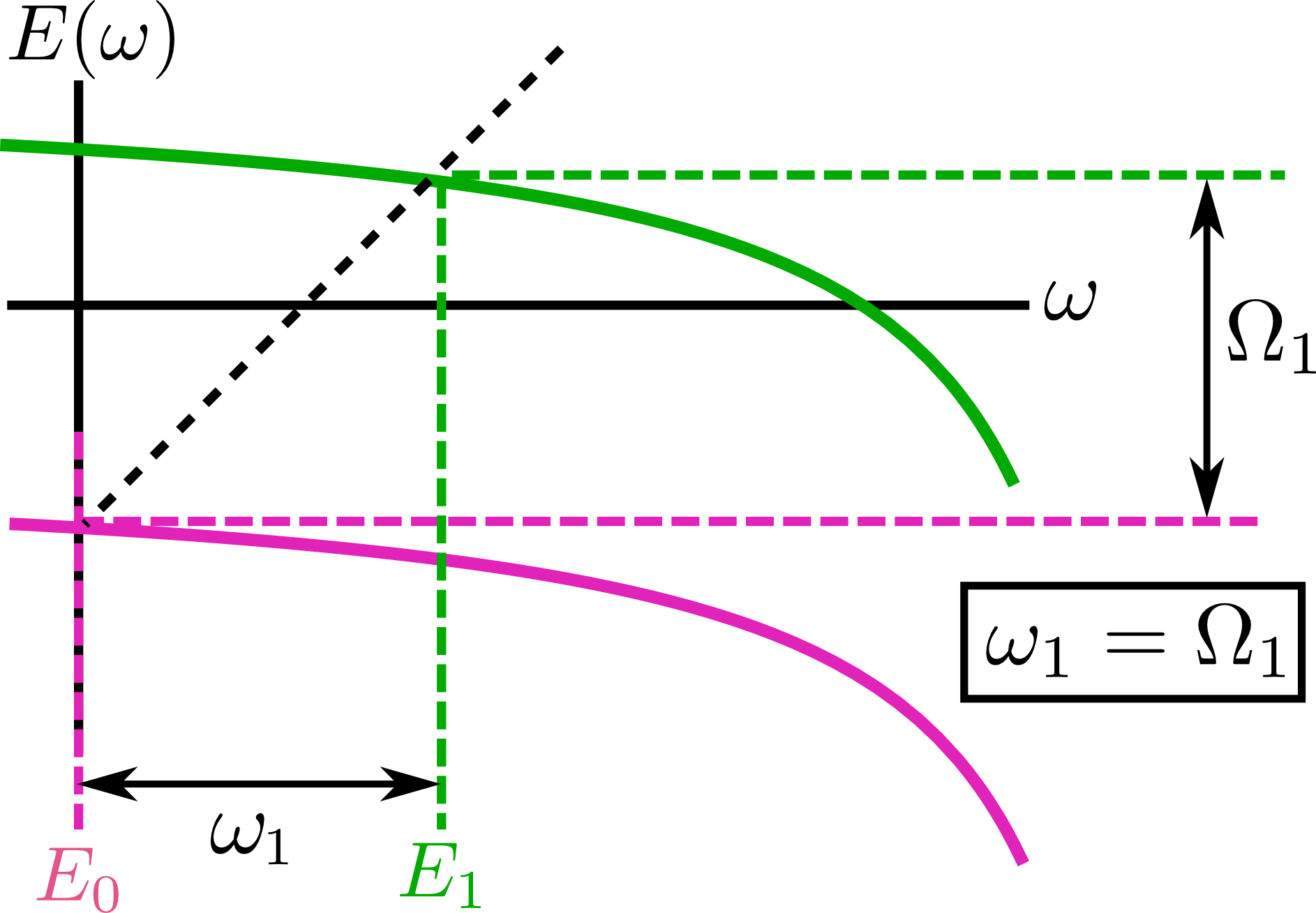}
\caption{Schematic (not real data) showing total energy curves for the ground (magenta) and first excited (green) states on a grid of $\omega$ values. Poles at the right side of the $\omega$-axis are derived from \sr excitations. For each excited state, the self-consistent solution intersects the diagonal line where $\omega = \Omega_i$.\label{self_consistent}}
\end{centering}
\end{figure} 

We treat the energy dependence of the resolvent exactly instead of approximating $E$ by an estimate of the energy. There is no self-consistency requirement on $\omega$ for the ground state. Instead, the self-consistency condition for the ground state is in the self-consistent calculation of $\tilde{E}_0$ to calculate $\Delta$. For each excited state, the excitation energy $\Omega_i$ is determined self-consistently by iterating Eq. \ref{final}. The self-consistency condition is on the excitation energy, updated at each iteration as $\Omega_i = E_i - E_0$. Here, $E_i$ is the $i^{\mathrm{th}}$ eigenvalue of Eq. \ref{final} (using $i=1$ for the first excited state) and $E_0$ is the lowest eigenvalue computed at $\omega=0$. There is no update of the \sr Hamiltonian during the self-consistency cycle. A self-consistent solution in \sr is allowed but should be done before diagonalizations of Eq. \ref{final} begin.

The self-consistency requirement on excited states is shown graphically in Fig. \ref{self_consistent}. Treating $\omega$ as a free parameter, one can compute total energy curves for each eigenstate on a grid of $\omega$ values. The ground state energy is always taken at the frequency $\omega=0$, and self-consistent excited states intersect the diagonal line in Fig. \ref{self_consistent}. All solutions are accessible, in principle, by solving the \sd space eigenvalue problem in Eq. \ref{final}. High energy solutions, derived principally from \sr configurations, exist as higher intersections of total energy curves with the diagonal line in Fig. \ref{self_consistent}.

$\Omega^{\mr}$ and $Z^{\mr}(\omega)$ are, in principle, matrices of dimension $\dim \mr$. However, our diagonal approximation reduces the inversion and matrix multiplication to a simple scalar multiplication. Importantly, we still use a proper treatment of correlation in the bath space. Our diagonal quasiparticle approximation is a better approximation than a diagonal matrix of $\mr H \mr$. The differences between the effective excitation energies in $\mr H \mr$ compared with $\Omega^{\mr}$ can be $\sim \hspace{-0.1cm}1$ Ha. Correlation is also included in the estimate of the energy $E_0^{\mr}$. The ability to describe dynamic correlation by correlating the full set of orbitals beyond the AS with the successful $GW$ approximation hopefully allows us to use small active spaces, even for large systems. With our approach, one should not need to enlarge the AS solely for the purpose of adding dynamic correlation.



\section{Procedure and numerical details}

\subsection{Electronic structure}
Our calculations are based on an initial restricted HF (RHF) mean-field calculation, followed by a single-shot perturbative $G_0W_{0,\mr}$ calculation. We use the FHI-AIMS electronic structure package based on numeric atomic orbitals (NAOs) and resolution-of-the-identity (RI) for one- and two-body matrix elements.\cite{blum_cpc_180,ren_njp_14,ihrig_njp_17,levchenko_cpc_192} Our atomic basis sets are chosen to either match benchmark theoretical calculations or approximate the complete basis set limit. Basis sets for individual calculations are specified with the results. We perform perturbative $G_0W_{0,\mr}$@HF calculations by constraining the polarization entering $W$ to the cRPA. After the $G_0W_{0,\mr}$ calculation, we write matrix elements of $t$, $v$, and $W_{0,\mr}$ to disk, to be used in the following DCI calculation. In order to match the basis for the $G_0W_{0,\mr}$ calculation, our single-particle basis for DCI is always the canonical RHF orbitals.

\subsection{Many-body basis}
Using a diagonal approximation to $Z^{\mr}(\omega)$ is a major simplification to the many-body basis. Our generation of the many-body basis is similar to the multi-reference configuration interaction singles doubles (MR-CISD) method.\cite{helgaker_molecular} If an \sr configuration does not couple directly to any state in $\md$, it does not need to be generated. This reduces the required number of configurations considerably.

We construct the many-body basis by first choosing the statically correlated orbitals, which are placed in the orbital AS. We adopt the quantum chemistry notation of $(\mathrm{els},\mathrm{orbs}) = (x,y)$ to denote $x$ electrons distributed in $y$ statically correlated orbitals. When feasible, \sd is constructed by exactly diagonalizing this AS, as in a CAS theory. It is also possible to truncate the excitation level in \sd with a selected level of CI. When possible, we construct \sr by distributing all $p$ electrons in the full spectrum of occupied and unoccupied states, $q$, omitting the configurations already in $\md$. In the present work, we never truncate the space of unoccupied orbitals or deep occupied states, so that \sd is embedded in the full set of transitions.

In matrix notation, the Hamiltonian is built with the matrix elements
\begin{eqnarray}
M_{II'}(\omega) &=& \sum_J \bra{I} H \ket{J} \frac{1}{( \omega - \Delta) - \Omega_J^{\mr}} \bra{J} H \ket{I'}  \nonumber  \\
H^{\mathrm{DCI}}_{II'}(\omega) &=& \bra{I} H \ket{I'} + M_{II'}(\omega)  \label{matrixform}
\end{eqnarray}
where $I$ and $I'$ are \sd configurations. Matrix elements of the many-body Hamiltonian $H$ are calculated with the Slater-Condon rules, which are located in Appendix \ref{slater_condon}.

\begin{figure}
\begin{centering}
\vspace{0.5cm}
\includegraphics[width=1.0\columnwidth]{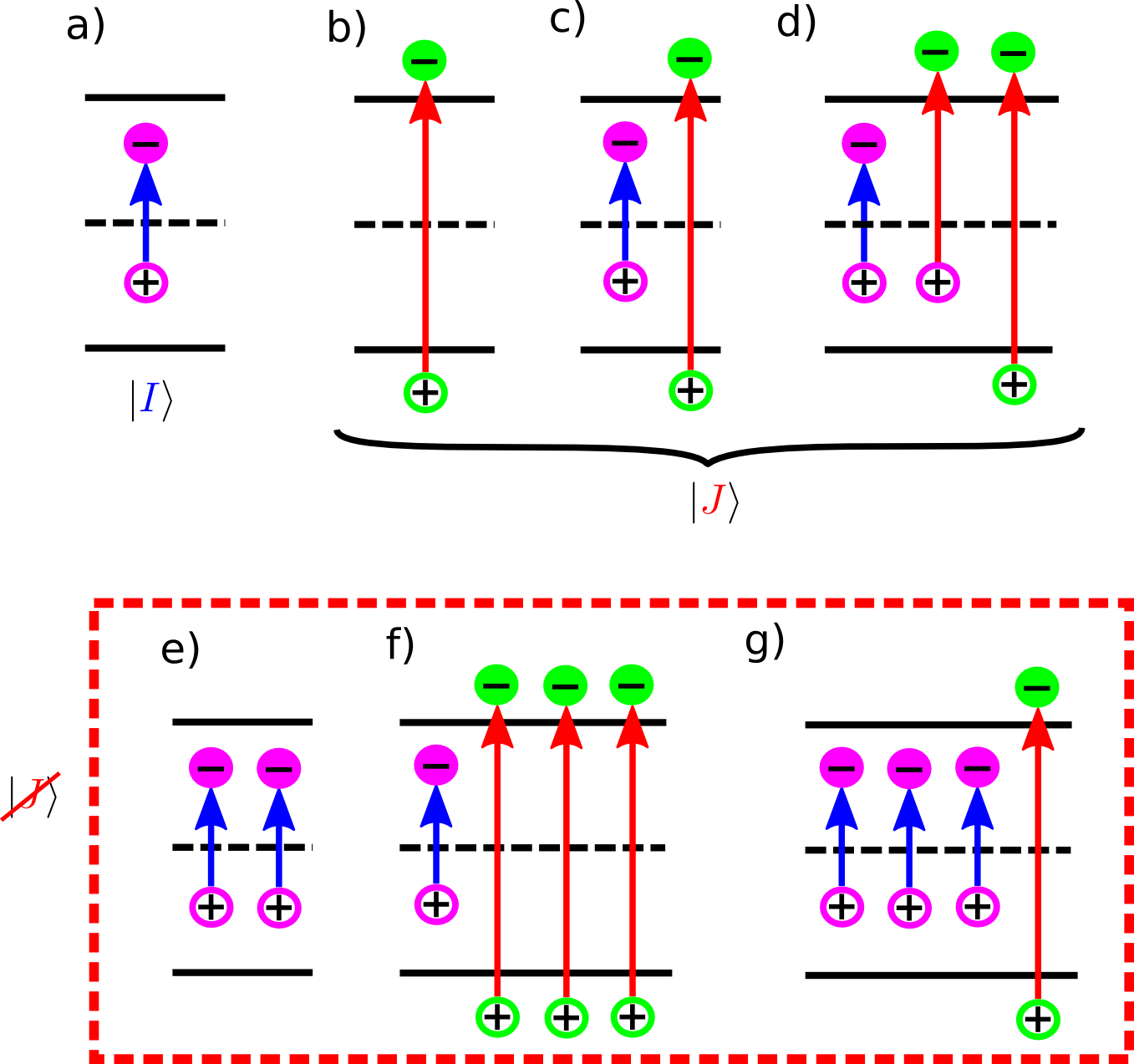}
\caption{Treating each $\ket{I}$ in \sd as a reference, we permute its occupation numbers to generate all single and double excitations to form its local set of \sr configurations, $\{ \ket{J} \}$. The configuration in panel (a) contains only AS excitations and belongs to $\md$. Panels (b-d) are generated as double, single, and double excitations from $\ket{I}$, respectively. The dashed red box encloses configurations which are not in the local set $\{ \ket{J} \}$. Panel (e) is a single excitation from $\ket{I}$ but is already included in $\md$, therefore it must be omitted from $\{ \ket{J} \}$ to avoid double-counting. Panels (f) and (g) are triple excitations from $\ket{I}$ and do not couple directly to $\ket{I}$. Therefore, (f) and (g) do not need to be generated or stored.\label{basis_generation}}
\end{centering}
\end{figure}

The matrix elements $M_{II'}$ obey their own selection rules, dictated by the matrix elements $\bra{I} H \ket{J}$. If $\ket{I}$ and $\ket{J}$ differ by more than two occupation numbers, $\bra{I} H \ket{J} = 0$ and the configuration $\ket{J}$ can be omitted altogether from the internal sum for $M_{II'}$. The corrections $M_{II'}$ are zero if $\ket{I}$ and $\ket{I'}$ differ by more than 4 occupation numbers. Based on this selection criterion, we treat each $\ket{I}$ as a reference configuration. We generate all single and double excitations from $\ket{I}$ to form its local set of configurations $\{ \ket{J} \}$ from the \sr space, as demonstrated in Fig. \ref{basis_generation}. The set $\{ \ket{J} \}$ is generated as an $\mathcal{O}(p^2 q^2)$ operation. The calculations are well-suited to parallel computers because there is no communication required between the different sets $\{ \ket{J} \}$ to compute the internal sums for $M(\omega)$.

Excitations of arbitrarily high level are naturally included in the many-body basis, depending on the highest excitation level in $\md$. For example, generating the local set of $\{\ket{J}\}$ for a $6\times$ excitation in \sd will include $8\times$ excitations in $\mr$. The procedure, therefore, gives a balanced treatment between ground and excited states since all \sd configurations couple to roughly the same number of \sr determinants. This balanced generation of determinants is the main attraction of MR-CISD. All reference configurations should be placed in $\md$ where they are treated exactly, including all off-diagonal coupling. The description of multi-reference states can be improved by systematically expanding the AS and adding configurations to $\md$. The diagonal approximation in \sr is only meant to add \textit{dynamic} correlation to $\md$.

\subsection{Computational workflow}
\begin{figure}
\begin{centering}
\includegraphics[width=1.0\columnwidth]{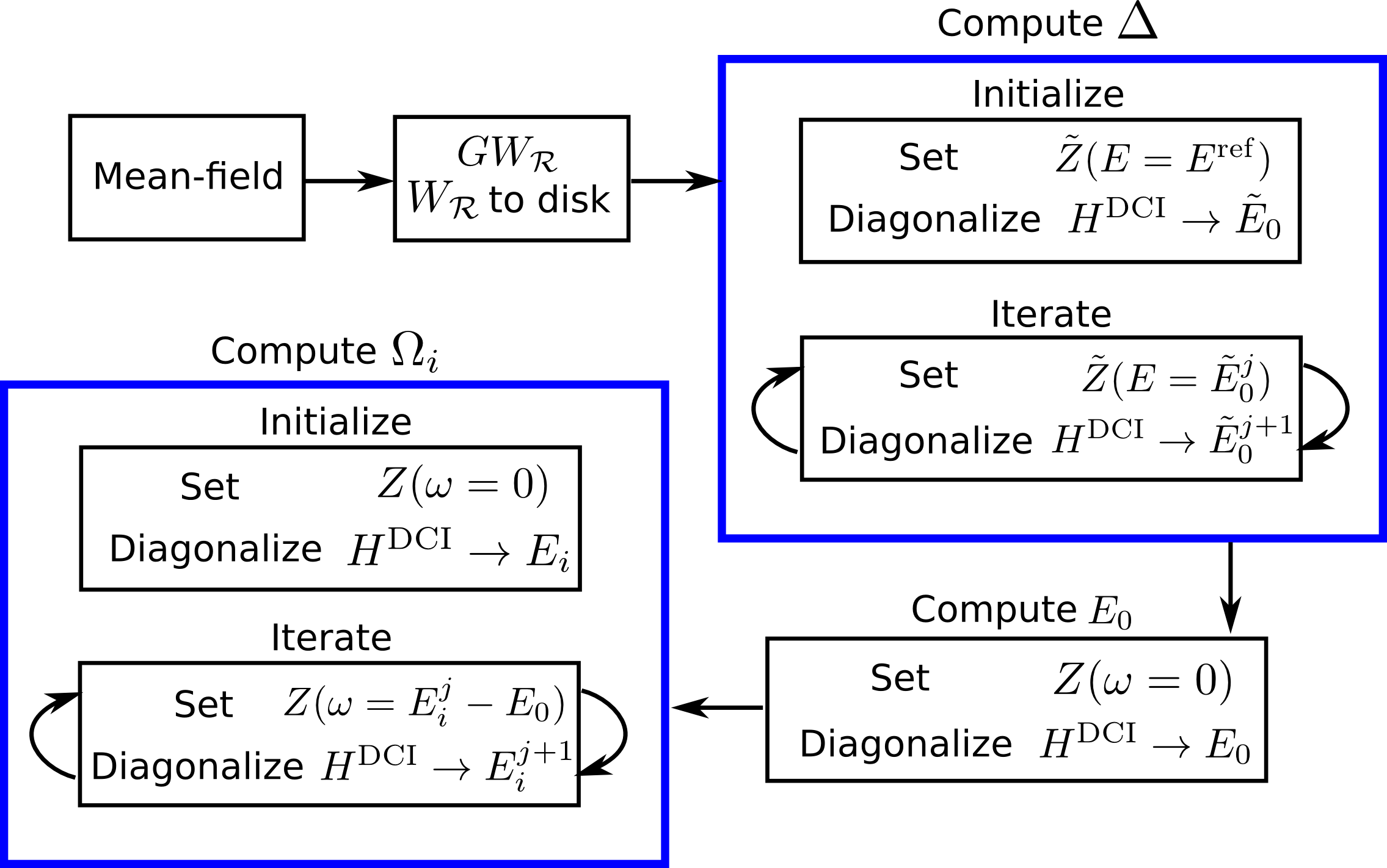}
\caption{Computational workflow for a DCI calculation. Self-consistency cycles are indicated by blue boxes. Initializing excitation energies at $\omega = 0$ is a reasonable starting point for self-consistency. The index $i$ refers to the state of interest, while $j$ is the $j^{\mathrm{th}}$ iteration in the self-consistency cycle for state $i$.\label{flow_chart}}
\end{centering}
\end{figure}

The computational procedure for a DCI calculation is a series of matrix constructions and diagonalizations. Depending on the step, we use the resolvent $Z^{\mr}$ or $\tilde{Z}^{\mr}$, and the Hamiltonian must be iterated to meet any self-consistency condition. The overall procedure is summarized in the flowchart shown in Fig. \ref{flow_chart}.

The first step is to determine the double-counting correction $\Delta$. The Hamiltonian is first constructed with the uncorrelated resolvent evaluated at the reference energy, $\tilde{Z}(E=E^{\mathrm{ref}})$. The lowest eigenvalue of the matrix diagonalization gives the first estimate to the total energy $\tilde{E}_0$. The evaluation energy is then set to $\tilde{E}_0$, and the procedure is iterated until self-consistent. The shift $\Delta$ is the calculated as $\Delta = E^{\mathrm{ref}} - \tilde{E}_0$ for the self-consistent $\tilde{E}_0$.

After calculating $\Delta$, we compute the ground state energy. We set $\omega = 0$ in Eq. \ref{final}, and the lowest eigenvalue of the matrix diagonalization is $E_0$. If one needs only the ground state energy, the calculation is finished.

An arbitrary number of excited states can be calculated in a similar way as the determination of $\Delta$. For excited states, $H^{\mathrm{DCI}}$ is built with the resolvent $Z^{\mr}(\omega)$ as in Eq. \ref{final}. Each excited state must be treated separately, one at a time, since $\omega$ is state dependent. The $i^{\mathrm{th}}$ eigenvalue of the matrix diagonalization (counting the first excited state as $i=1$) gives the total energy $E_i$. The excitation energy is calculated as $\Omega_i = E_i - E_0$, a new Hamiltonian is constructed at $\omega = \Omega_i$, and the procedure is iterated until self-consistent.

\FloatBarrier

\section{Results}
Molecular dissociation is a difficult multi-reference problem and a benchmark test of strongly-correlated methods. In the dissociation limit, the wave function with the correct symmetry cannot be written as a single Slater determinant. Mean-field theories which optimize single-particle states around a single Slater determinant, therefore, cannot correctly describe such dissociation. Even many-body methods which are biased towards a single reference may fail in bond breaking problems.

\begin{figure}
\begin{centering}
\includegraphics[width=1.0\columnwidth]{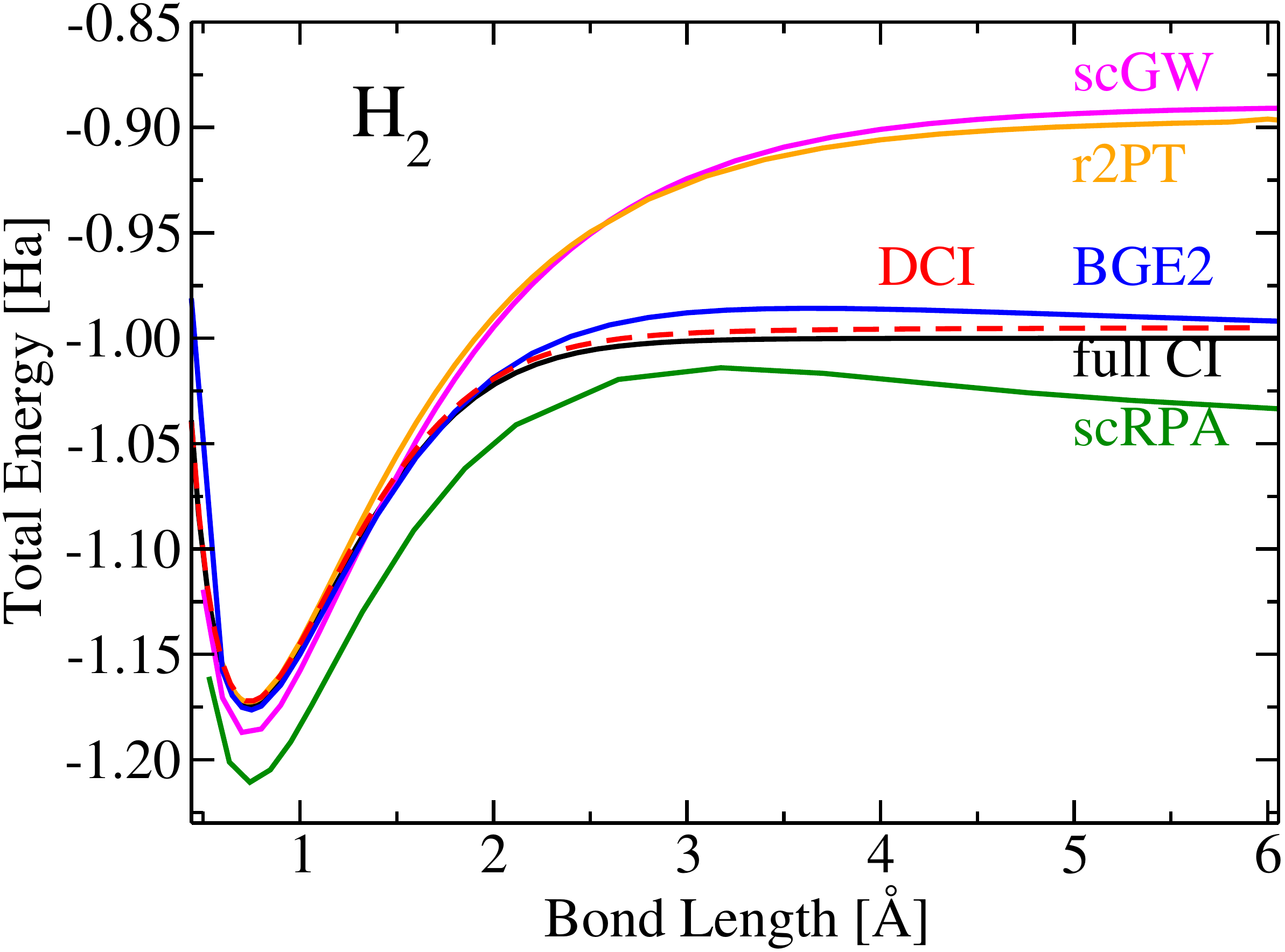}
\caption{H$_2$ dissociation for (2,6) DCI compared to other approaches (see text for definition of abbreviations). The black line is the full CI reference result.\cite{wolniewicz_jcp_99} We use the aug-cc-pVQZ basis set for comparison with reference data.\cite{hellgren_prb_91,caruso_prl_110}\label{h2_binding}}
\end{centering}
\end{figure}

We first test our theory by dissociating H$_2$ with a $(2,6)$ active space, shown in Fig. \ref{h2_binding}. While performing H$_2$ dissociation with a relatively large $(2,6)$ AS may seem ``easy'' for a wave function based method, it is still an important first test of the theory. Many approximate methods, including GF-based methods such as RPA or BSE,\cite{olsen_jcp_140} contain spurious maxima in the dissociation curve. It is not clear, just by examining the equations, if our underlying GF calculation in \sr based on a single reference introduces similar errors to a DCI calculation. The formalism for computing the correlation energy directly from the GF is different than ours, but DCI must still be checked for any similar errors. Additionally, our approximate double-counting correction $\Delta$ must be tested. Particularly in the dissociation limit, the behavior of $\Delta$ is unknown.

Our DCI calculation for H$_2$ is free of any spurious maxima or other obvious errors of the underlying single-reference $GW_{\mr}$ calculation. For a modest sized matrix diagonalization, DCI outperforms sc$GW$ and r2PT,\cite{ren_prb_88} a renormalized perturbation theory treatment up to second-order. H$_2$ dissocation based on the BSE improves upon the scRPA and sc$GW$ results, but still contains a maximum in the intermediate region\cite{olsen_jcp_140} (not shown here). Self-interaction errors are exaggerated in H$_2$ because of its small size and are a potential problem for GF methods dissociating H$_2$. DCI does not include any self-interaction effects in $\md$. As expected, DCI correctly describes the multi-reference character of the wave function in the dissociation limit. There is a slight overestimate of dynamic correlation near equilibrium, and overestimate of the total energy in dissociation. We discuss possible sources for these errors later. Our conclusion from Fig. \ref{h2_binding} is that the overall construction based on $GW_{\mr}$ quasiparticles, screened inter-quasiparticle interactions, and the double-counting correction $\Delta$ is on stable footing.

\begin{figure}
\begin{centering}
\includegraphics[width=1.0\columnwidth]{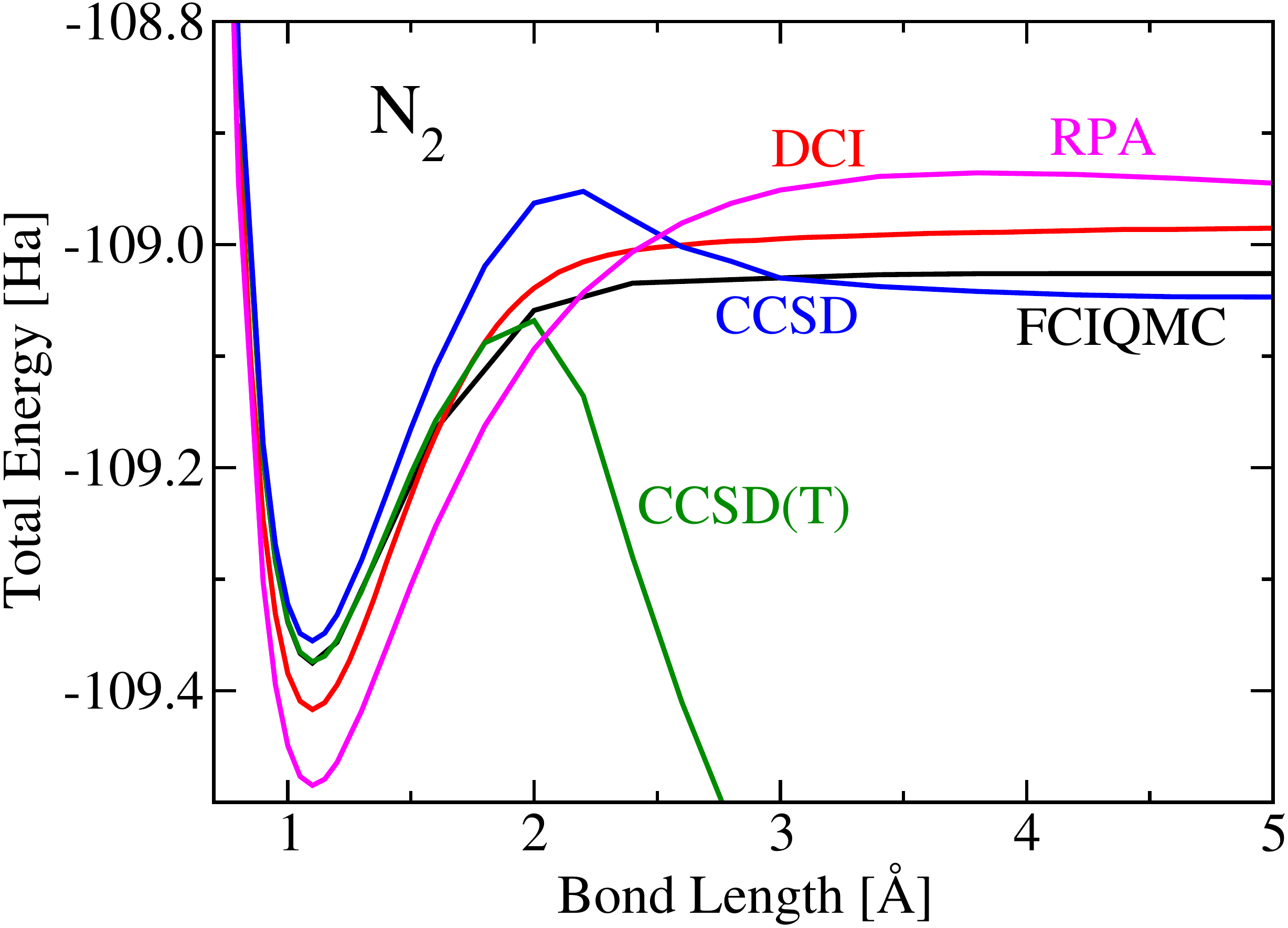}
\includegraphics[width=1.0\columnwidth]{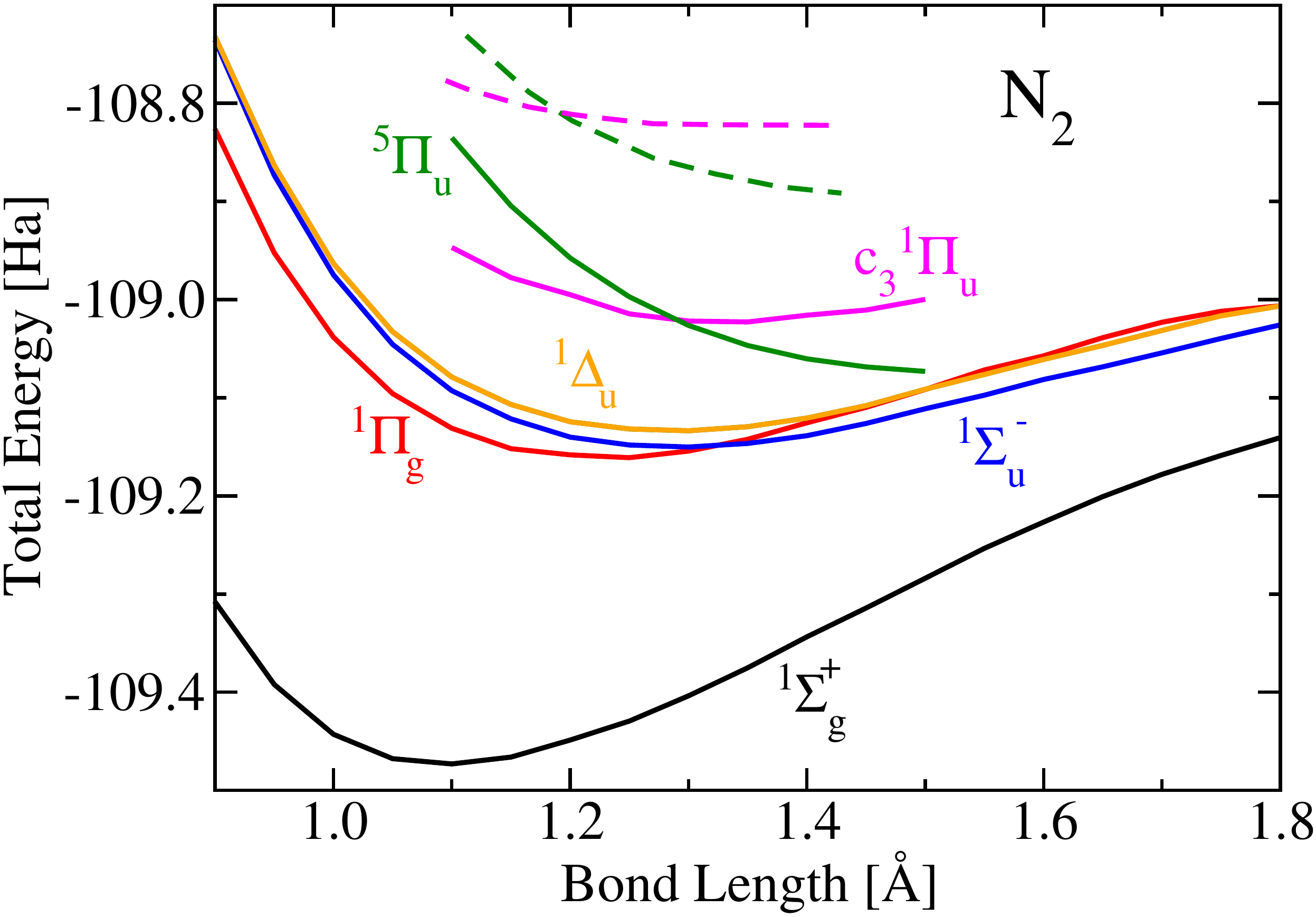}
\includegraphics[width=1.0\columnwidth]{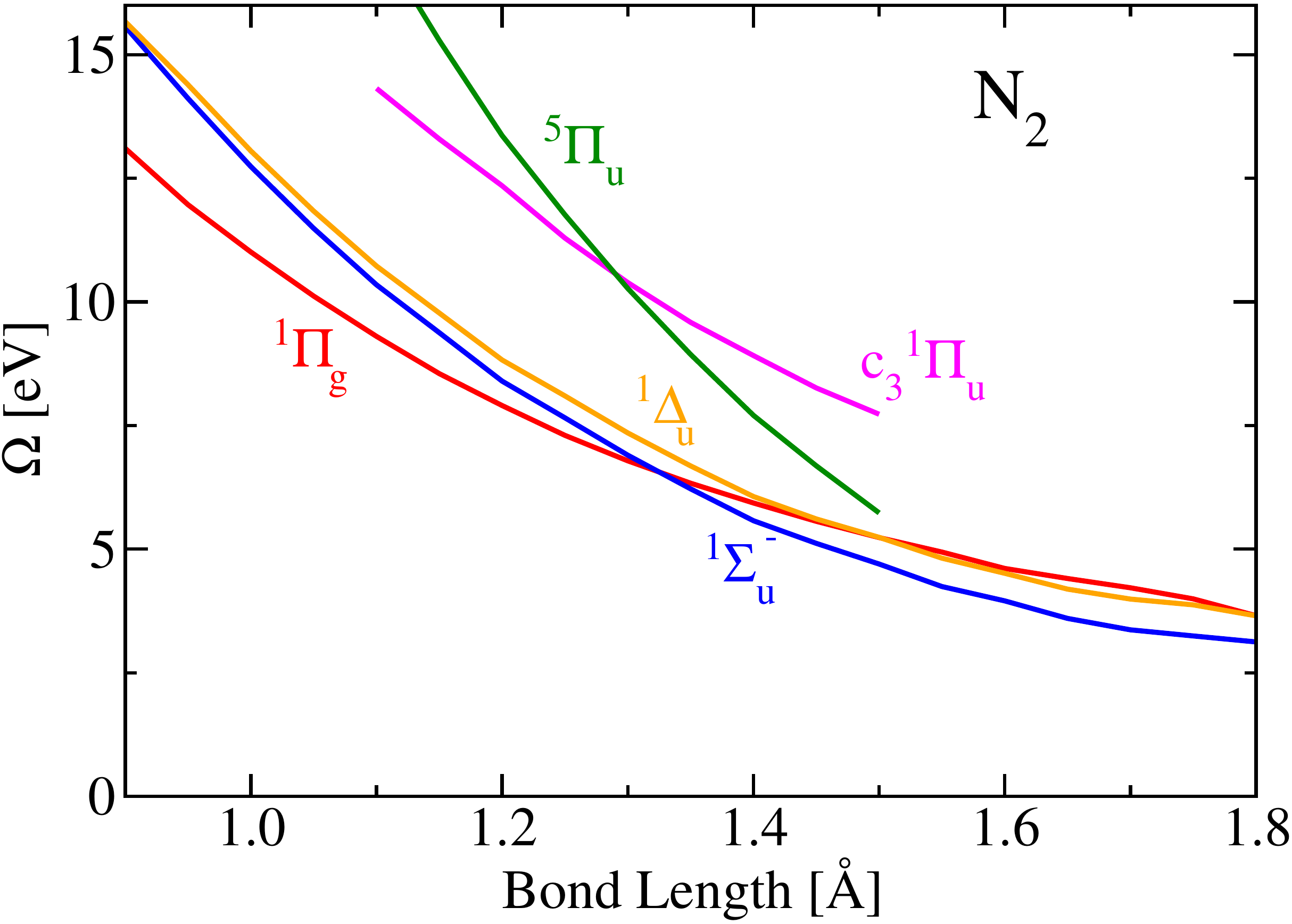}
\caption{N$_2$ dissociation for DCI $(6,6)$. For comparison with reference data,\cite{zhang_prl_117} we use the cc-pVTZ basis for the ground state energy (top). We also compute ground and excited state total energies in the cc-pVQZ basis (center). We show FCI reference data in the cc-pVDZ basis set from Ref.~\onlinecite{larsen_jcp_113}, shown in dashed lines. Only the shape of the curves and intersection are meant for comparison to our results. Plotting only the excitation energies removes the error in the absolute energy of the ground state (bottom). Our DCI calculations do not use spatial symmetries, and we take the state labeling from Ref.~\onlinecite{larsen_jcp_113}.\label{n2_binding}}
\end{centering}
\end{figure}

We next consider the more challenging problem of breaking the triple bond of N$_2$, with our results shown in Fig. \ref{n2_binding}. The nitrogen dimer presents even greater multi-reference character than H$_2$ while adding electrons and dynamic correlation to the dissociation. We use the cc-pVTZ basis set for comparison to our reference data.\cite{zhang_prl_117} To test the quality of the embedding, we use only the minimal $(6,6)$ AS derived from atomic $p$ states. Dissocation in the minimal AS of N$_2$ presents a much more difficult test than the H$_2$ dissociation shown in Fig. \ref{h2_binding}.

At equilibrium, DCI overestimates the dynamic correlation in N$_2$ compared to full configuration interaction quantum Monte Carlo (FCIQMC) results.\cite{zhang_prl_117} Here, we see an error introduced by the underlying MBPT calculation. The BSE with a static vertex underestimates the lowest optical excitation energies of N$_2$.\cite{hirose_prb_91} If \sr excitation energies are underestimated, the poles of $Z^{\mr}(\omega)$ are shifted too far in the $-\omega$ direction, exaggerating the corrections $M(\omega)$ and affecting the total energy. It is difficult to compare the multiple excitation energies in \sr to any benchmark data, but our present calculation suggests that dynamic correlation in \sr is overestimated. A different approximation in \sr may perform better, though improving upon the standard $GW$/BSE approximation in \sr is beyond the scope of our current research.

In the bond breaking regime, our calculation is free of spurious maxima or divergences that appear in single-reference theories. Conventional coupled-cluster with single and double excitations (CCSD), as well as including perturbative triples (CCSD(T)), both fail in this regard. These failures are well documented and understood. DCI correctly describes the multi-reference character of the triple bond. DCI also outperforms RPA across the full dissociation for a modest computational increase.

In the dissociation limit, DCI again overestimates the total energy. We identify two likely sources of error. The neglect of the initial de-excitation time ordering in $\mathcal{L}^{\mr}$ is likely a poor approximation for a multi-reference system. At dissociation, the ground state wave function has high weight on excited determinants so that an initial de-excitation of a noninteracting transition has a large effect. However, going beyond the TDA is not possible in a diagonal approximation. Nonetheless, our results demonstrate that most multi-reference character is contained in $\md$. Furthermore, our estimate of $\Delta$ may introduce some errors in this regime. Errors in either $\Omega_J^{\mr}$ or $\Delta$ have the same effect, which is to shift the poles of $Z^{\mr}(\omega)$ and misrepresent the dynamic correlation in $\mr$. Even at the present level of theory, however, expanding the AS is a clear and systematic route to improve the result. Without state dependent orbital optimization and using only a diagonal approximation in $\mr$, it should be possible to expand the AS to a relatively large size.


A major advantage of DCI is its treatment of excited states. Some strongly-correlated methods, particularly those based on a variational principle, do not have direct access to excited states. In practice, there are often techniques to project out the ground state and compute excitations. The physics behind these approaches is not necessarily transparent, however. Excited states are naturally included in the DCI concept. Furthermore, systematic errors in total energies may cancel with the ground state energy when iterating $\omega$ to compute excitation energies. The internally consistent treatment of correlation, as well as the balanced generation of determinants, may improve the performance of the theory for excitation energies compared to total energies.

To test our description of excited states, we compute the excited state potential energy surfaces (PES) of N$_2$ along the dissociation pathway. Properly describing ground and excited state PES is important for understanding reaction dynamics in quantum chemistry. PES feature avoided crossings and conical intersections along the reaction pathway, whose near-degeneracies and multi-reference character are an extremely demanding test of any theory.

Our N$_2$ PES are shown in Fig. \ref{n2_binding}. Qualitatively, our results are similar to the FCI results in Ref.~\onlinecite{larsen_jcp_113}. The lowest three excited states are relatively similar in shape to the ground state, with some crossings between different levels. These low energy surfaces are described reasonably well by CC (not shown here) in other work. However, higher excited PES feature a difficult conical intersection. The coupled cluster variants tested in Ref.~\onlinecite{larsen_jcp_113} miss this feature. The quintet $^5 \Pi_u$ state is primarily a double excitation, while the crossing $c_3 ^1\Pi_u$ state is a single excitation. Only a multi-reference theory treating both excitation levels equally, without any bias, can properly describe their crossing. In our calculation, all AS excitations up to $6\times$ are treated equally, with each bare excitation gaining a correction $M(\omega)$ from remaining degrees of freedom. Our DCI calculation closely mimics the FCI results of this intersection.

For a more quantitative comparison, we report the low-lying excitation energies at equilibrium in Table \ref{n2_vertical}. Again, we use the $(6,6)$ AS and test the theory against equation-of-motion coupled cluster with single and double excitations (EOM-CCSD), BSE, and experiment. For completeness, we report a small basis set convergence study in the Dunning basis sets\cite{dunning_jcp_90} to the cc-pVQZ level. $GW$/BSE noticeably underestimates the lowest excitation energies of N$_2$. As discussed previously, this is most likely the reason for the overestimate of dynamic correlation near equilibrium in Fig. \ref{n2_binding}. $GW$ relies heavily on the physical picture of screening, which is a relatively small effect in such small molecules, and could lead to large errors for $GW$/BSE in such small systems. In DCI, errors from the screening approximation only appear in the \sr subspace, minimizing the overall error in \sd excitation energies. EOM-CCSD gives good agreement with experiment. For a very small matrix diagonalization, DCI matches the benchmark results.

\begin{figure}
\begin{centering}
\includegraphics[width=1.0\columnwidth]{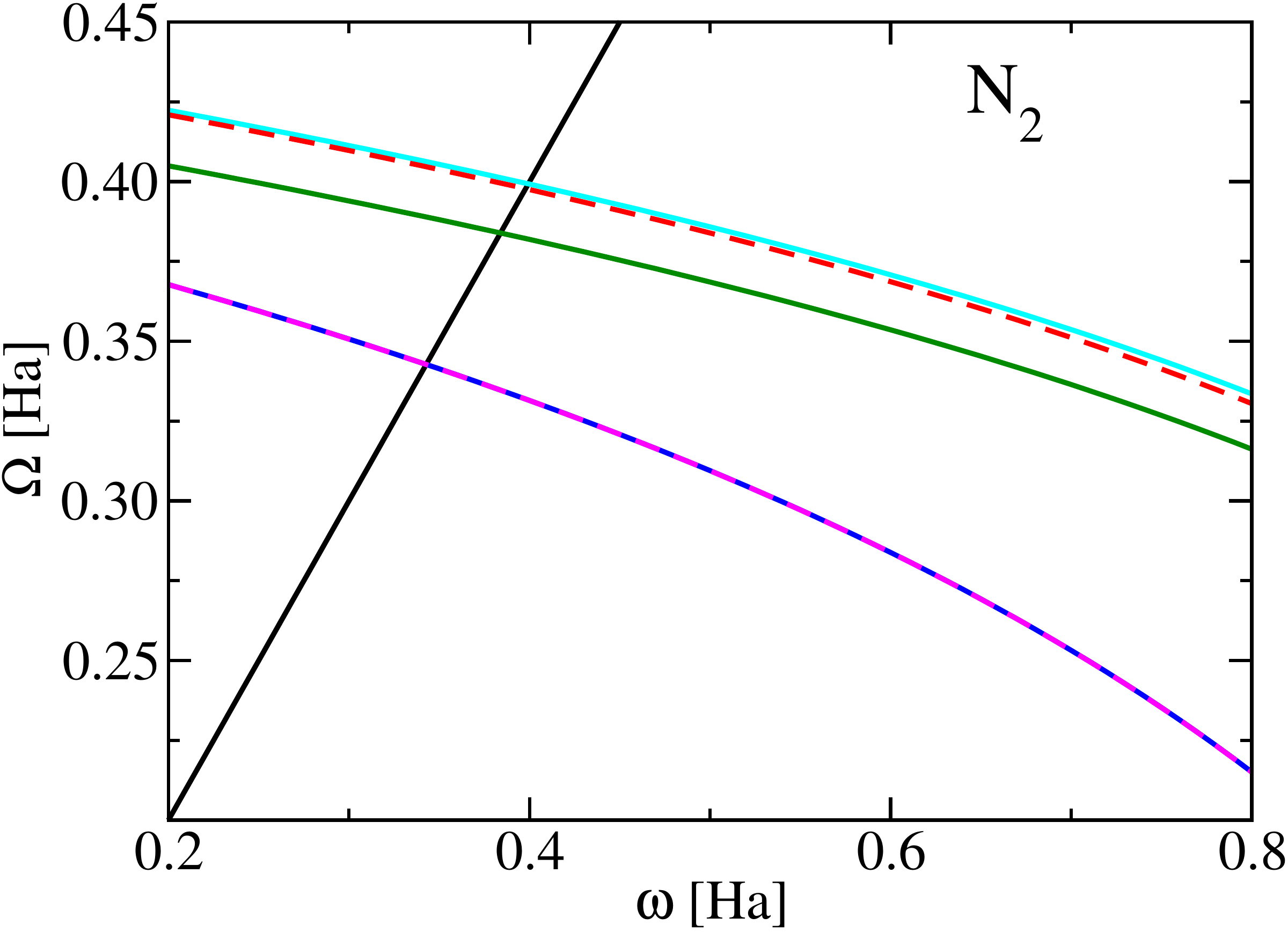}
\includegraphics[width=1.0\columnwidth]{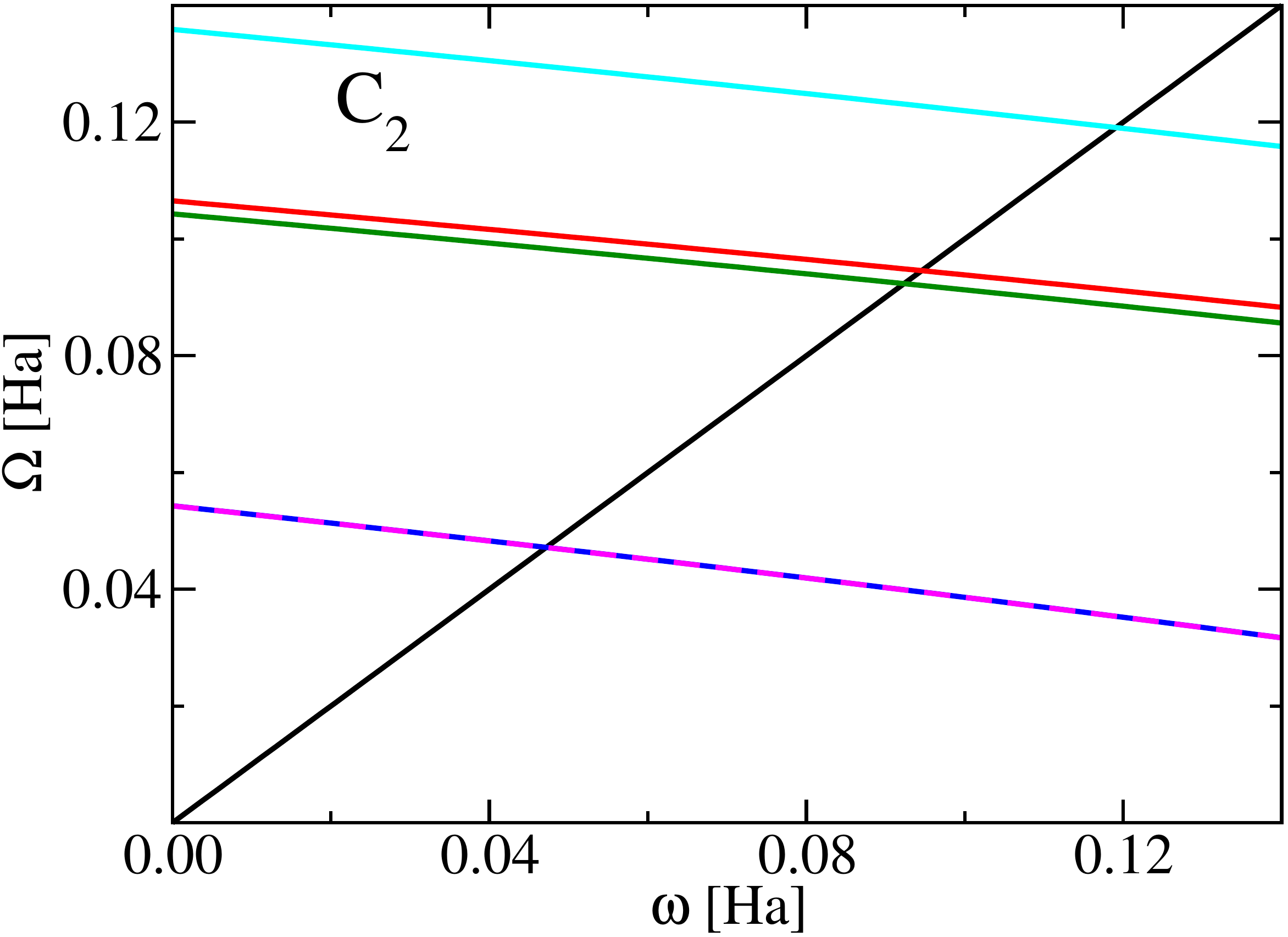}
\caption{Graphical solution for the 5 lowest singlet excitation energies of N$_2$ at the experimental bond length of 1.0977 \AA $\,$ (top). The 5 lowest singlet excitations of C$_2$ at the experimental bond length of 1.2425 \AA $\,$ are also shown (bottom). Both calculations are in the cc-pVQZ basis set. The self-consistent solution for each excitation energy is at its intersection with the diagonal, shown in black. Lines are a linear interpolation between data points which are sampled every $10^{-2}$ Ha.\label{n2graphical}}
\end{centering}
\end{figure}

We also compute the lowest singlet excitations in C$_2$. The carbon dimer presents a similar challenge of multi-configurational character combined with dynamic correlation. At the configuration interaction singles (CIS) level of theory, the first excitation energy in C$_2$ is even $<0$.\cite{head-gordon_jcp_103} DCI balances static and dynamic correlation to match the benchmark results for C$_2$. To elucidate the behavior of the theory, we show graphical solutions for excited states in Fig. \ref{n2graphical}. The graphical solutions show a stronger $\omega$ dependence for N$_2$ than C$_2$. The $\omega$ dependence for the $\Omega_2$ curve of N$_2$, shown in green in Fig. \ref{n2graphical}, is weaker than for $\Omega_1$. The $\Omega_2$ excitation energy is also overestimated with DCI, suggesting that the weaker $\omega$ dependence of $\Omega_2$ could be an error of the theory, AS, or basis set. By placing all high energy excitations in $\mr$, the excitation energies are well-behaved, monotonic functions of $\omega$.

\begin{table}
\caption{Vertical singlet excitation energies (eV) of N$_2$ computed with $GW$/BSE,\cite{hirose_prb_91} EOM-CCSD,\cite{hirose_prb_91,head-gordon_jcp_103} and DCI. Our $(6,6)$ DCI calculations are performed at the experimental bond length of 1.0977 \AA. EOM-CCSD calculations from Refs. ~\onlinecite{hirose_prb_91} and ~\onlinecite{head-gordon_jcp_103} are numerically close to each other for N$_2$.\label{n2_vertical}}
\begin{ruledtabular}
\begin{tabular}{ c c c c }
     N$_2$ & cc-pVDZ & cc-pVTZ & cc-pVQZ   \\
   \hline
  $\Omega_1$ & 9.00  & 9.14 & 9.33     \\
  $\Omega_2$ & 10.33  & 10.30 & 10.45   \\
\end{tabular}
\end{ruledtabular}
\vspace{0.3cm}
\begin{ruledtabular}
\begin{tabular}{ c c c c c }
     N$_2$ & $GW$/BSE & EOM-CCSD & \textbf{DCI} & Exp.\cite{oddershede_cp_97}  \\
   \hline
  $\Omega_1$ & 7.93  & 9.47 & \textbf{9.33} & 9.31    \\
  $\Omega_2$ & 8.29  & 10.08 & \textbf{10.45} & 9.97  \\
\end{tabular}
\end{ruledtabular}
\end{table}

\begin{table}
\caption{Vertical singlet excitation energies (eV) of C$_2$ computed with $GW$/BSE, EOM-CCSD,\cite{hirose_prb_91,head-gordon_jcp_103} and DCI. Our $(6,6)$ DCI calculations are performed at the experimental bond length of 1.2425 \AA. With no reference data available for $GW$/BSE, we perform our own $GW$/BSE calculation for C$_2$ at the $G_0 W_0$@HF level.}
\begin{ruledtabular}
\begin{tabular}{ c c c c }
     C$_2$ & cc-pVDZ & cc-pVTZ & cc-pVQZ   \\
   \hline
  $\Omega_1$ & 1.42  & 1.29 & 1.28     \\
  $\Omega_2$ & 2.70  & 2.52  & 2.51
   \\
\end{tabular}
\end{ruledtabular}
\vspace{0.3cm}
\begin{ruledtabular}
\begin{tabular}{ c c c c c }
   C$_2$ & $GW$/BSE & EOM-CCSD & \textbf{DCI} & Exp.\cite{head-gordon_jcp_103} \\ 
   \hline
   $\Omega_1$ & $< 0.1$ & 1.33 & \textbf{1.28}  & 1.23  \label{c2_vertical} 
\end{tabular}
\end{ruledtabular}
\end{table}

\section{Discussion}


We now discuss particular aspects of DCI and the accuracy of the underlying approximations. We first consider the static approximation to $\mathcal{L}^{\mr}$ and the neglect of $\mathrm{Im} \Sigma$. Both approximations assume that we are dealing with long-lived excitations. Long quasiparticle lifetimes are not guaranteed but often the case for weakly-correlated systems. Long-lived excitations in \sr is an \textit{easier} condition than for the physical excitations of $H$. \sr is effectively a large band gap system, which we expect to be less correlated than the full Hilbert space. By attaching \sr projectors to the interaction, we remove diagrams containing lines near $E_{\mathrm{F}}$ and energetically near each other. This restriction limits the number of possible decay channels and restricts transfer of spectral weight away from the single-particle peaks. Accordingly, subspace quasiparticle lifetimes are longer than their corresponding physical states. In the limit that $\mathcal{R} \rightarrow 0$, $GW_{\mr}$ quasiparticles reach their HF limit and have infinite lifetime. Therefore, the accuracy of a static approximation to $\Omega^{\mr}$ is controlled by the same systematic convergence that expands \sd and shrinks $\mr$.


We use only a diagrammatic construction for $G$ and avoid any mean-field starting point. Even though KS-DFT eigenvalues are often closer to quasiparticle energies than HF eigenvalues, using a mean-field GF ($G^{\mathrm{MF}}$) would introduce a double-counting problem similar to LDA+DMFT. \sr projectors apply only to configurations of noninteracting particles. Therefore, we begin the expansion for $\mathcal{L}^{\mr}$ from a true $\mathcal{L}_0$ based on noninteracting particles, then add only those self-energy insertions which we know can be constructed diagramatically. Under this principle, any insertion can always be decomposed into diagrams containing only $G_0$ lines. Both a self-consistent set of HF diagrams and a $G_0W_{0,\mathcal{R}}$ insertion obey this rule.

Our construction of the kernel $K^{\mr *}_{GW}$ and calculation of excitation energies with Eq. \ref{excitation} is motivated by a physically meaningful approximation. It may be possible that $K^{\mr *}_{GW}$ can be further reduced by cutting of $\mathcal{L}_0^{\mr}$ diagrams in the perturbation expansion. Even so, $K^{\mr *}_{GW}$ does not include \textit{every} possible diagram, and we can still use it as an effective building block for a series expansion. In a Dyson series based on $K^{\mr *}_{GW}$, each term depends on a different power of $K^{\mr *}_{GW}$ so that there are no repeated terms in the expansion. This is an unusual case where it could be advantageous to use a kernel which can be further reduced instead of the actual irreducible kernel. Any freedom in choosing the irreducible building block for a perturbation expansion in approximate theories is very interesting. In our case, we allow the system to propagate for a finite amount of time, gaining interactions and correlation at each time step, before we stop the expansion to form the irreducible part. Then, repeated applications of $K^{\mr *}_{GW}$ build up the full kernel $K^{\mr}_{GW}$.

The general embedding framework is flexible. For large systems, \sd can be generated with truncated CI, for example CISDT,\cite{dvorak_prl} instead of full CI. This approach could be advantageous for large, single-reference systems that require a beyond $GW$ or beyond $GW$/BSE treatment without resorting to expensive coupled cluster or vertex corrections. Hopefully, a small dynamical-CISDT problem can give accurate results at a lower cost. The embedding framework also allows us to go beyond $GW$ at the single quasiparticle level in $\mr$, in principle, as long as the correlation (i.e. the vertex) is properly constrained. Some type of self-consistency in \sr at the single quasiparticle level should also be possible.

A diagonal approximation to $Z^{\mr}$ has some limitations. For example, it cannot correctly treat degenerate states since the degenerate subspace is not diagonalized. Even for degenerate states at high energy, their treatment can be considered a statically correlated problem. However, our goal is to treat only dynamic correlation in $\mr$, which is still possible in a diagonal approximation. The diagonal approximation is best for eigenstates dominated by \sd configurations that are dressed by small contributions from surrounding \sr configurations. Going beyond the diagonal approximation to include initial de-excitations of the system (going beyond the TDA) could improve the performance for multi-reference systems.


As mentioned previously, $GW$/BSE relies heavily on the physical picture of screening, a concept which does not apply extremely well to H$_2$, N$_2$, and C$_2$. These molecules have both low charge density and highly multi-configurational character, which do not lend themselves to a simple screening interpretation. Because the errors of the screening approximation only enter in the \sr subspace, our results for \sd excitations are still in good agreement with other theory and experiment. However, DCI could perform better for larger, more polarizable systems for which $GW$/BSE is already a good approximation. This is more likely to be true in single-reference systems with excited states of a strong single excitation character. If the bath treatment at the $GW$/BSE level is already reasonably accurate, it may be possible to apply DCI to relatively large systems using only a very small AS. This could be the case, for example, in $d$-electron porphyrins or phthalocyanines where the screening concept should apply to the host molecule better than in small dimers. Applying DCI to the $d$-electron AS would treat $d$-electron correlation to all orders while remaining transitions on the host molecule are still described with good accuracy.

\section{Conclusion}
We have introduced a new quantum embedding, or active space, theory that combines aspects of quantum chemistry, GF embedding, and $GW$/BSE theory. The theory is parameter free, systematically improvable, and describes both ground and excited states. It is naturally suited for multi-reference ground states and multi-configurational excited states. Our initial calculations on dimers suggest that DCI could be competitive with high level quantum chemistry methods for excitation energies. Excited states of N$_2$ and C$_2$, which are difficult multi-configurational states, are already well described at the present level of theory. For a conical intersection along the N$_2$ dissociation PES, our theory outperforms conventional CC.


DCI offers several advantages compared to pure $GW$/BSE. Multiple excitations in the AS are easily accessible, and excitations are free of self-screening errors. Correlation in the AS is included to all orders, and the method can be systematically improved by expanding the AS. Compared with AS theories in quantum chemistry based on multi-configurational orbital optimization, DCI is an alternative approach for treating dynamic correlation. With our approach, dynamic correlation is built from the screened Coulomb interaction, a principle which has been extensively tested and validated on weakly-correlated systems. We believe that iterations of $H^{\mathrm{DCI}}$ are computationally and algorithmically simpler than MC-SCF optimizations. Finally, we do not embed the self-energy or vertex which greatly simplifies the frequency structure of the downfolding compared to GF embedding. The self-consistency condition is very simple and non-local correlation is naturally included in both spaces.


\section{Acknowledgements}

This work is supported by the Academy of Finland through grants No. 284621, 305632 and 316347. The authors acknowledge the CSC-IT Center for Science, Finland, for generous computational resources and the Aalto University School of Science ``Science-IT'' project for computational resources. We further acknowledge A. Harju for early discussions on the topic, as well as fruitful discussions with S. Biermann, R. van Leeuwen, and Y. Pavlyukh.

\appendix
\section{Slater-Condon rules}\label{slater_condon}
The Slater-Condon rules \cite{slater_pr_34,condon_pr_36} express matrix elements of the many-body Hamiltonian $H$ as sums over single-particle matrix elements. The rules are well known but we repeat them here for completeness. We assume a fermionic Hamiltonian, orthogonal orbitals, and normalized many-body configurations of $N$ particles. For the Hamiltonian of Eq. \ref{secondquant}, the relevant single-particle matrix elements are of the one-body operator $T$,
\begin{equation}
 \bra{i} t \ket{j} = t_{ij} = \int d\mathbf{r} \; \phi_i^*(\mathbf{r}) \; t \; \phi_j(\mathbf{r})
\end{equation}
where $t = -\frac{\nabla^2}{2} + U_{\mathrm{ext}}(\mathbf{r})$ and the Coulomb operator $v$,
\begin{eqnarray}
\bra{ij} v \ket{kl} &=& v_{ijkl}  \\ 
&=& \int d \mathbf{r} \; d \mathbf{r}' \; \phi_i^*(\mathbf{r}) \phi_j^*(\mathbf{r}') \frac{1}{| \mathbf{r} - \mathbf{r}'|} \phi_k(\mathbf{r}) \phi_l(\mathbf{r}') .  \nonumber
\end{eqnarray}
The direct electronic interaction for a pair of orbitals is given by $v_{ijij}$ while exchange is $v_{ijji}$.

Consider two many-body configurations $\ket{\Psi}$ and $\ket{\Psi_{mn}^{pq}}$ which differ by up to two occupation numbers. $p$ and $q$ denote the occupied states of $\ket{\Psi_{mn}^{pq}}$ that differ from $\ket{\Psi}$, while $m,n$ are the occupied states of $\ket{\Psi}$ which differ from $\ket{\Psi_{mn}^{pq}}$. The number of permutations required to arrange the occupation numbers of $\ket{\Psi}$ and $\ket{\Psi_{mn}^{pq}}$ in maximum coincidence is the number $S$, an integer.

The indices $i$ and $j$ contain both orbital and spin indices. The diagonal matrix element of $H$ is
\begin{equation}
\bra{\Psi} H \ket{\Psi} = \sum_{i \in \Psi}^N t_{ii} + \frac{1}{2} \sum_{i \in \Psi}^N \sum_{j \in \Psi}^N (v_{ijij} - \delta_{\sigma \sigma'} v_{ijji}).
\end{equation}
For off-diagonal matrix elements, each permutation of occupation introduces a sign change so that the total sign of the matrix element is $(-1)^S$. For one occupation number difference,
\begin{equation}
\bra{\Psi} H \ket{\Psi_m^p} =  (-1)^S \left( t_{mp} + \sum_{i \in \Psi}^N (v_{imip} - \delta_{\sigma \sigma'} v_{impi}) \right).
\end{equation}
For two occupation numbers different,
\begin{equation}
\bra{\Psi} H \ket{\Psi_{mn}^{pq}} = (-1)^S \left( v_{nmqp} - \delta_{\sigma \sigma'} v_{nmpq} \right).
\end{equation}
A matrix element between two configurations which differ by more than two occupation numbers is zero. With these rules, one can evaluate matrix elements described in the text of the type $\bra{I} H \ket{I'}$ or $\bra{I} H \ket{J}$.

\section{$\mr$ space with only one configuration}\label{one_conf}
One special case of the embedding is for one configuration in $\mr$. This is the opposite limit of the case $\mr \rightarrow \id$, which matches a normal calculation with MBPT as $W_{\mr} \rightarrow W$. The case of $\mr \rightarrow 0$ is less transparent, and we elaborate on it here. The case of $\mr = 0$ is too trivial since this obviously recovers a FCI calculation. The limits of the embedding are best demonstrated for one configuration in $\mr$, which we denote $\mr = \ket{J} \bra{J}$. For this case, we know the exact diagonalization of $\mr H \mr$ since it is only the single matrix element $\bra{J} H \ket{J}$.

Any diagonal matrix element of $H$, discussed in Appendix \ref{slater_condon}, can be rewritten in terms of the HF eigenvalues. The HF eigenvalues have meaning as particle addition/removal energies because of Koopman's theorem. Using this fact, the sums in Appendix \ref{slater_condon} can be rearranged so that the diagonal matrix element is
\begin{eqnarray}
\Omega_J^{\mathrm{Diag}} &=& \sum_{e \in J}^m \epsilon_e^{\mathrm{HF}} - \sum_{h \in J}^m \epsilon_h^{\mathrm{HF}}   \nonumber  \\  
&+& \sum_{e,h \in J}^m (-v_{eheh} + \delta_{\sigma_e \sigma_h} v_{ehhe} ) \nonumber  \\
&+& \sum_{\mathclap{\substack{e \in J \\
e \neq e'}}}^m \;\; (v_{ee'ee'} - \delta_{\sigma_e \sigma_{e'}} v_{ee'e'e} ) \nonumber \\  
&+& \sum_{\mathclap{\substack{h \in J   \label{exc_diag}  \\
h \neq h'}}}^m \;\; (v_{hh'hh'} - \delta_{\sigma_h \sigma_{h'}} v_{hh'h'h} ).   \\
\bra{J} H \ket{J} &=& E^{\mathrm{ref}} + \Omega_J^{\mathrm{Diag}}.  \label{h_diag}
\end{eqnarray}

This has a clear similarity to Eq. \ref{excitation}, except that the quasiparticle interactions are now unscreened and quasiparticle energies are replaced by HF eigenvalues. Importantly, Eq. \ref{exc_diag} is the \textit{actual limit} of the embeddded Hamiltonian $H^{\mr}$ based on the partially screened interaction $W_{\mr}$. For one configuration in $\mr$, there are no states available for screening and the interaction $W_{\mr}$ becomes the bare $v$. The dynamically screened self-energy $\Sigma = iGW_{\mr}$ approaches its bare exchange limit, and quasiparticle interactions are unscreened. 

Similarly, the ground state energy $E_0^{\mr}$ correctly becomes $E^{\mathrm{ref}}$ in this limit. The exact \sr correlation based on the ED result $E_0^{\mr}=E^{\mathrm{ref}}$ is $C_{\mr \mr}=0$. In our embedding, we estimate $C_{\mr \mr}$ as $E_0 - \tilde{E}_0$. For one configuration in \sr, the energy $\tilde{E}_0$ is very close to the true ground state energy $E_0$. Recall that $\tilde{E}_0$ is computed with the uncorrelated resolvent. In this limit, the uncorrelated resolvent is only a $1 \times 1$ matrix with a single matrix element of $H^{\mathrm{ref}}$ in place of the exact $H$. The effect of one high energy determinant on $E_0$ is very small, and $\tilde{E}_0$ will be very close to $E_0$. Therefore, the limit of the ground state for $\mr = \ket{J} \bra{J}$ is correctly $E_0^{\mr} \rightarrow E^{\mathrm{ref}}$. As with the excitation energy, the \textit{actual limit} of the embedding approaches the correct result.

We conclude that
\begin{equation}
H^{\mr} = \mr H \mr 
\end{equation}
when $\mr = \ket{J} \bra{J}$ and the second limit on the embedding ($W_{\mr} \rightarrow v$, $E_0^{\mr} \rightarrow E^{\mathrm{ref}}$) is both correct and satisfied by the theory. For the above equality to be true in this limit, we ignore self-screening and self-interaction effects that appear in MBPT and are known to be spurious. We also ignore ``self-correlation'' effects to the ground state that cause the small error $0<C_{\mr \mr} \ll 1$ because $\tilde{E}_0$ is not \textit{exactly} $\tilde{E}_0$ in this case. These self-interaction errors are expected to be extremely small.

\section{Resolvents and many-body Green's functions}\label{resolvents}
Here, we highlight the similarities and differences between the resolvent of the Hamiltonian and many-body Green's functions to improve understanding of the theory.

First, consider the resolvent of the full Hamiltonian.
\begin{equation}
Z(E) = \left( E - H \right) ^{-1} = \frac{1}{ E - H}
\end{equation}
In the eigenbasis of $H$ denoted $\{ \ket{N} \}$, both $\left( E-H \right)$ and its inverse are diagonal. In this diagonal basis, the effective amplitudes in the numerator are all one.
\begin{equation}
Z_{N N'} (E) = \frac{ \delta_{N N'} }{ E - E_N }
\end{equation}
$Z(E)$ has poles at the total energies of the system. By adding and subtracting the ground state energy $E_0$ to the denominator, we introduce the new argument $\omega = E - E_0$.
\begin{eqnarray}
Z_{NN'}(\omega) &=&  \frac{ \delta_{N N'} }{ (E - E_0) - (E_N - E_0) }  \nonumber  \\
&=& \frac{ \delta_{N N'} }{\omega - (E_N - E_0) }  \label{full_resolvent}
\end{eqnarray}
$\omega$ has the interpretation of an excitation energy, and poles of $Z$ are at exact excitation energies of the system.

Now examine the exact $\mathcal{L}(\omega)$. We again assume all the many-body eigenstates are known. We write the Lehmann representation for $\mathcal{L}(\omega)$ as
\begin{eqnarray}
\mathcal{L}_{JJ'}(\omega) &=& \sum_{N \ne 0} \frac{ \bra{\Psi} \widehat{\Omega}_J \ket{N} \bra{N} \widehat{\Omega}_{J'}^{\dagger} \ket{\Psi} }{ \omega - (E_{N} - E_0) + i \eta} \nonumber \\
&-& \sum_{N \ne 0} \frac{ \bra{\Psi} \widehat{\Omega}_{J'}^{\dagger} \ket{N} \bra{N} \widehat{\Omega}_{J} \ket{\Psi} }{ \omega + (E_{N} - E_0) - i \eta}  \label{bigl_lehmann_exact}
\end{eqnarray}

The resolvent $Z$ lacks both the Lehmann amplitudes and different time-orderings of the correlation function, but has the same frequency argument $\omega$ and energy differences in the denominator. Only on a heuristic level, we conclude that there is some correspondence between the resolvent $Z$ and the many-body GF $\mathcal{L}$. This naive correspondence is enough for the present discussion.


Next, we focus on the auxiliary eigenvalue problem in the \sr subspace. We must change the notation from the main text to accommodate offdiagonal matrix elements. $J$ and $J'$ refer to noninteracting excitations, while exact eigenstates of $\mr H \mr$ are labeled $E_{N}^{\mr}$.
\begin{equation}
\left[ \mr H \mr \right] \chi = E^{\mr} \chi \label{sub_schrodinger}
\end{equation}
that determines the eigenvalues $E_N^{\mr}$. Its resolvent is
\begin{equation}
Y^{\mr} ( E^{\mr} ) = \frac{1}{E^{\mr} - \mr H \mr}.
\end{equation}
Using the frequency $\omega^{\mr} = E^{\mr} - E_0^{\mr}$, we change the argument for the resolvent to $\omega^{\mr}$.
\begin{eqnarray}
Y^{\mr} (\omega^{\mr}) &=& \frac{1}{ (E^{\mr} - E_0^{\mr}) - (\mr H \mr - E_0^{\mr})} \nonumber \\
&=& \frac{1}{\omega^{\mr} - (\mr H \mr - E_0^{\mr}) }  
\end{eqnarray}
In the basis which diagonalizes $\mr H \mr$, $\{ \ket{N^{\mr}} \}$,
\begin{equation}
Y^{\mr}_{NN'}(\omega^{\mr}) = \frac{ \delta_{NN'} } { \omega^{\mr} - (E_N^{\mr} - E_0^{\mr}) }. \label{sub_resolvent}
\end{equation}

Consider the many-body correlation function $\mathcal{L}^{\mr}$ defined with respect to the \sr space ground state. Here, we mean particle addition/removal to the fictitious ground state, as described in the text, so that the physical problem emerges by increasing the size of $\mr$. Similar to the case of the full $\mathcal{L}$, the energy difference entering the Fourier transform of $\mathcal{L}^{\mr}$ is $\omega^{\mr} = E^{\mr} - E_0^{\mr}$. The poles of $\mathcal{L}^{\mr}$ are at the total energy differences $E_N^{\mr} - E_0^{\mr}$. It shares these characteristics with $Y^{\mr}$. $\mathcal{L}^{\mr}$ is 
\begin{eqnarray}
\mathcal{L}^{\mr}_{JJ'}(\omega^{\mr}) &=& \sum_{N^{\mr} \ne 0} \frac{ \bra{\Psi} \widehat{\Omega}_J^{\mr} \ket{N^{\mr}} \bra{N^{\mr}} \widehat{\Omega}_{J'}^{\mr \dagger} \ket{\Psi} }{ \omega^{\mr} - (E_N^{\mr} - E_0^{\mr}) + i \eta} \nonumber \\
&-& \sum_{N^{\mr} \ne 0} \frac{ \bra{\Psi} \widehat{\Omega}_{J'}^{\mr \dagger} \ket{N^{\mr}} \bra{N^{\mr}} \widehat{\Omega}_{J}^{\mr} \ket{\Psi} }{ \omega^{\mr} + (E_N^{\mr} - E_0^{\mr}) - i \eta}  \label{lehmann_subspace}
\end{eqnarray}
Comparing Eqs. \ref{sub_resolvent} and \ref{lehmann_subspace}, there is some correspondence between $Y_{\mr}$ and the correlation function $\mathcal{L}^{\mr}$. They share the frequency argument $\omega^{\mr}$ and pole positions $E_N^{\mr} - E_0^{\mr}$.

Finally, we examine the resolvent $Z^{\mr}$, as defined in the main text. This resolvent is of fundamentally different character than the previous two since the total energy $E$ cannot be an eigenvalue of the operator with which it shares the denominator $-$ this resolvent mixes the two spaces. The exact resolvent is
\begin{equation}
Z^{\mr}(E) = \frac{1}{  E - \mr H \mr}. 
\end{equation}
We again use the frequency $\omega = E - E_0$.
\begin{eqnarray}
Z^{\mr}(\omega) &=& \frac{1}{ (E-E_0) - (\mr H \mr - E_0)}   \nonumber  \\
&=& \frac{1}{ \omega - (E_N^{\mr} - E_0) } \label{fouriert}
\end{eqnarray}
in the eigenbasis of $\mr H \mr$. The poles of $Z^{\mr}$ are at the total energy differences $E_N^{\mr} - E_0$. Energy differences of this type, which mix spaces, are difficult to interpret.

Consider trying to construct a correlation function with this form. We assume there is an initial ground state in some space. If an excitation is created in the same space containing the initial ground state, only total energy differences of the type $E_N^{\mr} - E_0^{\mr}$ or $E_N - E_0$ can appear in the complex exponentials describing the interference among eigenstates. After Fourier transforming such a GF, these energy differences determine the pole positions in the denominator. This characteristic of the GF relies on the fundamental requirement that the excited state and ground state exist in the same space. This condition is \textit{not} met by the denominator of $Z^{\mr}$ in Eq. \ref{fouriert}.

Therefore, we see no readily apparent connection between the resolvent $Z^{\mr}$ and a many-body correlation function. $\mathcal{L}^{\mr}$ is \textit{not} a direct approximation to, or reinterpretation of, the resolvent $Z^{\mr}$. Our estimate of excitation energies with $\mathcal{L}^{\mr}$ is more similar to an order-by-order, energy dependent calculation of $Y^{\mr}$ than of $Z^{\mr}$. The energy dependence of an order-by-order construction of $Y^{\mr}$ is of $E^{\mr}$, not $E$.

This is no surprise if we reconsider the way the theory is constructed. The guiding principle of our theory is that $E_0^{\mr}$ and $\Omega_N^{\mr}$ should be chosen so that the equality
\begin{equation}
E_N^{\mr} = E_0^{\mr} + \Omega_N^{\mr}
\end{equation}
holds. $E_N^{\mr}$ are the eigenvalues of Eq. \ref{sub_schrodinger}, so it follows that the relevant excitation energies are derived from the resolvent of Eq. \ref{sub_schrodinger}. That resolvent is $Y^{\mr}$, or its corresponding correlation function $\mathcal{L}^{\mr}$, but not $Z^{\mr}$.

\section{Connection to Green's function embedding}\label{gf_embedding}
Having established some correspondence between certain resolvents and many-body GFs, we seek a stronger connection (at a heuristic level) to GF embedding theories.

We start by reconsidering the resolvent $Z^{\mr}$.
\begin{equation}
Z^{\mr} = \frac{1}{E - \mr H \mr} = \frac{1}{\left(\omega - \Delta \right) - \Omega^{\mr}}
\end{equation}
In general, the excitation matrix $\Omega^{\mr}$ is not known. It might be estimated with an order-by-order construction of the resolvent $Y^{\mr}$, similar to MBPT. Along this line, assume we know the resolvent instead of the excitation matrix. We search for a way to write $\Omega^{\mr}$ in terms of $Y^{\mr}$.
\begin{eqnarray}
Y^{\mr} &=& \frac{1}{ E^{\mr} - \mr H \mr}  \nonumber \\
&=& \frac{1}{ \omega^{\mr} - \Omega^{\mr} }    \nonumber  \\
\implies \Omega^{\mr} &=& \omega^{\mr} - (Y^{\mr})^{-1}
\end{eqnarray}
By inverting the eigenvalue problem for $\Omega^{\mr}$, we introduce the frequency $\omega^{\mr}$ to the problem. Written in terms of $Y^{\mr}$ instead of $\Omega^{\mr}$, the resolvent $Z^{\mr}$ gains a second frequency dependence.
\begin{eqnarray}
Z^{\mr} &=& \frac{1}{ \left( \omega - \Delta \right) - \Omega^{\mr} }  \label{freq_zr}  \nonumber  \\
Z^{\mr} &=& \frac{1}{ \left( \omega - \Delta \right) - \left( \omega^{\mr} - (Y^{\mr})^{-1} \right) } 
\end{eqnarray}
Had we chosen to work with the resolvent $Y^{\mr}$ instead of the Hamiltonian $\mr H \mr$ (or our approximation to it), our formalism would have acquired a second frequency dependence. The matrix elements which complete the energy dependent corrections, $\md H \mr$ and $\mr H \md$, still have no frequency dependence.

We take the connection to GF embedding one step further by re-examining the exact resolvent $Z$. We rewrite the exact resolvent based on the downfolded Hamiltonian.
\begin{equation}
Z = \frac{1}{E - H} = \frac{1}{ E - \left( \md H \md + M(\omega) \right) }
\end{equation}
Inserting the explicit form for the self-energy corrections $M(\omega)$,
\begin{eqnarray}
Z &=& \frac{1}{ E - \left( \md H \md + \left[ \md H \mr \right] Z^{\mr} \left[ \mr H \md \right] \right) }     \\
&=& \frac{1}{ E - \left( \md H \md + \left[ \md H \mr \right] \left[ \left( \omega - \Delta \right) - \Omega^{\mr} \right]^{-1}  \left[ \mr H \md \right] \right) }  \nonumber
\end{eqnarray}

Using the second line of Eq. \ref{freq_zr}, we can rewrite the exact resolvent $Z$ only in terms of $Y^{\mr}$, frequencies, and matrix elements of $H$.
\begin{widetext}
\begin{eqnarray}
Z &=& \frac{1}{ E - \left( \md H \md + \left[ \md H \mr \right] \left[ \left( \omega - \Delta \right) - \left(\omega^{\mr} - (Y^{\mr})^{-1}\right) \right]^{-1} \left[ \mr H \md \right] \right) } \nonumber  \\
Z &=& \frac{1}{ (E-E_0) - \left( \md H \md + \left[ \md H \mr \right] \left[ \left( \omega - \Delta \right) - \left(\omega^{\mr} - (Y^{\mr})^{-1}\right) \right]^{-1} \left[ \mr H \md \right] - E_0 \right) }    \nonumber  \\
Z( \omega, \omega^{\mr} ) &=& \frac{1}{ \omega - \left( \md H \md + \left[ \md H \mr \right] \left[ \left( \omega - \omega^{\mr} - \Delta \right) +  (Y^{\mr})^{-1} \right]^{-1} \left[ \mr H \md \right] - E_0 \right) }  \label{longz}
\end{eqnarray}
\end{widetext}
The mixed resolvent which cannot be related to a many-body GF, $Z^{\mr}$, has been eliminated. On a heuristic level, the embedded resolvent problem in Eq. \ref{longz} is similar to the expressions encountered in GF embedding theories. It has a double frequency dependence, and the full $Z$ is coupled to the subspace resolvent $Y^{\mr}$.

We identify the two equations for $Z^{\mr}$ in Eq. \ref{freq_zr} as a conceptual difference distinguishing our approach from GF embedding. Our theory is based on the first line of Eq. \ref{freq_zr}, while GF embedding is more closely related to the second. By introducing a second resolvent from the auxiliary eigenvalue problem into Eq. \ref{freq_zr}, the equations gain a second frequency dependence. This implies an outer self-consistency loop on the second frequency $\omega^{\mr}$. 

Since we choose to calculate $\Omega^{\mr}$ with GF, our approach \textit{does} have an $\omega^{\mr}$ dependence in $\mathcal{L}^{\mr}(\omega^{\mr})$. However, we choose a single frequency for each matrix element. We evaluate each matrix element of $\Omega^{\mr}$ at a \textit{different} frequency, determined by the quasiparticle energies for the excited particles. Because we do not need frequency dependent kernels to access multiple excitations, it is efficient to evaluate the $\omega^{\mr}$ dependence of $\mathcal{L}^{\mr}(\omega^{\mr})$ only once for each \sr excitation. We \textit{effectively} eliminate the $\omega^{\mr}$ dependence this way. A similar static approximation based on the single-particle $G$ or electron-hole $L$ would not give the same amount of information. If embedding $G$ or $L$, frequency integrals are necessary to couple to multiple excitations.

\bibliographystyle{apsrev4-1}
\bibliography{dci_theory}

\begin{thebibliography}{91}%
\makeatletter
\providecommand \@ifxundefined [1]{%
 \@ifx{#1\undefined}
}%
\providecommand \@ifnum [1]{%
 \ifnum #1\expandafter \@firstoftwo
 \else \expandafter \@secondoftwo
 \fi
}%
\providecommand \@ifx [1]{%
 \ifx #1\expandafter \@firstoftwo
 \else \expandafter \@secondoftwo
 \fi
}%
\providecommand \natexlab [1]{#1}%
\providecommand \enquote  [1]{``#1''}%
\providecommand \bibnamefont  [1]{#1}%
\providecommand \bibfnamefont [1]{#1}%
\providecommand \citenamefont [1]{#1}%
\providecommand \href@noop [0]{\@secondoftwo}%
\providecommand \href [0]{\begingroup \@sanitize@url \@href}%
\providecommand \@href[1]{\@@startlink{#1}\@@href}%
\providecommand \@@href[1]{\endgroup#1\@@endlink}%
\providecommand \@sanitize@url [0]{\catcode `\\12\catcode `\$12\catcode
  `\&12\catcode `\#12\catcode `\^12\catcode `\_12\catcode `\%12\relax}%
\providecommand \@@startlink[1]{}%
\providecommand \@@endlink[0]{}%
\providecommand \url  [0]{\begingroup\@sanitize@url \@url }%
\providecommand \@url [1]{\endgroup\@href {#1}{\urlprefix }}%
\providecommand \urlprefix  [0]{URL }%
\providecommand \Eprint [0]{\href }%
\providecommand \doibase [0]{http://dx.doi.org/}%
\providecommand \selectlanguage [0]{\@gobble}%
\providecommand \bibinfo  [0]{\@secondoftwo}%
\providecommand \bibfield  [0]{\@secondoftwo}%
\providecommand \translation [1]{[#1]}%
\providecommand \BibitemOpen [0]{}%
\providecommand \bibitemStop [0]{}%
\providecommand \bibitemNoStop [0]{.\EOS\space}%
\providecommand \EOS [0]{\spacefactor3000\relax}%
\providecommand \BibitemShut  [1]{\csname bibitem#1\endcsname}%
\let\auto@bib@innerbib\@empty
\bibitem [{\citenamefont {Fetter}\ and\ \citenamefont
  {Walecka}(1971)}]{fetter_quantum}%
  \BibitemOpen
  \bibfield  {author} {\bibinfo {author} {\bibfnamefont {A.~L.}\ \bibnamefont
  {Fetter}}\ and\ \bibinfo {author} {\bibfnamefont {J.~D.}\ \bibnamefont
  {Walecka}},\ }\href@noop {} {\emph {\bibinfo {title} {Quantum Theory of
  Many-Particle Systems}}}\ (\bibinfo  {publisher} {McGraw-Hill},\ \bibinfo
  {address} {Boston},\ \bibinfo {year} {1971})\BibitemShut {NoStop}%
\bibitem [{\citenamefont {Altland}\ and\ \citenamefont
  {Simons}(2010)}]{altland_field_theory}%
  \BibitemOpen
  \bibfield  {author} {\bibinfo {author} {\bibfnamefont {A.}~\bibnamefont
  {Altland}}\ and\ \bibinfo {author} {\bibfnamefont {B.~D.}\ \bibnamefont
  {Simons}},\ }\href@noop {} {\emph {\bibinfo {title} {Condensed matter field
  theory}}},\ \bibinfo {edition} {2nd}\ ed.\ (\bibinfo  {publisher} {Cambridge
  University Press},\ \bibinfo {year} {2010})\BibitemShut {NoStop}%
\bibitem [{\citenamefont {Bockstedte}\ \emph {et~al.}(2018)\citenamefont
  {Bockstedte}, \citenamefont {Sch\"utz}, \citenamefont {Garratt},
  \citenamefont {Iv\'ady},\ and\ \citenamefont {Gali}}]{bockstedte_npjqm_3}%
  \BibitemOpen
  \bibfield  {author} {\bibinfo {author} {\bibfnamefont {M.}~\bibnamefont
  {Bockstedte}}, \bibinfo {author} {\bibfnamefont {F.}~\bibnamefont
  {Sch\"utz}}, \bibinfo {author} {\bibfnamefont {T.}~\bibnamefont {Garratt}},
  \bibinfo {author} {\bibfnamefont {V.}~\bibnamefont {Iv\'ady}}, \ and\
  \bibinfo {author} {\bibfnamefont {A.}~\bibnamefont {Gali}},\ }\href@noop {}
  {\bibfield  {journal} {\bibinfo  {journal} {NPJ Quant. Mat.}\ }\textbf
  {\bibinfo {volume} {3}},\ \bibinfo {pages} {31} (\bibinfo {year}
  {2018})}\BibitemShut {NoStop}%
\bibitem [{\citenamefont {Sun}\ and\ \citenamefont {Chan}(2016)}]{chan_acr_49}%
  \BibitemOpen
  \bibfield  {author} {\bibinfo {author} {\bibfnamefont {Q.}~\bibnamefont
  {Sun}}\ and\ \bibinfo {author} {\bibfnamefont {G.~K.-L.}\ \bibnamefont
  {Chan}},\ }\href@noop {} {\bibfield  {journal} {\bibinfo  {journal} {Accounts
  of Chemical Research}\ }\textbf {\bibinfo {volume} {49}},\ \bibinfo {pages}
  {2705} (\bibinfo {year} {2016})}\BibitemShut {NoStop}%
\bibitem [{\citenamefont {Nolting}(1971)}]{nolting_quantum}%
  \BibitemOpen
  \bibfield  {author} {\bibinfo {author} {\bibfnamefont {W.}~\bibnamefont
  {Nolting}},\ }\href@noop {} {\emph {\bibinfo {title} {Fundamentals of
  many-body physics}}}\ (\bibinfo  {publisher} {Springer-Verlag Berlin
  Heidelberg},\ \bibinfo {address} {Berlin},\ \bibinfo {year}
  {1971})\BibitemShut {NoStop}%
\bibitem [{\citenamefont {Hedin}(1965)}]{hedin_pr_139}%
  \BibitemOpen
  \bibfield  {author} {\bibinfo {author} {\bibfnamefont {L.}~\bibnamefont
  {Hedin}},\ }\href@noop {} {\bibfield  {journal} {\bibinfo  {journal} {Phys.
  Rev.}\ }\textbf {\bibinfo {volume} {139}},\ \bibinfo {pages} {A796} (\bibinfo
  {year} {1965})}\BibitemShut {NoStop}%
\bibitem [{\citenamefont {Aryasetiawan}\ and\ \citenamefont
  {Gunnarsson}(1998)}]{aryasetiawan_rpp_61}%
  \BibitemOpen
  \bibfield  {author} {\bibinfo {author} {\bibfnamefont {F.}~\bibnamefont
  {Aryasetiawan}}\ and\ \bibinfo {author} {\bibfnamefont {O.}~\bibnamefont
  {Gunnarsson}},\ }\href@noop {} {\bibfield  {journal} {\bibinfo  {journal}
  {Reports on Progress in Physics}\ }\textbf {\bibinfo {volume} {61}},\
  \bibinfo {pages} {237} (\bibinfo {year} {1998})}\BibitemShut {NoStop}%
\bibitem [{\citenamefont {Hybertsen}\ and\ \citenamefont
  {Louie}(1986)}]{louie_prb_34}%
  \BibitemOpen
  \bibfield  {author} {\bibinfo {author} {\bibfnamefont {M.~S.}\ \bibnamefont
  {Hybertsen}}\ and\ \bibinfo {author} {\bibfnamefont {S.~G.}\ \bibnamefont
  {Louie}},\ }\href@noop {} {\bibfield  {journal} {\bibinfo  {journal} {Phys.
  Rev. B}\ }\textbf {\bibinfo {volume} {34}},\ \bibinfo {pages} {5390}
  (\bibinfo {year} {1986})}\BibitemShut {NoStop}%
\bibitem [{\citenamefont {Salpeter}\ and\ \citenamefont
  {Bethe}(1951)}]{salpeter_pr_84}%
  \BibitemOpen
  \bibfield  {author} {\bibinfo {author} {\bibfnamefont {E.~E.}\ \bibnamefont
  {Salpeter}}\ and\ \bibinfo {author} {\bibfnamefont {H.~A.}\ \bibnamefont
  {Bethe}},\ }\href@noop {} {\bibfield  {journal} {\bibinfo  {journal} {Phys.
  Rev.}\ }\textbf {\bibinfo {volume} {84}},\ \bibinfo {pages} {1232} (\bibinfo
  {year} {1951})}\BibitemShut {NoStop}%
\bibitem [{\citenamefont {Rohlfing}\ and\ \citenamefont
  {Louie}(2000)}]{rohlfing_prb_62}%
  \BibitemOpen
  \bibfield  {author} {\bibinfo {author} {\bibfnamefont {M.}~\bibnamefont
  {Rohlfing}}\ and\ \bibinfo {author} {\bibfnamefont {S.~G.}\ \bibnamefont
  {Louie}},\ }\href@noop {} {\bibfield  {journal} {\bibinfo  {journal} {Phys.
  Rev. B}\ }\textbf {\bibinfo {volume} {62}},\ \bibinfo {pages} {4927}
  (\bibinfo {year} {2000})}\BibitemShut {NoStop}%
\bibitem [{\citenamefont {Blase}\ \emph {et~al.}(2018)\citenamefont {Blase},
  \citenamefont {Duchemin},\ and\ \citenamefont {Jacquemin}}]{blase_csr_47}%
  \BibitemOpen
  \bibfield  {author} {\bibinfo {author} {\bibfnamefont {X.}~\bibnamefont
  {Blase}}, \bibinfo {author} {\bibfnamefont {I.}~\bibnamefont {Duchemin}}, \
  and\ \bibinfo {author} {\bibfnamefont {D.}~\bibnamefont {Jacquemin}},\
  }\href@noop {} {\bibfield  {journal} {\bibinfo  {journal} {Chem. Soc. Rev.}\
  }\textbf {\bibinfo {volume} {47}},\ \bibinfo {pages} {1022} (\bibinfo {year}
  {2018})}\BibitemShut {NoStop}%
\bibitem [{\citenamefont {Albrecht}\ \emph {et~al.}(1998)\citenamefont
  {Albrecht}, \citenamefont {Reining}, \citenamefont {Del~Sole},\ and\
  \citenamefont {Onida}}]{albrecht_prl_80}%
  \BibitemOpen
  \bibfield  {author} {\bibinfo {author} {\bibfnamefont {S.}~\bibnamefont
  {Albrecht}}, \bibinfo {author} {\bibfnamefont {L.}~\bibnamefont {Reining}},
  \bibinfo {author} {\bibfnamefont {R.}~\bibnamefont {Del~Sole}}, \ and\
  \bibinfo {author} {\bibfnamefont {G.}~\bibnamefont {Onida}},\ }\href@noop {}
  {\bibfield  {journal} {\bibinfo  {journal} {Phys. Rev. Lett.}\ }\textbf
  {\bibinfo {volume} {80}},\ \bibinfo {pages} {4510} (\bibinfo {year}
  {1998})}\BibitemShut {NoStop}%
\bibitem [{\citenamefont {Sangalli}\ \emph {et~al.}(2011)\citenamefont
  {Sangalli}, \citenamefont {Romaniello}, \citenamefont {Onida},\ and\
  \citenamefont {Marini}}]{marini_jcp_134}%
  \BibitemOpen
  \bibfield  {author} {\bibinfo {author} {\bibfnamefont {D.}~\bibnamefont
  {Sangalli}}, \bibinfo {author} {\bibfnamefont {P.}~\bibnamefont
  {Romaniello}}, \bibinfo {author} {\bibfnamefont {G.}~\bibnamefont {Onida}}, \
  and\ \bibinfo {author} {\bibfnamefont {A.}~\bibnamefont {Marini}},\
  }\href@noop {} {\bibfield  {journal} {\bibinfo  {journal} {J. Chem. Phys.}\
  }\textbf {\bibinfo {volume} {134}},\ \bibinfo {pages} {034115} (\bibinfo
  {year} {2011})}\BibitemShut {NoStop}%
\bibitem [{\citenamefont {Bruneval}\ \emph {et~al.}(2015)\citenamefont
  {Bruneval}, \citenamefont {Hamed},\ and\ \citenamefont
  {Neaton}}]{bruneval_jcp_142}%
  \BibitemOpen
  \bibfield  {author} {\bibinfo {author} {\bibfnamefont {F.}~\bibnamefont
  {Bruneval}}, \bibinfo {author} {\bibfnamefont {S.~M.}\ \bibnamefont {Hamed}},
  \ and\ \bibinfo {author} {\bibfnamefont {J.~B.}\ \bibnamefont {Neaton}},\
  }\href@noop {} {\bibfield  {journal} {\bibinfo  {journal} {J. Chem. Phys.}\
  }\textbf {\bibinfo {volume} {142}},\ \bibinfo {eid} {244101} (\bibinfo {year}
  {2015})}\BibitemShut {NoStop}%
\bibitem [{\citenamefont {Romaniello}\ \emph
  {et~al.}(2009{\natexlab{a}})\citenamefont {Romaniello}, \citenamefont
  {Sangalli}, \citenamefont {Berger}, \citenamefont {Sottile}, \citenamefont
  {Molinari}, \citenamefont {Reining},\ and\ \citenamefont
  {Onida}}]{romaniello_jcp_130}%
  \BibitemOpen
  \bibfield  {author} {\bibinfo {author} {\bibfnamefont {P.}~\bibnamefont
  {Romaniello}}, \bibinfo {author} {\bibfnamefont {D.}~\bibnamefont
  {Sangalli}}, \bibinfo {author} {\bibfnamefont {J.~A.}\ \bibnamefont
  {Berger}}, \bibinfo {author} {\bibfnamefont {F.}~\bibnamefont {Sottile}},
  \bibinfo {author} {\bibfnamefont {L.~G.}\ \bibnamefont {Molinari}}, \bibinfo
  {author} {\bibfnamefont {L.}~\bibnamefont {Reining}}, \ and\ \bibinfo
  {author} {\bibfnamefont {G.}~\bibnamefont {Onida}},\ }\href@noop {}
  {\bibfield  {journal} {\bibinfo  {journal} {J. Chem. Phys.}\ }\textbf
  {\bibinfo {volume} {130}},\ \bibinfo {pages} {044108} (\bibinfo {year}
  {2009}{\natexlab{a}})}\BibitemShut {NoStop}%
\bibitem [{\citenamefont {Attaccalite}\ \emph {et~al.}(2011)\citenamefont
  {Attaccalite}, \citenamefont {Gr\"uning},\ and\ \citenamefont
  {Marini}}]{attaccalite_prb_84}%
  \BibitemOpen
  \bibfield  {author} {\bibinfo {author} {\bibfnamefont {C.}~\bibnamefont
  {Attaccalite}}, \bibinfo {author} {\bibfnamefont {M.}~\bibnamefont
  {Gr\"uning}}, \ and\ \bibinfo {author} {\bibfnamefont {A.}~\bibnamefont
  {Marini}},\ }\href@noop {} {\bibfield  {journal} {\bibinfo  {journal} {Phys.
  Rev. B}\ }\textbf {\bibinfo {volume} {84}},\ \bibinfo {pages} {245110}
  (\bibinfo {year} {2011})}\BibitemShut {NoStop}%
\bibitem [{\citenamefont {Zhang}\ \emph {et~al.}(2013)\citenamefont {Zhang},
  \citenamefont {Steinmann},\ and\ \citenamefont {Yang}}]{zhang_jcp_139}%
  \BibitemOpen
  \bibfield  {author} {\bibinfo {author} {\bibfnamefont {D.}~\bibnamefont
  {Zhang}}, \bibinfo {author} {\bibfnamefont {S.~N.}\ \bibnamefont
  {Steinmann}}, \ and\ \bibinfo {author} {\bibfnamefont {W.}~\bibnamefont
  {Yang}},\ }\href@noop {} {\bibfield  {journal} {\bibinfo  {journal} {J. Chem.
  Phys.}\ }\textbf {\bibinfo {volume} {139}},\ \bibinfo {pages} {154109}
  (\bibinfo {year} {2013})}\BibitemShut {NoStop}%
\bibitem [{\citenamefont {Lischner}\ \emph {et~al.}(2012)\citenamefont
  {Lischner}, \citenamefont {Deslippe}, \citenamefont {Jain},\ and\
  \citenamefont {Louie}}]{lischner_prl_109}%
  \BibitemOpen
  \bibfield  {author} {\bibinfo {author} {\bibfnamefont {J.}~\bibnamefont
  {Lischner}}, \bibinfo {author} {\bibfnamefont {J.}~\bibnamefont {Deslippe}},
  \bibinfo {author} {\bibfnamefont {M.}~\bibnamefont {Jain}}, \ and\ \bibinfo
  {author} {\bibfnamefont {S.~G.}\ \bibnamefont {Louie}},\ }\href@noop {}
  {\bibfield  {journal} {\bibinfo  {journal} {Phys. Rev. Lett.}\ }\textbf
  {\bibinfo {volume} {109}},\ \bibinfo {pages} {036406} (\bibinfo {year}
  {2012})}\BibitemShut {NoStop}%
\bibitem [{\citenamefont {Maggio}\ and\ \citenamefont
  {Kresse}(2017)}]{kresse_jctc_13}%
  \BibitemOpen
  \bibfield  {author} {\bibinfo {author} {\bibfnamefont {E.}~\bibnamefont
  {Maggio}}\ and\ \bibinfo {author} {\bibfnamefont {G.}~\bibnamefont
  {Kresse}},\ }\href@noop {} {\bibfield  {journal} {\bibinfo  {journal} {J.
  Chem. Theory Comput.}\ }\textbf {\bibinfo {volume} {13}},\ \bibinfo {pages}
  {4765} (\bibinfo {year} {2017})}\BibitemShut {NoStop}%
\bibitem [{\citenamefont {Starke}\ and\ \citenamefont
  {Kresse}(2012)}]{kresse_prb_85}%
  \BibitemOpen
  \bibfield  {author} {\bibinfo {author} {\bibfnamefont {R.}~\bibnamefont
  {Starke}}\ and\ \bibinfo {author} {\bibfnamefont {G.}~\bibnamefont
  {Kresse}},\ }\href@noop {} {\bibfield  {journal} {\bibinfo  {journal} {Phys.
  Rev. B}\ }\textbf {\bibinfo {volume} {85}},\ \bibinfo {pages} {075119}
  (\bibinfo {year} {2012})}\BibitemShut {NoStop}%
\bibitem [{\citenamefont {Hung}\ \emph {et~al.}(2016)\citenamefont {Hung},
  \citenamefont {da~Jornada}, \citenamefont {Souto-Casares}, \citenamefont
  {Chelikowsky}, \citenamefont {Louie},\ and\ \citenamefont
  {\"O\ifmmode~\breve{g}\else \u{g}\fi{}\"ut}}]{louie_prb_94}%
  \BibitemOpen
  \bibfield  {author} {\bibinfo {author} {\bibfnamefont {L.}~\bibnamefont
  {Hung}}, \bibinfo {author} {\bibfnamefont {F.~H.}\ \bibnamefont
  {da~Jornada}}, \bibinfo {author} {\bibfnamefont {J.}~\bibnamefont
  {Souto-Casares}}, \bibinfo {author} {\bibfnamefont {J.~R.}\ \bibnamefont
  {Chelikowsky}}, \bibinfo {author} {\bibfnamefont {S.~G.}\ \bibnamefont
  {Louie}}, \ and\ \bibinfo {author} {\bibfnamefont {S.}~\bibnamefont
  {\"O\ifmmode~\breve{g}\else \u{g}\fi{}\"ut}},\ }\href@noop {} {\bibfield
  {journal} {\bibinfo  {journal} {Phys. Rev. B}\ }\textbf {\bibinfo {volume}
  {94}},\ \bibinfo {pages} {085125} (\bibinfo {year} {2016})}\BibitemShut
  {NoStop}%
\bibitem [{\citenamefont {Pavlyukh}\ \emph {et~al.}(2016)\citenamefont
  {Pavlyukh}, \citenamefont {Uimonen}, \citenamefont {Stefanucci},\ and\
  \citenamefont {van Leeuwen}}]{pavlyukh_prl_117}%
  \BibitemOpen
  \bibfield  {author} {\bibinfo {author} {\bibfnamefont {Y.}~\bibnamefont
  {Pavlyukh}}, \bibinfo {author} {\bibfnamefont {A.-M.}\ \bibnamefont
  {Uimonen}}, \bibinfo {author} {\bibfnamefont {G.}~\bibnamefont {Stefanucci}},
  \ and\ \bibinfo {author} {\bibfnamefont {R.}~\bibnamefont {van Leeuwen}},\
  }\href@noop {} {\bibfield  {journal} {\bibinfo  {journal} {Phys. Rev. Lett.}\
  }\textbf {\bibinfo {volume} {117}},\ \bibinfo {pages} {206402} (\bibinfo
  {year} {2016})}\BibitemShut {NoStop}%
\bibitem [{\citenamefont {Isobe}\ \emph {et~al.}(2018)\citenamefont {Isobe},
  \citenamefont {Kuwahara},\ and\ \citenamefont {Ohno}}]{isobe_pra_97}%
  \BibitemOpen
  \bibfield  {author} {\bibinfo {author} {\bibfnamefont {T.}~\bibnamefont
  {Isobe}}, \bibinfo {author} {\bibfnamefont {R.}~\bibnamefont {Kuwahara}}, \
  and\ \bibinfo {author} {\bibfnamefont {K.}~\bibnamefont {Ohno}},\ }\href@noop
  {} {\bibfield  {journal} {\bibinfo  {journal} {Phys. Rev. A}\ }\textbf
  {\bibinfo {volume} {97}},\ \bibinfo {pages} {060502} (\bibinfo {year}
  {2018})}\BibitemShut {NoStop}%
\bibitem [{\citenamefont {Kuwahara}\ \emph {et~al.}(2016)\citenamefont
  {Kuwahara}, \citenamefont {Noguchi},\ and\ \citenamefont
  {Ohno}}]{kwahara_prb_94}%
  \BibitemOpen
  \bibfield  {author} {\bibinfo {author} {\bibfnamefont {R.}~\bibnamefont
  {Kuwahara}}, \bibinfo {author} {\bibfnamefont {Y.}~\bibnamefont {Noguchi}}, \
  and\ \bibinfo {author} {\bibfnamefont {K.}~\bibnamefont {Ohno}},\ }\href@noop
  {} {\bibfield  {journal} {\bibinfo  {journal} {Phys. Rev. B}\ }\textbf
  {\bibinfo {volume} {94}},\ \bibinfo {pages} {121116} (\bibinfo {year}
  {2016})}\BibitemShut {NoStop}%
\bibitem [{\citenamefont {Gr\"uneis}\ \emph {et~al.}(2014)\citenamefont
  {Gr\"uneis}, \citenamefont {Kresse}, \citenamefont {Hinuma},\ and\
  \citenamefont {Oba}}]{gruneis_prl_112}%
  \BibitemOpen
  \bibfield  {author} {\bibinfo {author} {\bibfnamefont {A.}~\bibnamefont
  {Gr\"uneis}}, \bibinfo {author} {\bibfnamefont {G.}~\bibnamefont {Kresse}},
  \bibinfo {author} {\bibfnamefont {Y.}~\bibnamefont {Hinuma}}, \ and\ \bibinfo
  {author} {\bibfnamefont {F.}~\bibnamefont {Oba}},\ }\href@noop {} {\bibfield
  {journal} {\bibinfo  {journal} {Phys. Rev. Lett.}\ }\textbf {\bibinfo
  {volume} {112}},\ \bibinfo {pages} {096401} (\bibinfo {year}
  {2014})}\BibitemShut {NoStop}%
\bibitem [{\citenamefont {Romaniello}\ \emph
  {et~al.}(2009{\natexlab{b}})\citenamefont {Romaniello}, \citenamefont
  {Guyot},\ and\ \citenamefont {Reining}}]{romaniello_jcp_131}%
  \BibitemOpen
  \bibfield  {author} {\bibinfo {author} {\bibfnamefont {P.}~\bibnamefont
  {Romaniello}}, \bibinfo {author} {\bibfnamefont {S.}~\bibnamefont {Guyot}}, \
  and\ \bibinfo {author} {\bibfnamefont {L.}~\bibnamefont {Reining}},\
  }\href@noop {} {\bibfield  {journal} {\bibinfo  {journal} {J. Chem. Phys.}\
  }\textbf {\bibinfo {volume} {131}},\ \bibinfo {pages} {154111} (\bibinfo
  {year} {2009}{\natexlab{b}})}\BibitemShut {NoStop}%
\bibitem [{\citenamefont {Peng}\ and\ \citenamefont
  {Kowalski}(2018)}]{peng_jctc_14}%
  \BibitemOpen
  \bibfield  {author} {\bibinfo {author} {\bibfnamefont {B.}~\bibnamefont
  {Peng}}\ and\ \bibinfo {author} {\bibfnamefont {K.}~\bibnamefont
  {Kowalski}},\ }\href@noop {} {\bibfield  {journal} {\bibinfo  {journal} {J.
  Chem. Theory Comput.}\ }\textbf {\bibinfo {volume} {14}},\ \bibinfo {pages}
  {4335} (\bibinfo {year} {2018})}\BibitemShut {NoStop}%
\bibitem [{\citenamefont {Lange}\ and\ \citenamefont
  {Berkelbach}(2018)}]{berkelbach_jctc_14}%
  \BibitemOpen
  \bibfield  {author} {\bibinfo {author} {\bibfnamefont {M.~F.}\ \bibnamefont
  {Lange}}\ and\ \bibinfo {author} {\bibfnamefont {T.~C.}\ \bibnamefont
  {Berkelbach}},\ }\href@noop {} {\bibfield  {journal} {\bibinfo  {journal} {J.
  Chem. Theory Comput.}\ }\textbf {\bibinfo {volume} {14}},\ \bibinfo {pages}
  {4224} (\bibinfo {year} {2018})}\BibitemShut {NoStop}%
\bibitem [{\citenamefont {Georges}\ and\ \citenamefont
  {Kotliar}(1992)}]{georges_prb_45}%
  \BibitemOpen
  \bibfield  {author} {\bibinfo {author} {\bibfnamefont {A.}~\bibnamefont
  {Georges}}\ and\ \bibinfo {author} {\bibfnamefont {G.}~\bibnamefont
  {Kotliar}},\ }\href@noop {} {\bibfield  {journal} {\bibinfo  {journal} {Phys.
  Rev. B}\ }\textbf {\bibinfo {volume} {45}},\ \bibinfo {pages} {6479}
  (\bibinfo {year} {1992})}\BibitemShut {NoStop}%
\bibitem [{\citenamefont {Held}\ \emph {et~al.}(2008)\citenamefont {Held},
  \citenamefont {Andersen}, \citenamefont {Feldbacher}, \citenamefont
  {Yamasaki},\ and\ \citenamefont {Yang}}]{held_jpcm_20}%
  \BibitemOpen
  \bibfield  {author} {\bibinfo {author} {\bibfnamefont {K.}~\bibnamefont
  {Held}}, \bibinfo {author} {\bibfnamefont {O.~K.}\ \bibnamefont {Andersen}},
  \bibinfo {author} {\bibfnamefont {M.}~\bibnamefont {Feldbacher}}, \bibinfo
  {author} {\bibfnamefont {A.}~\bibnamefont {Yamasaki}}, \ and\ \bibinfo
  {author} {\bibfnamefont {Y.-F.}\ \bibnamefont {Yang}},\ }\href@noop {}
  {\bibfield  {journal} {\bibinfo  {journal} {J Phys: Cond. Matt.}\ }\textbf
  {\bibinfo {volume} {20}},\ \bibinfo {pages} {064202} (\bibinfo {year}
  {2008})}\BibitemShut {NoStop}%
\bibitem [{\citenamefont {Kotliar}\ \emph {et~al.}(2006)\citenamefont
  {Kotliar}, \citenamefont {Savrasov}, \citenamefont {Haule}, \citenamefont
  {Oudovenko}, \citenamefont {Parcollet},\ and\ \citenamefont
  {Marianetti}}]{kotliar_rmp_78}%
  \BibitemOpen
  \bibfield  {author} {\bibinfo {author} {\bibfnamefont {G.}~\bibnamefont
  {Kotliar}}, \bibinfo {author} {\bibfnamefont {S.~Y.}\ \bibnamefont
  {Savrasov}}, \bibinfo {author} {\bibfnamefont {K.}~\bibnamefont {Haule}},
  \bibinfo {author} {\bibfnamefont {V.~S.}\ \bibnamefont {Oudovenko}}, \bibinfo
  {author} {\bibfnamefont {O.}~\bibnamefont {Parcollet}}, \ and\ \bibinfo
  {author} {\bibfnamefont {C.~A.}\ \bibnamefont {Marianetti}},\ }\href@noop {}
  {\bibfield  {journal} {\bibinfo  {journal} {Rev. Mod. Phys.}\ }\textbf
  {\bibinfo {volume} {78}},\ \bibinfo {pages} {865} (\bibinfo {year}
  {2006})}\BibitemShut {NoStop}%
\bibitem [{\citenamefont {Kotliar}\ \emph {et~al.}(2001)\citenamefont
  {Kotliar}, \citenamefont {Savrasov}, \citenamefont {P\'alsson},\ and\
  \citenamefont {Biroli}}]{kotliar_prl_87}%
  \BibitemOpen
  \bibfield  {author} {\bibinfo {author} {\bibfnamefont {G.}~\bibnamefont
  {Kotliar}}, \bibinfo {author} {\bibfnamefont {S.~Y.}\ \bibnamefont
  {Savrasov}}, \bibinfo {author} {\bibfnamefont {G.}~\bibnamefont {P\'alsson}},
  \ and\ \bibinfo {author} {\bibfnamefont {G.}~\bibnamefont {Biroli}},\
  }\href@noop {} {\bibfield  {journal} {\bibinfo  {journal} {Phys. Rev. Lett.}\
  }\textbf {\bibinfo {volume} {87}},\ \bibinfo {pages} {186401} (\bibinfo
  {year} {2001})}\BibitemShut {NoStop}%
\bibitem [{\citenamefont {Maier}\ \emph {et~al.}(2005)\citenamefont {Maier},
  \citenamefont {Jarrell}, \citenamefont {Pruschke},\ and\ \citenamefont
  {Hettler}}]{maier_rmp_77}%
  \BibitemOpen
  \bibfield  {author} {\bibinfo {author} {\bibfnamefont {T.}~\bibnamefont
  {Maier}}, \bibinfo {author} {\bibfnamefont {M.}~\bibnamefont {Jarrell}},
  \bibinfo {author} {\bibfnamefont {T.}~\bibnamefont {Pruschke}}, \ and\
  \bibinfo {author} {\bibfnamefont {M.~H.}\ \bibnamefont {Hettler}},\
  }\href@noop {} {\bibfield  {journal} {\bibinfo  {journal} {Rev. Mod. Phys.}\
  }\textbf {\bibinfo {volume} {77}},\ \bibinfo {pages} {1027} (\bibinfo {year}
  {2005})}\BibitemShut {NoStop}%
\bibitem [{\citenamefont {Biermann}\ \emph {et~al.}(2003)\citenamefont
  {Biermann}, \citenamefont {Aryasetiawan},\ and\ \citenamefont
  {Georges}}]{biermann_prl_90}%
  \BibitemOpen
  \bibfield  {author} {\bibinfo {author} {\bibfnamefont {S.}~\bibnamefont
  {Biermann}}, \bibinfo {author} {\bibfnamefont {F.}~\bibnamefont
  {Aryasetiawan}}, \ and\ \bibinfo {author} {\bibfnamefont {A.}~\bibnamefont
  {Georges}},\ }\href@noop {} {\bibfield  {journal} {\bibinfo  {journal} {Phys.
  Rev. Lett.}\ }\textbf {\bibinfo {volume} {90}},\ \bibinfo {pages} {086402}
  (\bibinfo {year} {2003})}\BibitemShut {NoStop}%
\bibitem [{\citenamefont {Tomczak}\ \emph {et~al.}(2012)\citenamefont
  {Tomczak}, \citenamefont {Casula}, \citenamefont {Miyake}, \citenamefont
  {Aryasetiawan},\ and\ \citenamefont {Biermann}}]{tomczak_epl_100}%
  \BibitemOpen
  \bibfield  {author} {\bibinfo {author} {\bibfnamefont {J.~M.}\ \bibnamefont
  {Tomczak}}, \bibinfo {author} {\bibfnamefont {M.}~\bibnamefont {Casula}},
  \bibinfo {author} {\bibfnamefont {T.}~\bibnamefont {Miyake}}, \bibinfo
  {author} {\bibfnamefont {F.}~\bibnamefont {Aryasetiawan}}, \ and\ \bibinfo
  {author} {\bibfnamefont {S.}~\bibnamefont {Biermann}},\ }\href@noop {}
  {\bibfield  {journal} {\bibinfo  {journal} {EPL (Europhysics Letters)}\
  }\textbf {\bibinfo {volume} {100}},\ \bibinfo {pages} {67001} (\bibinfo
  {year} {2012})}\BibitemShut {NoStop}%
\bibitem [{\citenamefont {Biermann}(2014)}]{biermann_jpcm_26}%
  \BibitemOpen
  \bibfield  {author} {\bibinfo {author} {\bibfnamefont {S.}~\bibnamefont
  {Biermann}},\ }\href@noop {} {\bibfield  {journal} {\bibinfo  {journal} {J.
  of Phys: Cond. Matt.}\ }\textbf {\bibinfo {volume} {26}},\ \bibinfo {pages}
  {173202} (\bibinfo {year} {2014})}\BibitemShut {NoStop}%
\bibitem [{\citenamefont {Boehnke}\ \emph {et~al.}(2016)\citenamefont
  {Boehnke}, \citenamefont {Nilsson}, \citenamefont {Aryasetiawan},\ and\
  \citenamefont {Werner}}]{boehnke_prb_94}%
  \BibitemOpen
  \bibfield  {author} {\bibinfo {author} {\bibfnamefont {L.}~\bibnamefont
  {Boehnke}}, \bibinfo {author} {\bibfnamefont {F.}~\bibnamefont {Nilsson}},
  \bibinfo {author} {\bibfnamefont {F.}~\bibnamefont {Aryasetiawan}}, \ and\
  \bibinfo {author} {\bibfnamefont {P.}~\bibnamefont {Werner}},\ }\href@noop {}
  {\bibfield  {journal} {\bibinfo  {journal} {Phys. Rev. B}\ }\textbf {\bibinfo
  {volume} {94}},\ \bibinfo {pages} {201106} (\bibinfo {year}
  {2016})}\BibitemShut {NoStop}%
\bibitem [{\citenamefont {Chibani}\ \emph {et~al.}(2016)\citenamefont
  {Chibani}, \citenamefont {Ren}, \citenamefont {Scheffler},\ and\
  \citenamefont {Rinke}}]{rinke_prb_93}%
  \BibitemOpen
  \bibfield  {author} {\bibinfo {author} {\bibfnamefont {W.}~\bibnamefont
  {Chibani}}, \bibinfo {author} {\bibfnamefont {X.}~\bibnamefont {Ren}},
  \bibinfo {author} {\bibfnamefont {M.}~\bibnamefont {Scheffler}}, \ and\
  \bibinfo {author} {\bibfnamefont {P.}~\bibnamefont {Rinke}},\ }\href@noop {}
  {\bibfield  {journal} {\bibinfo  {journal} {Phys. Rev. B}\ }\textbf {\bibinfo
  {volume} {93}},\ \bibinfo {pages} {165106} (\bibinfo {year}
  {2016})}\BibitemShut {NoStop}%
\bibitem [{\citenamefont {Aryasetiawan}\ \emph {et~al.}(2009)\citenamefont
  {Aryasetiawan}, \citenamefont {Tomczak}, \citenamefont {Miyake},\ and\
  \citenamefont {Sakuma}}]{aryasetiawan_prl_102}%
  \BibitemOpen
  \bibfield  {author} {\bibinfo {author} {\bibfnamefont {F.}~\bibnamefont
  {Aryasetiawan}}, \bibinfo {author} {\bibfnamefont {J.~M.}\ \bibnamefont
  {Tomczak}}, \bibinfo {author} {\bibfnamefont {T.}~\bibnamefont {Miyake}}, \
  and\ \bibinfo {author} {\bibfnamefont {R.}~\bibnamefont {Sakuma}},\
  }\href@noop {} {\bibfield  {journal} {\bibinfo  {journal} {Phys. Rev. Lett.}\
  }\textbf {\bibinfo {volume} {102}},\ \bibinfo {pages} {176402} (\bibinfo
  {year} {2009})}\BibitemShut {NoStop}%
\bibitem [{\citenamefont {Zgid}\ and\ \citenamefont
  {Gull}(2017)}]{zgid_njp_19}%
  \BibitemOpen
  \bibfield  {author} {\bibinfo {author} {\bibfnamefont {D.}~\bibnamefont
  {Zgid}}\ and\ \bibinfo {author} {\bibfnamefont {E.}~\bibnamefont {Gull}},\
  }\href@noop {} {\bibfield  {journal} {\bibinfo  {journal} {New J. Phys.}\
  }\textbf {\bibinfo {volume} {19}},\ \bibinfo {pages} {023047} (\bibinfo
  {year} {2017})}\BibitemShut {NoStop}%
\bibitem [{\citenamefont {Lan}\ \emph {et~al.}(2015)\citenamefont {Lan},
  \citenamefont {Kananenka},\ and\ \citenamefont {Zgid}}]{zgid_jcp_143}%
  \BibitemOpen
  \bibfield  {author} {\bibinfo {author} {\bibfnamefont {T.~N.}\ \bibnamefont
  {Lan}}, \bibinfo {author} {\bibfnamefont {A.~A.}\ \bibnamefont {Kananenka}},
  \ and\ \bibinfo {author} {\bibfnamefont {D.}~\bibnamefont {Zgid}},\
  }\href@noop {} {\bibfield  {journal} {\bibinfo  {journal} {J. Chem. Phys.}\
  }\textbf {\bibinfo {volume} {143}},\ \bibinfo {pages} {241102} (\bibinfo
  {year} {2015})}\BibitemShut {NoStop}%
\bibitem [{\citenamefont {Kananenka}\ \emph {et~al.}(2015)\citenamefont
  {Kananenka}, \citenamefont {Gull},\ and\ \citenamefont {Zgid}}]{zgid_prb_91}%
  \BibitemOpen
  \bibfield  {author} {\bibinfo {author} {\bibfnamefont {A.~A.}\ \bibnamefont
  {Kananenka}}, \bibinfo {author} {\bibfnamefont {E.}~\bibnamefont {Gull}}, \
  and\ \bibinfo {author} {\bibfnamefont {D.}~\bibnamefont {Zgid}},\ }\href@noop
  {} {\bibfield  {journal} {\bibinfo  {journal} {Phys. Rev. B}\ }\textbf
  {\bibinfo {volume} {91}},\ \bibinfo {pages} {121111} (\bibinfo {year}
  {2015})}\BibitemShut {NoStop}%
\bibitem [{\citenamefont {Nguyen~Lan}\ \emph {et~al.}(2016)\citenamefont
  {Nguyen~Lan}, \citenamefont {Kananenka},\ and\ \citenamefont
  {Zgid}}]{lan_jctc_12}%
  \BibitemOpen
  \bibfield  {author} {\bibinfo {author} {\bibfnamefont {T.}~\bibnamefont
  {Nguyen~Lan}}, \bibinfo {author} {\bibfnamefont {A.~A.}\ \bibnamefont
  {Kananenka}}, \ and\ \bibinfo {author} {\bibfnamefont {D.}~\bibnamefont
  {Zgid}},\ }\href@noop {} {\bibfield  {journal} {\bibinfo  {journal} {J. Chem.
  Theory Comput.}\ }\textbf {\bibinfo {volume} {12}},\ \bibinfo {pages} {4856}
  (\bibinfo {year} {2016})}\BibitemShut {NoStop}%
\bibitem [{\citenamefont {Szabo}\ and\ \citenamefont
  {Ostlund}(1996)}]{szabo_qchem}%
  \BibitemOpen
  \bibfield  {author} {\bibinfo {author} {\bibfnamefont {A.}~\bibnamefont
  {Szabo}}\ and\ \bibinfo {author} {\bibfnamefont {N.~S.}\ \bibnamefont
  {Ostlund}},\ }\href@noop {} {\emph {\bibinfo {title} {Modern Quantum
  Chemistry: Introduction to Advanced Electronic Structure Theory}}},\ \bibinfo
  {edition} {1st}\ ed.\ (\bibinfo  {publisher} {Dover Publications, Inc.},\
  \bibinfo {address} {Mineola},\ \bibinfo {year} {1996})\BibitemShut {NoStop}%
\bibitem [{\citenamefont {Helgaker}\ \emph {et~al.}(2014)\citenamefont
  {Helgaker}, \citenamefont {J{\o}rgensen},\ and\ \citenamefont
  {Olsen}}]{helgaker_molecular}%
  \BibitemOpen
  \bibfield  {author} {\bibinfo {author} {\bibfnamefont {T.}~\bibnamefont
  {Helgaker}}, \bibinfo {author} {\bibfnamefont {P.}~\bibnamefont
  {J{\o}rgensen}}, \ and\ \bibinfo {author} {\bibfnamefont {J.}~\bibnamefont
  {Olsen}},\ }\href@noop {} {\emph {\bibinfo {title} {Molecular electronic
  structure theory}}},\ \bibinfo {edition} {1st}\ ed.\ (\bibinfo  {publisher}
  {John Wiley \& Sons, Ltd},\ \bibinfo {year} {2014})\BibitemShut {NoStop}%
\bibitem [{\citenamefont {Loos}\ \emph {et~al.}(2018)\citenamefont {Loos},
  \citenamefont {Scemama}, \citenamefont {Blondel}, \citenamefont {Garniron},
  \citenamefont {Caffarel},\ and\ \citenamefont {Jacquemin}}]{loos_jctc_14}%
  \BibitemOpen
  \bibfield  {author} {\bibinfo {author} {\bibfnamefont {P.-F.}\ \bibnamefont
  {Loos}}, \bibinfo {author} {\bibfnamefont {A.}~\bibnamefont {Scemama}},
  \bibinfo {author} {\bibfnamefont {A.}~\bibnamefont {Blondel}}, \bibinfo
  {author} {\bibfnamefont {Y.}~\bibnamefont {Garniron}}, \bibinfo {author}
  {\bibfnamefont {M.}~\bibnamefont {Caffarel}}, \ and\ \bibinfo {author}
  {\bibfnamefont {D.}~\bibnamefont {Jacquemin}},\ }\href@noop {} {\bibfield
  {journal} {\bibinfo  {journal} {J. Chem. Theory Comput.}\ }\textbf {\bibinfo
  {volume} {14}},\ \bibinfo {pages} {4360} (\bibinfo {year}
  {2018})}\BibitemShut {NoStop}%
\bibitem [{\citenamefont {Evangelista}(2011)}]{evangelista_jcp_134}%
  \BibitemOpen
  \bibfield  {author} {\bibinfo {author} {\bibfnamefont {F.~A.}\ \bibnamefont
  {Evangelista}},\ }\href@noop {} {\bibfield  {journal} {\bibinfo  {journal}
  {J. Chem. Phys.}\ }\textbf {\bibinfo {volume} {134}},\ \bibinfo {pages}
  {224102} (\bibinfo {year} {2011})}\BibitemShut {NoStop}%
\bibitem [{\citenamefont {Szalay}\ \emph {et~al.}(2012)\citenamefont {Szalay},
  \citenamefont {M\"uller}, \citenamefont {Gidofalvi}, \citenamefont
  {Lischka},\ and\ \citenamefont {Shepard}}]{szalay_cr_112}%
  \BibitemOpen
  \bibfield  {author} {\bibinfo {author} {\bibfnamefont {P.~G.}\ \bibnamefont
  {Szalay}}, \bibinfo {author} {\bibfnamefont {T.}~\bibnamefont {M\"uller}},
  \bibinfo {author} {\bibfnamefont {G.}~\bibnamefont {Gidofalvi}}, \bibinfo
  {author} {\bibfnamefont {H.}~\bibnamefont {Lischka}}, \ and\ \bibinfo
  {author} {\bibfnamefont {R.}~\bibnamefont {Shepard}},\ }\href@noop {}
  {\bibfield  {journal} {\bibinfo  {journal} {Chemical Reviews}\ }\textbf
  {\bibinfo {volume} {112}},\ \bibinfo {pages} {108} (\bibinfo {year}
  {2012})}\BibitemShut {NoStop}%
\bibitem [{\citenamefont {Olsen}(2011)}]{olsen_ijqc_111}%
  \BibitemOpen
  \bibfield  {author} {\bibinfo {author} {\bibfnamefont {J.}~\bibnamefont
  {Olsen}},\ }\href@noop {} {\bibfield  {journal} {\bibinfo  {journal} {Int. J.
  Quan. Chem.}\ }\textbf {\bibinfo {volume} {111}},\ \bibinfo {pages} {3267}
  (\bibinfo {year} {2011})}\BibitemShut {NoStop}%
\bibitem [{\citenamefont {Li~Manni}\ \emph {et~al.}(2016)\citenamefont
  {Li~Manni}, \citenamefont {Smart},\ and\ \citenamefont
  {Alavi}}]{alavi_jctc_12}%
  \BibitemOpen
  \bibfield  {author} {\bibinfo {author} {\bibfnamefont {G.}~\bibnamefont
  {Li~Manni}}, \bibinfo {author} {\bibfnamefont {S.~D.}\ \bibnamefont {Smart}},
  \ and\ \bibinfo {author} {\bibfnamefont {A.}~\bibnamefont {Alavi}},\
  }\href@noop {} {\bibfield  {journal} {\bibinfo  {journal} {J. Chem. Theory
  Comput.}\ }\textbf {\bibinfo {volume} {12}},\ \bibinfo {pages} {1245}
  (\bibinfo {year} {2016})}\BibitemShut {NoStop}%
\bibitem [{\citenamefont {Choe}\ \emph {et~al.}(2001)\citenamefont {Choe},
  \citenamefont {Witek}, \citenamefont {Finley},\ and\ \citenamefont
  {Hirao}}]{choe_jcp_114}%
  \BibitemOpen
  \bibfield  {author} {\bibinfo {author} {\bibfnamefont {Y.-K.}\ \bibnamefont
  {Choe}}, \bibinfo {author} {\bibfnamefont {H.~A.}\ \bibnamefont {Witek}},
  \bibinfo {author} {\bibfnamefont {J.~P.}\ \bibnamefont {Finley}}, \ and\
  \bibinfo {author} {\bibfnamefont {K.}~\bibnamefont {Hirao}},\ }\href@noop {}
  {\bibfield  {journal} {\bibinfo  {journal} {J. Chem. Phys.}\ }\textbf
  {\bibinfo {volume} {114}},\ \bibinfo {pages} {3913} (\bibinfo {year}
  {2001})}\BibitemShut {NoStop}%
\bibitem [{\citenamefont {Ma}\ \emph {et~al.}(2016)\citenamefont {Ma},
  \citenamefont {Li~Manni}, \citenamefont {Olsen},\ and\ \citenamefont
  {Gagliardi}}]{ma_jctc_12}%
  \BibitemOpen
  \bibfield  {author} {\bibinfo {author} {\bibfnamefont {D.}~\bibnamefont
  {Ma}}, \bibinfo {author} {\bibfnamefont {G.}~\bibnamefont {Li~Manni}},
  \bibinfo {author} {\bibfnamefont {J.}~\bibnamefont {Olsen}}, \ and\ \bibinfo
  {author} {\bibfnamefont {L.}~\bibnamefont {Gagliardi}},\ }\href@noop {}
  {\bibfield  {journal} {\bibinfo  {journal} {J. Chem. Theory Comput.}\
  }\textbf {\bibinfo {volume} {12}},\ \bibinfo {pages} {3208} (\bibinfo {year}
  {2016})}\BibitemShut {NoStop}%
\bibitem [{\citenamefont {Manni}\ \emph {et~al.}(2011)\citenamefont {Manni},
  \citenamefont {Aquilante},\ and\ \citenamefont
  {Gagliardi}}]{gagliardi_jcp_134}%
  \BibitemOpen
  \bibfield  {author} {\bibinfo {author} {\bibfnamefont {G.~L.}\ \bibnamefont
  {Manni}}, \bibinfo {author} {\bibfnamefont {F.}~\bibnamefont {Aquilante}}, \
  and\ \bibinfo {author} {\bibfnamefont {L.}~\bibnamefont {Gagliardi}},\
  }\href@noop {} {\bibfield  {journal} {\bibinfo  {journal} {J. Chem. Phys.}\
  }\textbf {\bibinfo {volume} {134}},\ \bibinfo {pages} {034114} (\bibinfo
  {year} {2011})}\BibitemShut {NoStop}%
\bibitem [{\citenamefont {Li~Manni}\ \emph {et~al.}(2013)\citenamefont
  {Li~Manni}, \citenamefont {Ma}, \citenamefont {Aquilante}, \citenamefont
  {Olsen},\ and\ \citenamefont {Gagliardi}}]{gagliardi_jctc_9}%
  \BibitemOpen
  \bibfield  {author} {\bibinfo {author} {\bibfnamefont {G.}~\bibnamefont
  {Li~Manni}}, \bibinfo {author} {\bibfnamefont {D.}~\bibnamefont {Ma}},
  \bibinfo {author} {\bibfnamefont {F.}~\bibnamefont {Aquilante}}, \bibinfo
  {author} {\bibfnamefont {J.}~\bibnamefont {Olsen}}, \ and\ \bibinfo {author}
  {\bibfnamefont {L.}~\bibnamefont {Gagliardi}},\ }\href@noop {} {\bibfield
  {journal} {\bibinfo  {journal} {J. Chem. Theory Comput.}\ }\textbf {\bibinfo
  {volume} {9}},\ \bibinfo {pages} {3375} (\bibinfo {year} {2013})}\BibitemShut
  {NoStop}%
\bibitem [{\citenamefont {Zgid}\ \emph {et~al.}(2012)\citenamefont {Zgid},
  \citenamefont {Gull},\ and\ \citenamefont {Chan}}]{zgid_prb_86}%
  \BibitemOpen
  \bibfield  {author} {\bibinfo {author} {\bibfnamefont {D.}~\bibnamefont
  {Zgid}}, \bibinfo {author} {\bibfnamefont {E.}~\bibnamefont {Gull}}, \ and\
  \bibinfo {author} {\bibfnamefont {G.~K.-L.}\ \bibnamefont {Chan}},\
  }\href@noop {} {\bibfield  {journal} {\bibinfo  {journal} {Phys. Rev. B}\
  }\textbf {\bibinfo {volume} {86}},\ \bibinfo {pages} {165128} (\bibinfo
  {year} {2012})}\BibitemShut {NoStop}%
\bibitem [{\citenamefont {Pavlyukh}\ and\ \citenamefont
  {H\"ubner}(2007)}]{pavlyukh_prb_75}%
  \BibitemOpen
  \bibfield  {author} {\bibinfo {author} {\bibfnamefont {Y.}~\bibnamefont
  {Pavlyukh}}\ and\ \bibinfo {author} {\bibfnamefont {W.}~\bibnamefont
  {H\"ubner}},\ }\href@noop {} {\bibfield  {journal} {\bibinfo  {journal}
  {Phys. Rev. B}\ }\textbf {\bibinfo {volume} {75}},\ \bibinfo {pages} {205129}
  (\bibinfo {year} {2007})}\BibitemShut {NoStop}%
\bibitem [{\citenamefont {Dzuba}\ \emph {et~al.}(1996)\citenamefont {Dzuba},
  \citenamefont {Flambaum},\ and\ \citenamefont {Kozlov}}]{dzuba_pra_54}%
  \BibitemOpen
  \bibfield  {author} {\bibinfo {author} {\bibfnamefont {V.~A.}\ \bibnamefont
  {Dzuba}}, \bibinfo {author} {\bibfnamefont {V.~V.}\ \bibnamefont {Flambaum}},
  \ and\ \bibinfo {author} {\bibfnamefont {M.~G.}\ \bibnamefont {Kozlov}},\
  }\href@noop {} {\bibfield  {journal} {\bibinfo  {journal} {Phys. Rev. A}\
  }\textbf {\bibinfo {volume} {54}},\ \bibinfo {pages} {3948} (\bibinfo {year}
  {1996})}\BibitemShut {NoStop}%
\bibitem [{\citenamefont {Dzuba}\ \emph {et~al.}(2017)\citenamefont {Dzuba},
  \citenamefont {Berengut}, \citenamefont {Harabati},\ and\ \citenamefont
  {Flambaum}}]{dzuba_pra_95}%
  \BibitemOpen
  \bibfield  {author} {\bibinfo {author} {\bibfnamefont {V.~A.}\ \bibnamefont
  {Dzuba}}, \bibinfo {author} {\bibfnamefont {J.~C.}\ \bibnamefont {Berengut}},
  \bibinfo {author} {\bibfnamefont {C.}~\bibnamefont {Harabati}}, \ and\
  \bibinfo {author} {\bibfnamefont {V.~V.}\ \bibnamefont {Flambaum}},\
  }\href@noop {} {\bibfield  {journal} {\bibinfo  {journal} {Phys. Rev. A}\
  }\textbf {\bibinfo {volume} {95}},\ \bibinfo {pages} {012503} (\bibinfo
  {year} {2017})}\BibitemShut {NoStop}%
\bibitem [{\citenamefont {Pavlyukh}\ \emph {et~al.}(2015)\citenamefont
  {Pavlyukh}, \citenamefont {Sch\"uler},\ and\ \citenamefont
  {Berakdar}}]{pavlyukh_prb_91}%
  \BibitemOpen
  \bibfield  {author} {\bibinfo {author} {\bibfnamefont {Y.}~\bibnamefont
  {Pavlyukh}}, \bibinfo {author} {\bibfnamefont {M.}~\bibnamefont {Sch\"uler}},
  \ and\ \bibinfo {author} {\bibfnamefont {J.}~\bibnamefont {Berakdar}},\
  }\href@noop {} {\bibfield  {journal} {\bibinfo  {journal} {Phys. Rev. B}\
  }\textbf {\bibinfo {volume} {91}},\ \bibinfo {pages} {155116} (\bibinfo
  {year} {2015})}\BibitemShut {NoStop}%
\bibitem [{\citenamefont {L\"owdin}(1962)}]{lowdin_jmp_3}%
  \BibitemOpen
  \bibfield  {author} {\bibinfo {author} {\bibfnamefont {P.}~\bibnamefont
  {L\"owdin}},\ }\href@noop {} {\bibfield  {journal} {\bibinfo  {journal} {J.
  Math. Phys.}\ }\textbf {\bibinfo {volume} {3}},\ \bibinfo {pages} {969}
  (\bibinfo {year} {1962})}\BibitemShut {NoStop}%
\bibitem [{\citenamefont {L\"owdin}(1968)}]{lowdin_ijqc_2}%
  \BibitemOpen
  \bibfield  {author} {\bibinfo {author} {\bibfnamefont {P.}~\bibnamefont
  {L\"owdin}},\ }\href@noop {} {\bibfield  {journal} {\bibinfo  {journal} {Int.
  J. Quan. Chem.}\ }\textbf {\bibinfo {volume} {2}},\ \bibinfo {pages} {867}
  (\bibinfo {year} {1968})}\BibitemShut {NoStop}%
\bibitem [{\citenamefont {Slater}(1929)}]{slater_pr_34}%
  \BibitemOpen
  \bibfield  {author} {\bibinfo {author} {\bibfnamefont {J.~C.}\ \bibnamefont
  {Slater}},\ }\href@noop {} {\bibfield  {journal} {\bibinfo  {journal} {Phys.
  Rev.}\ }\textbf {\bibinfo {volume} {34}},\ \bibinfo {pages} {1293} (\bibinfo
  {year} {1929})}\BibitemShut {NoStop}%
\bibitem [{\citenamefont {Condon}(1930)}]{condon_pr_36}%
  \BibitemOpen
  \bibfield  {author} {\bibinfo {author} {\bibfnamefont {E.~U.}\ \bibnamefont
  {Condon}},\ }\href@noop {} {\bibfield  {journal} {\bibinfo  {journal} {Phys.
  Rev.}\ }\textbf {\bibinfo {volume} {36}},\ \bibinfo {pages} {1121} (\bibinfo
  {year} {1930})}\BibitemShut {NoStop}%
\bibitem [{\citenamefont {Jacquemin}\ \emph {et~al.}(2015)\citenamefont
  {Jacquemin}, \citenamefont {Duchemin},\ and\ \citenamefont
  {Blase}}]{jacquemin_jctc_11}%
  \BibitemOpen
  \bibfield  {author} {\bibinfo {author} {\bibfnamefont {D.}~\bibnamefont
  {Jacquemin}}, \bibinfo {author} {\bibfnamefont {I.}~\bibnamefont {Duchemin}},
  \ and\ \bibinfo {author} {\bibfnamefont {X.}~\bibnamefont {Blase}},\
  }\href@noop {} {\bibfield  {journal} {\bibinfo  {journal} {J. Chem. Theory
  Comput.}\ }\textbf {\bibinfo {volume} {11}},\ \bibinfo {pages} {3290}
  (\bibinfo {year} {2015})}\BibitemShut {NoStop}%
\bibitem [{\citenamefont {Tiago}\ and\ \citenamefont
  {Chelikowsky}(2006)}]{chelikowsky_prb_73}%
  \BibitemOpen
  \bibfield  {author} {\bibinfo {author} {\bibfnamefont {M.~L.}\ \bibnamefont
  {Tiago}}\ and\ \bibinfo {author} {\bibfnamefont {J.~R.}\ \bibnamefont
  {Chelikowsky}},\ }\href@noop {} {\bibfield  {journal} {\bibinfo  {journal}
  {Phys. Rev. B}\ }\textbf {\bibinfo {volume} {73}},\ \bibinfo {pages} {205334}
  (\bibinfo {year} {2006})}\BibitemShut {NoStop}%
\bibitem [{\citenamefont {K\"orbel}\ \emph {et~al.}(2014)\citenamefont
  {K\"orbel}, \citenamefont {Boulanger}, \citenamefont {Duchemin},
  \citenamefont {Blase}, \citenamefont {Marques},\ and\ \citenamefont
  {Botti}}]{botti_jctc_10}%
  \BibitemOpen
  \bibfield  {author} {\bibinfo {author} {\bibfnamefont {S.}~\bibnamefont
  {K\"orbel}}, \bibinfo {author} {\bibfnamefont {P.}~\bibnamefont {Boulanger}},
  \bibinfo {author} {\bibfnamefont {I.}~\bibnamefont {Duchemin}}, \bibinfo
  {author} {\bibfnamefont {X.}~\bibnamefont {Blase}}, \bibinfo {author}
  {\bibfnamefont {M.~A.~L.}\ \bibnamefont {Marques}}, \ and\ \bibinfo {author}
  {\bibfnamefont {S.}~\bibnamefont {Botti}},\ }\href@noop {} {\bibfield
  {journal} {\bibinfo  {journal} {J. Chem. Theory Comput.}\ }\textbf {\bibinfo
  {volume} {10}},\ \bibinfo {pages} {3934} (\bibinfo {year}
  {2014})}\BibitemShut {NoStop}%
\bibitem [{\citenamefont {Onida}\ \emph {et~al.}(2002)\citenamefont {Onida},
  \citenamefont {Reining},\ and\ \citenamefont {Rubio}}]{reining_rmp_74}%
  \BibitemOpen
  \bibfield  {author} {\bibinfo {author} {\bibfnamefont {G.}~\bibnamefont
  {Onida}}, \bibinfo {author} {\bibfnamefont {L.}~\bibnamefont {Reining}}, \
  and\ \bibinfo {author} {\bibfnamefont {A.}~\bibnamefont {Rubio}},\
  }\href@noop {} {\bibfield  {journal} {\bibinfo  {journal} {Rev. Mod. Phys.}\
  }\textbf {\bibinfo {volume} {74}},\ \bibinfo {pages} {601} (\bibinfo {year}
  {2002})}\BibitemShut {NoStop}%
\bibitem [{\citenamefont {Aryasetiawan}\ \emph {et~al.}(2004)\citenamefont
  {Aryasetiawan}, \citenamefont {Imada}, \citenamefont {Georges}, \citenamefont
  {Kotliar}, \citenamefont {Biermann},\ and\ \citenamefont
  {Lichtenstein}}]{aryasetiawan_prb_70}%
  \BibitemOpen
  \bibfield  {author} {\bibinfo {author} {\bibfnamefont {F.}~\bibnamefont
  {Aryasetiawan}}, \bibinfo {author} {\bibfnamefont {M.}~\bibnamefont {Imada}},
  \bibinfo {author} {\bibfnamefont {A.}~\bibnamefont {Georges}}, \bibinfo
  {author} {\bibfnamefont {G.}~\bibnamefont {Kotliar}}, \bibinfo {author}
  {\bibfnamefont {S.}~\bibnamefont {Biermann}}, \ and\ \bibinfo {author}
  {\bibfnamefont {A.~I.}\ \bibnamefont {Lichtenstein}},\ }\href@noop {}
  {\bibfield  {journal} {\bibinfo  {journal} {Phys. Rev. B}\ }\textbf {\bibinfo
  {volume} {70}},\ \bibinfo {pages} {195104} (\bibinfo {year}
  {2004})}\BibitemShut {NoStop}%
\bibitem [{\citenamefont {\ifmmode \mbox{\c{S}}\else \c{S}\fi{}a\ifmmode
  \mbox{\c{s}}\else \c{s}\fi{}\ifmmode \imath \else \i
  \fi{}o\ifmmode~\breve{g}\else \u{g}\fi{}lu}\ \emph
  {et~al.}(2011)\citenamefont {\ifmmode \mbox{\c{S}}\else \c{S}\fi{}a\ifmmode
  \mbox{\c{s}}\else \c{s}\fi{}\ifmmode \imath \else \i
  \fi{}o\ifmmode~\breve{g}\else \u{g}\fi{}lu}, \citenamefont {Friedrich},\ and\
  \citenamefont {Bl\"ugel}}]{sasioglu_prb_83}%
  \BibitemOpen
  \bibfield  {author} {\bibinfo {author} {\bibfnamefont {E.}~\bibnamefont
  {\ifmmode \mbox{\c{S}}\else \c{S}\fi{}a\ifmmode \mbox{\c{s}}\else
  \c{s}\fi{}\ifmmode \imath \else \i \fi{}o\ifmmode~\breve{g}\else
  \u{g}\fi{}lu}}, \bibinfo {author} {\bibfnamefont {C.}~\bibnamefont
  {Friedrich}}, \ and\ \bibinfo {author} {\bibfnamefont {S.}~\bibnamefont
  {Bl\"ugel}},\ }\href@noop {} {\bibfield  {journal} {\bibinfo  {journal}
  {Phys. Rev. B}\ }\textbf {\bibinfo {volume} {83}},\ \bibinfo {pages} {121101}
  (\bibinfo {year} {2011})}\BibitemShut {NoStop}%
\bibitem [{\citenamefont {Vaugier}\ \emph {et~al.}(2012)\citenamefont
  {Vaugier}, \citenamefont {Jiang},\ and\ \citenamefont
  {Biermann}}]{biermann_prb_86}%
  \BibitemOpen
  \bibfield  {author} {\bibinfo {author} {\bibfnamefont {L.}~\bibnamefont
  {Vaugier}}, \bibinfo {author} {\bibfnamefont {H.}~\bibnamefont {Jiang}}, \
  and\ \bibinfo {author} {\bibfnamefont {S.}~\bibnamefont {Biermann}},\
  }\href@noop {} {\bibfield  {journal} {\bibinfo  {journal} {Phys. Rev. B}\
  }\textbf {\bibinfo {volume} {86}},\ \bibinfo {pages} {165105} (\bibinfo
  {year} {2012})}\BibitemShut {NoStop}%
\bibitem [{\citenamefont {Hirayama}\ \emph {et~al.}(2017)\citenamefont
  {Hirayama}, \citenamefont {Miyake}, \citenamefont {Imada},\ and\
  \citenamefont {Biermann}}]{biermann_prb_96}%
  \BibitemOpen
  \bibfield  {author} {\bibinfo {author} {\bibfnamefont {M.}~\bibnamefont
  {Hirayama}}, \bibinfo {author} {\bibfnamefont {T.}~\bibnamefont {Miyake}},
  \bibinfo {author} {\bibfnamefont {M.}~\bibnamefont {Imada}}, \ and\ \bibinfo
  {author} {\bibfnamefont {S.}~\bibnamefont {Biermann}},\ }\href@noop {}
  {\bibfield  {journal} {\bibinfo  {journal} {Phys. Rev. B}\ }\textbf {\bibinfo
  {volume} {96}},\ \bibinfo {pages} {075102} (\bibinfo {year}
  {2017})}\BibitemShut {NoStop}%
\bibitem [{\citenamefont {Shinaoka}\ \emph {et~al.}(2015)\citenamefont
  {Shinaoka}, \citenamefont {Troyer},\ and\ \citenamefont
  {Werner}}]{werner_prb_91}%
  \BibitemOpen
  \bibfield  {author} {\bibinfo {author} {\bibfnamefont {H.}~\bibnamefont
  {Shinaoka}}, \bibinfo {author} {\bibfnamefont {M.}~\bibnamefont {Troyer}}, \
  and\ \bibinfo {author} {\bibfnamefont {P.}~\bibnamefont {Werner}},\
  }\href@noop {} {\bibfield  {journal} {\bibinfo  {journal} {Phys. Rev. B}\
  }\textbf {\bibinfo {volume} {91}},\ \bibinfo {pages} {245156} (\bibinfo
  {year} {2015})}\BibitemShut {NoStop}%
\bibitem [{\citenamefont {Martin}\ \emph {et~al.}(2016)\citenamefont {Martin},
  \citenamefont {Reining},\ and\ \citenamefont
  {Ceperley}}]{martin_reining_ceperley_2016}%
  \BibitemOpen
  \bibfield  {author} {\bibinfo {author} {\bibfnamefont {R.~M.}\ \bibnamefont
  {Martin}}, \bibinfo {author} {\bibfnamefont {L.}~\bibnamefont {Reining}}, \
  and\ \bibinfo {author} {\bibfnamefont {D.~M.}\ \bibnamefont {Ceperley}},\
  }\href {\doibase 10.1017/CBO9781139050807} {\emph {\bibinfo {title}
  {Interacting Electrons: Theory and Computational Approaches}}}\ (\bibinfo
  {publisher} {Cambridge University Press},\ \bibinfo {year}
  {2016})\BibitemShut {NoStop}%
\bibitem [{\citenamefont {Deilmann}\ \emph {et~al.}(2016)\citenamefont
  {Deilmann}, \citenamefont {Dr\"uppel},\ and\ \citenamefont
  {Rohlfing}}]{rohlfing_prl_116}%
  \BibitemOpen
  \bibfield  {author} {\bibinfo {author} {\bibfnamefont {T.}~\bibnamefont
  {Deilmann}}, \bibinfo {author} {\bibfnamefont {M.}~\bibnamefont {Dr\"uppel}},
  \ and\ \bibinfo {author} {\bibfnamefont {M.}~\bibnamefont {Rohlfing}},\
  }\href@noop {} {\bibfield  {journal} {\bibinfo  {journal} {Phys. Rev. Lett.}\
  }\textbf {\bibinfo {volume} {116}},\ \bibinfo {pages} {196804} (\bibinfo
  {year} {2016})}\BibitemShut {NoStop}%
\bibitem [{\citenamefont {Deilmann}\ and\ \citenamefont
  {Thygesen}(2017)}]{deilmann_prb_96}%
  \BibitemOpen
  \bibfield  {author} {\bibinfo {author} {\bibfnamefont {T.}~\bibnamefont
  {Deilmann}}\ and\ \bibinfo {author} {\bibfnamefont {K.~S.}\ \bibnamefont
  {Thygesen}},\ }\href@noop {} {\bibfield  {journal} {\bibinfo  {journal}
  {Phys. Rev. B}\ }\textbf {\bibinfo {volume} {96}},\ \bibinfo {pages} {201113}
  (\bibinfo {year} {2017})}\BibitemShut {NoStop}%
\bibitem [{\citenamefont {Blum}\ \emph {et~al.}(2009)\citenamefont {Blum},
  \citenamefont {Gehrke}, \citenamefont {Hanke}, \citenamefont {Havu},
  \citenamefont {Havu}, \citenamefont {Ren}, \citenamefont {Reuter},\ and\
  \citenamefont {Scheffler}}]{blum_cpc_180}%
  \BibitemOpen
  \bibfield  {author} {\bibinfo {author} {\bibfnamefont {V.}~\bibnamefont
  {Blum}}, \bibinfo {author} {\bibfnamefont {R.}~\bibnamefont {Gehrke}},
  \bibinfo {author} {\bibfnamefont {F.}~\bibnamefont {Hanke}}, \bibinfo
  {author} {\bibfnamefont {P.}~\bibnamefont {Havu}}, \bibinfo {author}
  {\bibfnamefont {V.}~\bibnamefont {Havu}}, \bibinfo {author} {\bibfnamefont
  {X.}~\bibnamefont {Ren}}, \bibinfo {author} {\bibfnamefont {K.}~\bibnamefont
  {Reuter}}, \ and\ \bibinfo {author} {\bibfnamefont {M.}~\bibnamefont
  {Scheffler}},\ }\href@noop {} {\bibfield  {journal} {\bibinfo  {journal}
  {Comp. Phys. Comm.}\ }\textbf {\bibinfo {volume} {180}},\ \bibinfo {pages}
  {2175 } (\bibinfo {year} {2009})}\BibitemShut {NoStop}%
\bibitem [{\citenamefont {Ren}\ \emph {et~al.}(2012)\citenamefont {Ren},
  \citenamefont {Rinke}, \citenamefont {Blum}, \citenamefont {Wieferink},
  \citenamefont {Tkatchenko}, \citenamefont {Sanfilippo}, \citenamefont
  {Reuter},\ and\ \citenamefont {Scheffler}}]{ren_njp_14}%
  \BibitemOpen
  \bibfield  {author} {\bibinfo {author} {\bibfnamefont {X.}~\bibnamefont
  {Ren}}, \bibinfo {author} {\bibfnamefont {P.}~\bibnamefont {Rinke}}, \bibinfo
  {author} {\bibfnamefont {V.}~\bibnamefont {Blum}}, \bibinfo {author}
  {\bibfnamefont {J.}~\bibnamefont {Wieferink}}, \bibinfo {author}
  {\bibfnamefont {A.}~\bibnamefont {Tkatchenko}}, \bibinfo {author}
  {\bibfnamefont {A.}~\bibnamefont {Sanfilippo}}, \bibinfo {author}
  {\bibfnamefont {K.}~\bibnamefont {Reuter}}, \ and\ \bibinfo {author}
  {\bibfnamefont {M.}~\bibnamefont {Scheffler}},\ }\href@noop {} {\bibfield
  {journal} {\bibinfo  {journal} {N. J. Phys.}\ }\textbf {\bibinfo {volume}
  {14}},\ \bibinfo {pages} {053020} (\bibinfo {year} {2012})}\BibitemShut
  {NoStop}%
\bibitem [{\citenamefont {Ihrig}\ \emph {et~al.}(2015)\citenamefont {Ihrig},
  \citenamefont {Wieferink}, \citenamefont {Zhang}, \citenamefont {Ropo},
  \citenamefont {Ren}, \citenamefont {Rinke}, \citenamefont {Scheffler},\ and\
  \citenamefont {Blum}}]{ihrig_njp_17}%
  \BibitemOpen
  \bibfield  {author} {\bibinfo {author} {\bibfnamefont {A.~C.}\ \bibnamefont
  {Ihrig}}, \bibinfo {author} {\bibfnamefont {J.}~\bibnamefont {Wieferink}},
  \bibinfo {author} {\bibfnamefont {I.~Y.}\ \bibnamefont {Zhang}}, \bibinfo
  {author} {\bibfnamefont {M.}~\bibnamefont {Ropo}}, \bibinfo {author}
  {\bibfnamefont {X.}~\bibnamefont {Ren}}, \bibinfo {author} {\bibfnamefont
  {P.}~\bibnamefont {Rinke}}, \bibinfo {author} {\bibfnamefont
  {M.}~\bibnamefont {Scheffler}}, \ and\ \bibinfo {author} {\bibfnamefont
  {V.}~\bibnamefont {Blum}},\ }\href@noop {} {\bibfield  {journal} {\bibinfo
  {journal} {N. J. Phys.}\ }\textbf {\bibinfo {volume} {17}},\ \bibinfo {pages}
  {093020} (\bibinfo {year} {2015})}\BibitemShut {NoStop}%
\bibitem [{\citenamefont {Levchenko}\ \emph {et~al.}(2015)\citenamefont
  {Levchenko}, \citenamefont {Ren}, \citenamefont {Wieferink}, \citenamefont
  {Johanni}, \citenamefont {Rinke}, \citenamefont {Blum},\ and\ \citenamefont
  {Scheffler}}]{levchenko_cpc_192}%
  \BibitemOpen
  \bibfield  {author} {\bibinfo {author} {\bibfnamefont {S.~V.}\ \bibnamefont
  {Levchenko}}, \bibinfo {author} {\bibfnamefont {X.}~\bibnamefont {Ren}},
  \bibinfo {author} {\bibfnamefont {J.}~\bibnamefont {Wieferink}}, \bibinfo
  {author} {\bibfnamefont {R.}~\bibnamefont {Johanni}}, \bibinfo {author}
  {\bibfnamefont {P.}~\bibnamefont {Rinke}}, \bibinfo {author} {\bibfnamefont
  {V.}~\bibnamefont {Blum}}, \ and\ \bibinfo {author} {\bibfnamefont
  {M.}~\bibnamefont {Scheffler}},\ }\href@noop {} {\bibfield  {journal}
  {\bibinfo  {journal} {Comp. Phys. Comm.}\ }\textbf {\bibinfo {volume}
  {192}},\ \bibinfo {pages} {60 } (\bibinfo {year} {2015})}\BibitemShut
  {NoStop}%
\bibitem [{\citenamefont {Olsen}\ and\ \citenamefont
  {Thygesen}(2014)}]{olsen_jcp_140}%
  \BibitemOpen
  \bibfield  {author} {\bibinfo {author} {\bibfnamefont {T.}~\bibnamefont
  {Olsen}}\ and\ \bibinfo {author} {\bibfnamefont {K.~S.}\ \bibnamefont
  {Thygesen}},\ }\href@noop {} {\bibfield  {journal} {\bibinfo  {journal} {J.
  Chem. Phys.}\ }\textbf {\bibinfo {volume} {140}},\ \bibinfo {pages} {164116}
  (\bibinfo {year} {2014})}\BibitemShut {NoStop}%
\bibitem [{\citenamefont {Ren}\ \emph {et~al.}(2013)\citenamefont {Ren},
  \citenamefont {Rinke}, \citenamefont {Scuseria},\ and\ \citenamefont
  {Scheffler}}]{ren_prb_88}%
  \BibitemOpen
  \bibfield  {author} {\bibinfo {author} {\bibfnamefont {X.}~\bibnamefont
  {Ren}}, \bibinfo {author} {\bibfnamefont {P.}~\bibnamefont {Rinke}}, \bibinfo
  {author} {\bibfnamefont {G.~E.}\ \bibnamefont {Scuseria}}, \ and\ \bibinfo
  {author} {\bibfnamefont {M.}~\bibnamefont {Scheffler}},\ }\href@noop {}
  {\bibfield  {journal} {\bibinfo  {journal} {Phys. Rev. B}\ }\textbf {\bibinfo
  {volume} {88}},\ \bibinfo {pages} {035120} (\bibinfo {year}
  {2013})}\BibitemShut {NoStop}%
\bibitem [{\citenamefont {Wolniewicz}(1993)}]{wolniewicz_jcp_99}%
  \BibitemOpen
  \bibfield  {author} {\bibinfo {author} {\bibfnamefont {L.}~\bibnamefont
  {Wolniewicz}},\ }\href@noop {} {\bibfield  {journal} {\bibinfo  {journal}
  {The Journal of Chemical Physics}\ }\textbf {\bibinfo {volume} {99}},\
  \bibinfo {pages} {1851} (\bibinfo {year} {1993})}\BibitemShut {NoStop}%
\bibitem [{\citenamefont {Hellgren}\ \emph {et~al.}(2015)\citenamefont
  {Hellgren}, \citenamefont {Caruso}, \citenamefont {Rohr}, \citenamefont
  {Ren}, \citenamefont {Rubio}, \citenamefont {Scheffler},\ and\ \citenamefont
  {Rinke}}]{hellgren_prb_91}%
  \BibitemOpen
  \bibfield  {author} {\bibinfo {author} {\bibfnamefont {M.}~\bibnamefont
  {Hellgren}}, \bibinfo {author} {\bibfnamefont {F.}~\bibnamefont {Caruso}},
  \bibinfo {author} {\bibfnamefont {D.~R.}\ \bibnamefont {Rohr}}, \bibinfo
  {author} {\bibfnamefont {X.}~\bibnamefont {Ren}}, \bibinfo {author}
  {\bibfnamefont {A.}~\bibnamefont {Rubio}}, \bibinfo {author} {\bibfnamefont
  {M.}~\bibnamefont {Scheffler}}, \ and\ \bibinfo {author} {\bibfnamefont
  {P.}~\bibnamefont {Rinke}},\ }\href@noop {} {\bibfield  {journal} {\bibinfo
  {journal} {Phys. Rev. B}\ }\textbf {\bibinfo {volume} {91}},\ \bibinfo
  {pages} {165110} (\bibinfo {year} {2015})}\BibitemShut {NoStop}%
\bibitem [{\citenamefont {Caruso}\ \emph {et~al.}(2013)\citenamefont {Caruso},
  \citenamefont {Rohr}, \citenamefont {Hellgren}, \citenamefont {Ren},
  \citenamefont {Rinke}, \citenamefont {Rubio},\ and\ \citenamefont
  {Scheffler}}]{caruso_prl_110}%
  \BibitemOpen
  \bibfield  {author} {\bibinfo {author} {\bibfnamefont {F.}~\bibnamefont
  {Caruso}}, \bibinfo {author} {\bibfnamefont {D.~R.}\ \bibnamefont {Rohr}},
  \bibinfo {author} {\bibfnamefont {M.}~\bibnamefont {Hellgren}}, \bibinfo
  {author} {\bibfnamefont {X.}~\bibnamefont {Ren}}, \bibinfo {author}
  {\bibfnamefont {P.}~\bibnamefont {Rinke}}, \bibinfo {author} {\bibfnamefont
  {A.}~\bibnamefont {Rubio}}, \ and\ \bibinfo {author} {\bibfnamefont
  {M.}~\bibnamefont {Scheffler}},\ }\href@noop {} {\bibfield  {journal}
  {\bibinfo  {journal} {Phys. Rev. Lett.}\ }\textbf {\bibinfo {volume} {110}},\
  \bibinfo {pages} {146403} (\bibinfo {year} {2013})}\BibitemShut {NoStop}%
\bibitem [{\citenamefont {Zhang}\ \emph {et~al.}(2016)\citenamefont {Zhang},
  \citenamefont {Rinke}, \citenamefont {Perdew},\ and\ \citenamefont
  {Scheffler}}]{zhang_prl_117}%
  \BibitemOpen
  \bibfield  {author} {\bibinfo {author} {\bibfnamefont {I.~Y.}\ \bibnamefont
  {Zhang}}, \bibinfo {author} {\bibfnamefont {P.}~\bibnamefont {Rinke}},
  \bibinfo {author} {\bibfnamefont {J.~P.}\ \bibnamefont {Perdew}}, \ and\
  \bibinfo {author} {\bibfnamefont {M.}~\bibnamefont {Scheffler}},\ }\href@noop
  {} {\bibfield  {journal} {\bibinfo  {journal} {Phys. Rev. Lett.}\ }\textbf
  {\bibinfo {volume} {117}},\ \bibinfo {pages} {133002} (\bibinfo {year}
  {2016})}\BibitemShut {NoStop}%
\bibitem [{\citenamefont {Hirose}\ \emph {et~al.}(2015)\citenamefont {Hirose},
  \citenamefont {Noguchi},\ and\ \citenamefont {Sugino}}]{hirose_prb_91}%
  \BibitemOpen
  \bibfield  {author} {\bibinfo {author} {\bibfnamefont {D.}~\bibnamefont
  {Hirose}}, \bibinfo {author} {\bibfnamefont {Y.}~\bibnamefont {Noguchi}}, \
  and\ \bibinfo {author} {\bibfnamefont {O.}~\bibnamefont {Sugino}},\
  }\href@noop {} {\bibfield  {journal} {\bibinfo  {journal} {Phys. Rev. B}\
  }\textbf {\bibinfo {volume} {91}},\ \bibinfo {pages} {205111} (\bibinfo
  {year} {2015})}\BibitemShut {NoStop}%
\bibitem [{\citenamefont {Larsen}\ \emph {et~al.}(2000)\citenamefont {Larsen},
  \citenamefont {Olsen}, \citenamefont {J{\o}rgensen},\ and\ \citenamefont
  {Christiansen}}]{larsen_jcp_113}%
  \BibitemOpen
  \bibfield  {author} {\bibinfo {author} {\bibfnamefont {H.}~\bibnamefont
  {Larsen}}, \bibinfo {author} {\bibfnamefont {J.}~\bibnamefont {Olsen}},
  \bibinfo {author} {\bibfnamefont {P.}~\bibnamefont {J{\o}rgensen}}, \ and\
  \bibinfo {author} {\bibfnamefont {O.}~\bibnamefont {Christiansen}},\
  }\href@noop {} {\bibfield  {journal} {\bibinfo  {journal} {J. Chem. Phys.}\
  }\textbf {\bibinfo {volume} {113}},\ \bibinfo {pages} {6677} (\bibinfo {year}
  {2000})}\BibitemShut {NoStop}%
\bibitem [{\citenamefont {Dunning}(1989)}]{dunning_jcp_90}%
  \BibitemOpen
  \bibfield  {author} {\bibinfo {author} {\bibfnamefont {T.~H.}\ \bibnamefont
  {Dunning}},\ }\href@noop {} {\bibfield  {journal} {\bibinfo  {journal} {J.
  Chem. Phys.}\ }\textbf {\bibinfo {volume} {90}},\ \bibinfo {pages} {1007}
  (\bibinfo {year} {1989})}\BibitemShut {NoStop}%
\bibitem [{\citenamefont {Stanton}\ \emph {et~al.}(1995)\citenamefont
  {Stanton}, \citenamefont {Gauss}, \citenamefont {Ishikawa},\ and\
  \citenamefont {Head{-}Gordon}}]{head-gordon_jcp_103}%
  \BibitemOpen
  \bibfield  {author} {\bibinfo {author} {\bibfnamefont {J.~F.}\ \bibnamefont
  {Stanton}}, \bibinfo {author} {\bibfnamefont {J.}~\bibnamefont {Gauss}},
  \bibinfo {author} {\bibfnamefont {N.}~\bibnamefont {Ishikawa}}, \ and\
  \bibinfo {author} {\bibfnamefont {M.}~\bibnamefont {Head{-}Gordon}},\
  }\href@noop {} {\bibfield  {journal} {\bibinfo  {journal} {J. Chem. Phys.}\
  }\textbf {\bibinfo {volume} {103}},\ \bibinfo {pages} {4160} (\bibinfo {year}
  {1995})}\BibitemShut {NoStop}%
\bibitem [{\citenamefont {Oddershede}\ \emph {et~al.}(1985)\citenamefont
  {Oddershede}, \citenamefont {Gruner},\ and\ \citenamefont
  {Diercksen}}]{oddershede_cp_97}%
  \BibitemOpen
  \bibfield  {author} {\bibinfo {author} {\bibfnamefont {J.}~\bibnamefont
  {Oddershede}}, \bibinfo {author} {\bibfnamefont {N.~E.}\ \bibnamefont
  {Gruner}}, \ and\ \bibinfo {author} {\bibfnamefont {G.~H.}\ \bibnamefont
  {Diercksen}},\ }\href@noop {} {\bibfield  {journal} {\bibinfo  {journal}
  {Chem. Phys.}\ }\textbf {\bibinfo {volume} {97}},\ \bibinfo {pages} {303 }
  (\bibinfo {year} {1985})}\BibitemShut {NoStop}%
\bibitem [{\citenamefont {Dvorak}\ \emph {et~al.}(2018)\citenamefont {Dvorak},
  \citenamefont {Golze},\ and\ \citenamefont {Rinke}}]{dvorak_prl}%
  \BibitemOpen
  \bibfield  {author} {\bibinfo {author} {\bibfnamefont {M.}~\bibnamefont
  {Dvorak}}, \bibinfo {author} {\bibfnamefont {D.}~\bibnamefont {Golze}}, \
  and\ \bibinfo {author} {\bibfnamefont {P.}~\bibnamefont {Rinke}},\ }\href
  {arXiv.com} {\bibfield  {journal} {\bibinfo  {journal} {arXiv}\ } (\bibinfo
  {year} {2018})}\BibitemShut {NoStop}%
\end{thebibliography}%

\end{document}